%% file: thesis.tex
\documentclass{wsu-thesis}
\pdfoutput=1
\usepackage{graphicx}
\graphicspath{{./../figures/}}
\usepackage{tensor}
\usepackage{siunitx}
\usepackage{float}
\usepackage[numbers]{natbib}
\usepackage{xcolor} 

\title{Scalable and Cost-Effective Data Flow \\
Analysis for Distributed Software: \\
Algorithms and Applications}
\author{Xiaoqin Fu}
\thesistype{Dissertation} 
\degree{Doctor of Philosophy}
\department{School of Electrical Engineering and Computer Science}
\submitmonth{July}
\submityear{2022} 

\newcommand{\phd}{Ph.D.}

\newcommand{\referencedir}{./references}

\usepackage{verbatim}

\usepackage{makecell}

\usepackage{wrapfig}
\usepackage{pifont}
\newcommand{\xmark}{\ding{55}}%

\newcommand{\distmeasure}{\textsc{DistMeasure}}
\newcommand{\distfax}{\textsc{DistFax}}
\newcommand{\flowdist}{\mbox{\textsc{FlowDist}}}
\newcommand{\disttaint}{\mbox{\textsc{DistTaint}}}
\newcommand{\seads}{\mbox{\textsc{Seads}}}
\newcommand{\dads}{\mbox{\textsc{Dads}}}

\newcommand{\diveronline}{\mbox{\textsc{DiverOnline}}}
\newcommand{\diapro}{\textsc{DiaPro}}
\newcommand{\diver}{\mbox{\textsc{Diver}}}
\newcommand{\distia}{\mbox{\textsc{DistIa}}}
\newcommand{\distea}{\mbox{\textsc{DistEa}}}

\newcommand{\pieas}{\textsc{PI/EAS}}

\newcommand{\doda}{\mbox{\textsc{Doda}}}

\newcommand{\flowdistsim}{\mbox{\textsc{FlowDist$sim$}}}
\newcommand{\flowdistmul}{\mbox{\textsc{FlowDist$mul$}}}

\usepackage{multirow}
\usepackage{xcolor}
\definecolor{mygray}{gray}{.9}
\definecolor{mypink}{rgb}{.99,.91,.95}
\definecolor{mycyan}{cmyk}{.3,0,0,0}

\usepackage{algorithm,algpseudocode}
\usepackage{algorithmicx}
\usepackage{caption}
\usepackage{geometry}
\mathchardef\mhyphen="2D

\begin{document}
\maketitle

\begin{signaturepage}
  \signature{Haipeng Cai}{\phd}[Chair]
  \signature{Zhe Dang}{\phd}
  \signature{Dingwen Tao}{\phd}
\end{signaturepage} 

\begin{acknowledgements}
  \input{\frontmatterdir/acknowledgements}
\end{acknowledgements} 

\begin{abstract}{\phd}{Haipeng Cai}
  \input{\frontmatterdir/abstract}

\end{abstract} 
 
\tableofcontents
\listoffigures
\listoftables

\begin{dedication}
  \input{\frontmatterdir/dedication}
\end{dedication}

\begin{mainchapters}
  \input{\chapterdir/chapter1}
  \input{\chapterdir/chapter2}

  \input{\chapterdir/chapter3}

  \input{\chapterdir/chapter4}

  \input{\chapterdir/chapter5}
  \input{\chapterdir/chapter6}

\end{mainchapters}


\makebibliography{\referencedir/references}


\end{document}

%% file: front-matter/acknowledgements.tex
I want to use this space to express my deepest gratitude and appreciation to those who have made it possible for me to complete my PhD.

First and foremost, I would like to thank my advisor and the chair of my committee, Dr. Haipeng Cai.
He has been supportive of me throughout my WSU pursuit of a PhD in Computer Science.
He has continuously encouraged me throughout my doctoral studies, as a foundation of strength during difficult times.
Without his care, help, and support, I would never have finished my studies. He is a true advisor and has become a truthful friend.

I am grateful to Dr. Zhe Dang for his assistance throughout this journey.
In the past four years, I learned from him about how to become a successful scientist.
He will be my role model in my future research and career.
I wish to thank my committee member Dr. Dingwen Tao, who agreed to be a part of this journey with me.
He is more than generous with his expertise and precious time throughout the whole process.

I would also like to thank the other faculty and staff of the School of Electrical Engineering \& Computer Science at Washington State University for their support and encouragement throughout this process.
The chair, Professor Partha Pratim Pande, is very kind and supportive.
Dr. Noel Schulz, Dr. Venera Arnaoudova, Dr. Sakire Arslan Ay, Dr. Kung-Chi Wang, Dr. David Bakken, Dr. Janardhan Rao Doppa, and Dr. Assefaw Gebremedhin, they have treated me as a friend and colleague.
I would like to particularly thank Jessica Cross for her encouragement, suggestion, and unconditional support from the first day I met her.
A special thanks goes to Dr. Li Li at Monash University, for his generous assistance to my research and career.

Finally, I want to acknowledge the National Science Foundation (NSF) for the grant, number CCF-1936522, supporting me to pursue my doctoral research.

%% file: front-matter/abstract.tex
More and more distributed software systems are being developed and deployed today.
Like other software, distributed software systems also need very strong quality assurance support.
Distributed software is often very large/complex, has distributed components, and does not have a global clock.
All these characteristics make it very challenging to analyze the information flow of such systems to support the software quality assurance.
One challenge is that existing dynamic analysis techniques hardly scale to large distributed software systems in the real world.
It is also challenging to develop cost-effective dynamic analysis approaches.
There are also applicability and portability challenges for dynamic analysis algorithms/applications of distributed software.
  
My dissertation addresses these challenges via three novel approaches to data flow analysis for distributed software.
My first approach is based on measuring interprocess communications to understand distributed software behaviors and predict distributed software quality.
Then, I developed a particular approach that can actually pinpoint sensitive information via multi-staged and refinement-based dynamic information flow analysis for distributed software.
Finally, I explored dynamic dependence analysis for distributed systems, utilizing reinforcement learning to automatically adjust analysis configurations for scalability and better cost-effectiveness tradeoffs.

%% file: front-matter/dedication.tex
\center\emph{dedicated to ...}

\center{My supportive wife, parents, and friends.}

%% file: chapters/chapter1.tex
\chapter{Introduction}
Due to increasing requirements for computational scalability and performance, there are more and more real-world software systems designed as {\em distributed}
today~\cite{Coulouris2011DSC}.
Software systems, which perform general-purpose distributed computations, were defined in~\cite{Coulouris2011DSC}, called {\em common} distributed systems.
Rather than common distributed systems, there are specialized distributed systems, such as RMI-based systems~\cite{sharp2006static}, distributed event-based (DEB) systems~\cite{muhl2006distributed}, cloud systems~\cite{madakam2015internet}, Internet of Things (IoT) systems~\cite{wang2010distributed}, etc.

Compared with single-process programs, distributed software systems have multiple unique characteristics:
\begin{enumerate}
\item Components/processes in a distributed system are concurrent.
\item Multiple components/processes are generally autonomous in nature.
\item Hardware and software resources can be shared.
\item Distributed software typically lacks a global clock.
\item The price/performance ratio may be much better.
\end{enumerate}

On the other hand, distributed software systems widely serve critical application domains (e.g., aviation, banking, medical, social media).

Like other software systems, distributed software also needs very strong quality assurance support (e.g., performance monitoring, program optimization, software testing, vulnerability detection~\cite{kreindl2019towards}), where the quality includes security and other factors, such as performance efficiency, maintainability, and functional suitability~\cite{iso25010}.
However, the characteristics of distributed systems not only complicate security issues~\cite{top10v,cvesite,cve20140085,cve20153254,cve20175637} but also bring about severe challenges for analyzing the data flow (i.e., information flows) in these systems.

One of the major challenges is that existing data analysis techniques (e.g., \cite{newsome2005dynamic,cheng2006tainttrace,qin2006lift,ming2015taintpipe,dfsan,she2020neutaint,ashouri2019hybrid,banerjee2019iodine}) for single-process programs, scarcely scaled to distributed systems (e.g., Apache Zookeeper~\cite{zookeeper}, a distributed coordination service
for achieving synchronization and consistency as used by Apache Hadoop and Yahoo).
The reason is that they rely on explicit dependencies among program entities and dismiss implicit dependencies across processes~\cite{fu2019measuring}.

Developing a cost-effective dynamic analysis is also challenging, since the cost and effectiveness often counteract and compete within a specific analysis.
Developing such an analysis approach for most distributed systems in the real world is even more challenging because of their
typically larger scale and greater complexity than single-process programs.
Nondeterministic, varying, and typically unbounded executions of distributed systems further exacerbate such challenges.

The main goal of this dissertation is to explore and study data flow analysis approaches for distributed software security, overcoming scalability, cost-effectiveness, and other (i.e., applicability and portability) challenges.
In general, this work provides fundamental support for quality assurance of distributed software.
In particular, this work aims to predict and understand the quality of distributed software systems, related to their run-time behaviors and execution dynamics.
Moreover, this work mainly targets dynamic information flow security, dynamic program dependencies, and corresponding applications that could achieve
practical scalability and balance analysis cost-effectiveness via various ways,
such as a principled, multi-phase analysis strategy, a reinforcement learning strategy, and two self-adaptation strategies utilizing optimal reinforcement learning and deep reinforcement learning, respectively.

\section{Motivation and Problem Statement}
Among a variety of security threats (e.g., code injection) distributed software suffers, a major type lies in
assorted vulnerabilities in information flow paths in distributed programs.
In these programs, sensitive information (e.g., username or password) might leak and cause serious losses/damage.
To defend against such information flow threats, it is crucial to check sensitive data that passes throughout the entire system (across its distributed components and corresponding processes).
Effective information flow analysis (i.e., data flow analysis) often requires fine-grained (e.g., statement-level) computation of control and data flow paths.
However, precise, fine-grained data flow analysis is usually very expensive.
The great complexity of distributed systems is a major reason that most existing relevant approaches are not applicable (e.g., due to scalability barriers) or very limited utility (e.g., only for single components/processes).
For many distributed systems (e.g., online/cloud services) that are normally running continuously, it is desirable to
keep monitoring them against security threats. In these scenarios, scalability can be even more difficult to achieve and maintain.

Moreover, besides the {\em scalability} problem, there are multiple additional issues of developing data flow analysis solutions for distributed software systems.
For example, developing a {\em cost-effective} dynamic dependence analysis, however, is challenging, especially given the known
substantial overheads of dynamic analysis in general.
Prior research has demonstrated the difficulties and complexity of balancing the cost and effectiveness
in dynamic dependence analysis for single-process programs~\cite{cai2016diapro}.
Executive non-determinism, the variety of and uncertainties in run-time environments in the real world,
and the unbounded executions (due to their continuously-running nature) further exacerbate such challenges for most distributed software systems.

In addition, traditional dynamic analysis approaches are hardly {\em applicable} to multi-process programs, such as distributed systems.
The reason is that
they
rely on explicit dependencies among program entities and
dismiss
implicit dependencies across processes~\cite{fu2019measuring}.

A few existing dynamic analysis tools (e.g.,~\cite{jiang2017kakute,sun2016pileus}) overcome their {\em applicability challenge} by working at system level with platform customization.
Yet these tools typically face a {\em portability challenge} due to their customization with diverse and rapidly evolving platforms, which would be time-consuming and even infeasible~\cite{fu2021flowdist}.

\section{Contributions}
\begin{figure}[htbp]
  \centering
    \captionsetup{justification=centering}
	\caption{The overview of my research}
  \includegraphics[width=0.8\textwidth]{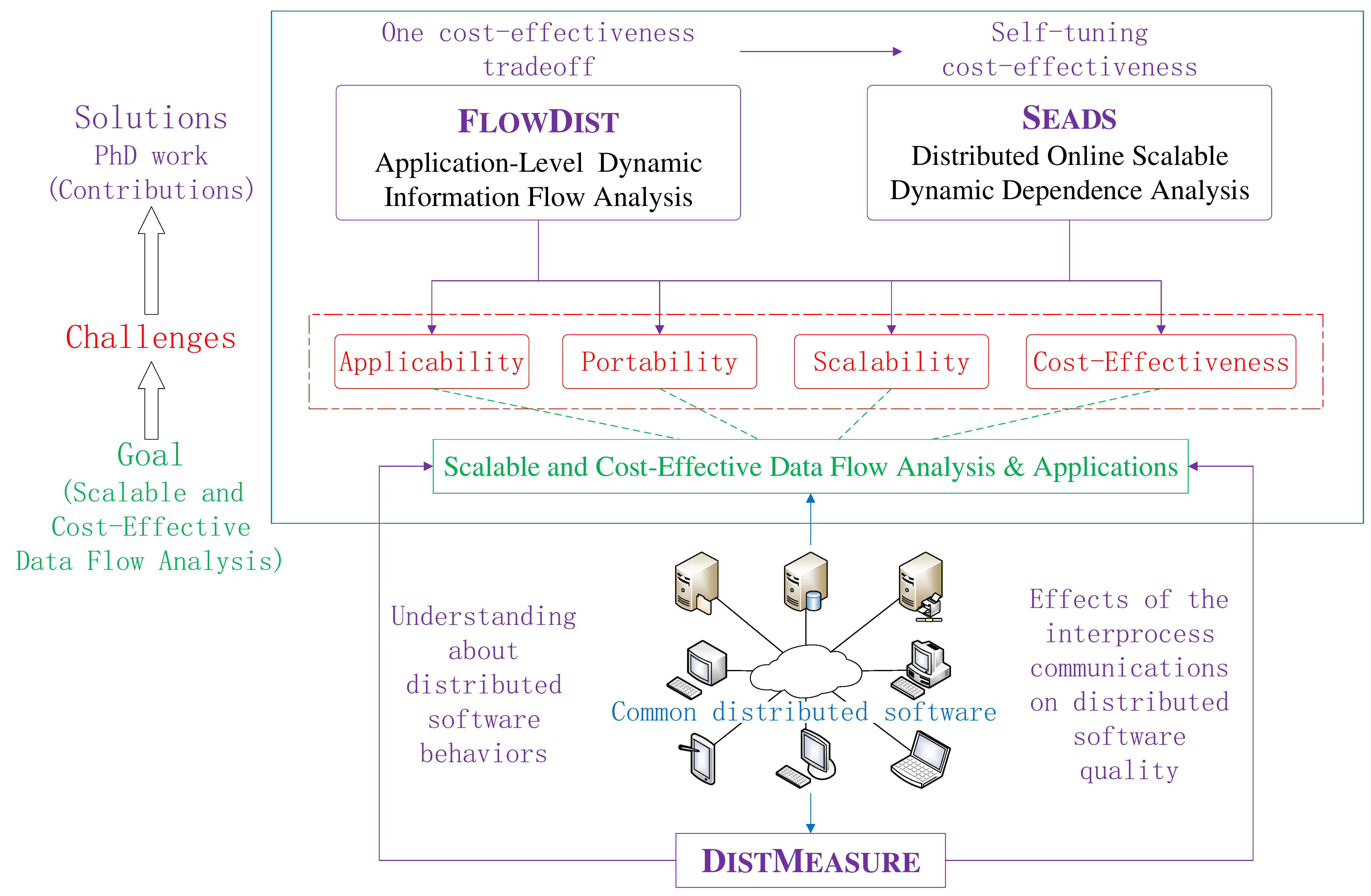}
  \label{fig:projects}
\end{figure}
In this dissertation, I explored how to design realistic solutions to deal with the scalability, cost-effectiveness, and other (i.e., applicability and portability) challenges in data flow analysis for large and complex distributed software systems.
Accordingly, as depicted in Figure~\ref{fig:projects}, my research has been focused on three connected themes: {\distmeasure}, {\disttaint}, and {\seads}.

First, I studied {\distmeasure} (the expansion of \cite{fu2019measuring}), including a novel set of interprocess communications (IPC) metrics to measure common distributed systems for understanding system behaviors and effects of system IPCs. Based on {\distmeasure}, DistFax~\cite{fu2022distfax}, a toolkit for measuring IPCs and quality of distributed software, was implemented.
Using {\distmeasure}, I demonstrated the usefulness and practicality of IPC metrics against 11 real-world distributed software systems and their diverse execution scenarios. My experiments revealed that higher IPC coupling between distributed processes tended to be negatively associated with distributed software quality,
while individual processes' cohesion gave its positive quality implications. The evaluation of {\distmeasure}'s learning-based quality-level classification showed promising merits of IPC measurement for understanding distributed system behaviors in terms of their statistical and predictive relationships with various aspects of the quality of distributed software.

Second, I explored a dynamic information flow analysis framework {\flowdist}~\cite{fu2021flowdist} that overcomes multiple technical (applicability, portability, and scalability) challenges through a principled multi-phase analysis scheme, scaling traditional dynamic information flow analysis to distributed systems. 
The corresponding tool, {\disttaint} \cite{fu2019dynamic}, a dynamic taint analyzer for distributed software, was implemented.
The evaluation of 12 real-world distributed systems against two baselines revealed the superior effectiveness, practical efficiency and scalability of {\flowdist}.
It found 24 existing vulnerabilities and 24 new vulnerabilities, 17 of which were confirmed and 2 of which were fixed. 
Two alternative designs of {\flowdist} for diverse subject accommodations were also presented and evaluated.

Since {\flowdist} provides only one cost-effectiveness tradeoff,
I developed {\seads} \cite{fu2020seads} that is a distributed and online dynamic dependence analysis framework for continuously running distributed systems, offering
self-tuning cost-effectiveness (tradeoffs).
This was my attempt and the first step to achieve better scalability and cost-effectiveness on the fly through automatically and continually adjusting analysis configurations during the execution(s), using reinforcement learning, according to previous configurations, corresponding costs, and user-defined budgets.
I also implemented {\dads} \cite{fu2020dads}, as a distributed and online dynamic slicer for continuously-running distributed software with respect to user-specified budget constraints, achieving and maintaining practical scalability and cost-effectiveness tradeoffs
according to given budgets on analysis time by continually and automatically adjusting analysis configurations on the fly via reinforcement learning.
The empirical results revealed the scalability and efficiency advantages of {\seads} over a conventional dynamic analysis approach, at least for computing dynamic dependencies at method level.

\section{Dissertation Organization}
In the rest of this dissertation, I discuss some techniques and key concepts of my research in Chapter 2 ($\S$\ref{ch:background}).
Based on measuring interprocess communications of distributed systems, I start with {\distmeasure} (the expansion of \cite{fu2019measuring}) for understanding/predicting the behaviors and quality of distributed software, in Chapter 3 ($\S$\ref{ch:distmeasure}).
Then, I present multi-Staged refinement-based dynamic information flow analysis {\flowdist}~\cite{fu2021flowdist} for distributed software, and scalable/cost-effective dynamic dependence analysis {\seads}~\cite{fu2020seads} of distributed systems via reinforcement learning, in Chapters 4 ($\S$\ref{ch:flowdist}) and 5 ($\S$\ref{ch:seads}), respectively.
Lastly, Chapter 6 ($\S$\ref{ch:fwk}) summarizes this dissertation and discusses several research directions for the future.

%% file: chapters/chapter2.tex
\chapter{Background} \label{ch:background}
In the section, I discuss some techniques and key concepts of my research, including {\em software metrics}, {\em distributed system architectures}, {\em logic clocks in distributed systems}, {\em dependence analysis}, {\em dynamic information flow analysis and dynamic taint analysis}, {\em reinforcement learning}, and {\em analysis with variable cost-effectiveness}.

\section{Software Metrics}
Measuring software systems in terms of properly chosen metrics is an integral step
in software quality assurance~\cite{galin2004software}.
Defining appropriate software metrics is essential for both software process quality
and product quality, throughout the entire software development lifecycle~\cite{kan2002metrics}.
Prior to the implementation phase, software metrics provide a means for specifying quality
requirements with respect to relevant quality factors.
After implementation, the metrics further serve as crucial guidance for evaluating
the software product with respect to the specification of quality requirements.
Software metrics also play a vital role in software project management as a whole (e.g., for
cost and effort estimation)~\cite{fenton2014software}.

Two main classes of metrics can be used in software measurement: static and dynamic~\cite{tahir2012systematic}.
In comparison, static metrics are generally easier to compute relative to dynamic
counterparts~\cite{chhabra2010survey,dufour2003dynamic}. Additionally, static metrics are not subject to
limited code coverage or generalizability, as are dynamic metrics.
On the other hand, static metrics are not sufficient for measuring and interpreting
dynamic behaviors of software, for which dynamic metrics offer much more precise indicators.
In fact, concerning quality factors that are ultimately attested
at runtime (e.g., performance~\cite{dufour2003dynamic}, reliability~\cite{yacoub2002methodology},
and testability~\cite{gosain2016predicting}), dynamic metrics are much more preferable.
Meanwhile, understanding a software behavior does not always need complete code coverage~\cite{richner1999recovering},
thus a limited execution that dynamic metrics address does not necessarily constitute a constraint of dynamic measurement.
Dynamic metrics cannot simply be (over-)approximated
by corresponding static metrics either---in some cases, they are not even correlated~\cite{geetika2014empirical}.

Software coupling is the strength of the relationships among software modules, for measuring how closely connected the modules are~\cite{myers1975reliable,myers1975reliable}.
Coupling metrics have been well studied for single-process systems~\cite{chhabra2010survey,kan2002metrics,tahir2012systematic}.
For example, Arisholm et al.~\cite{arisholm2004dynamic} defined a set of dynamic coupling metrics for object-oriented software and studied the relationship between dynamic coupling measures and software change-proneness. 
Dynamic coupling metrics also have been used to estimate architectural risks~\cite{yacoub2002methodology}
and complexity~\cite{gosain2017object} in relation to quality metrics such as maintainability~\cite{perepletchikov2011controlled,gosain2016predicting}.
Most of these metrics were defined under the assumption that there exists an explicit reference/invocation between
the entities (e.g., object, method, and class) involved in the coupling measure.

The cohesion of a software component refers to the extent to which the elements of a component are related~\cite{bieman1994measuring}.
A highly cohesive component performs a set of closely relevant actions,
and it is difficult to be split into separate components~\cite{yacoub2002methodology}.
Static cohesion has been widely explored in software measurement. 
More recent relevant works increasingly focused on run-time (i.e., dynamic)
cohesion~\cite{mitchell2004run,gupta2011dynamic,zheng2012software,desouky2014object}.

\section{Distributed System Architectures}\label{subsubsec:arch}
\begin{figure}[htbp]
	\centering
	\caption{The architectures of a distributed systems: (a) left:
	{\em client-server}, (b) median: {\em peer-to-peer}, and (c) right: {\em n-tier} (3-tier)}
	\includegraphics[width=1\textwidth]{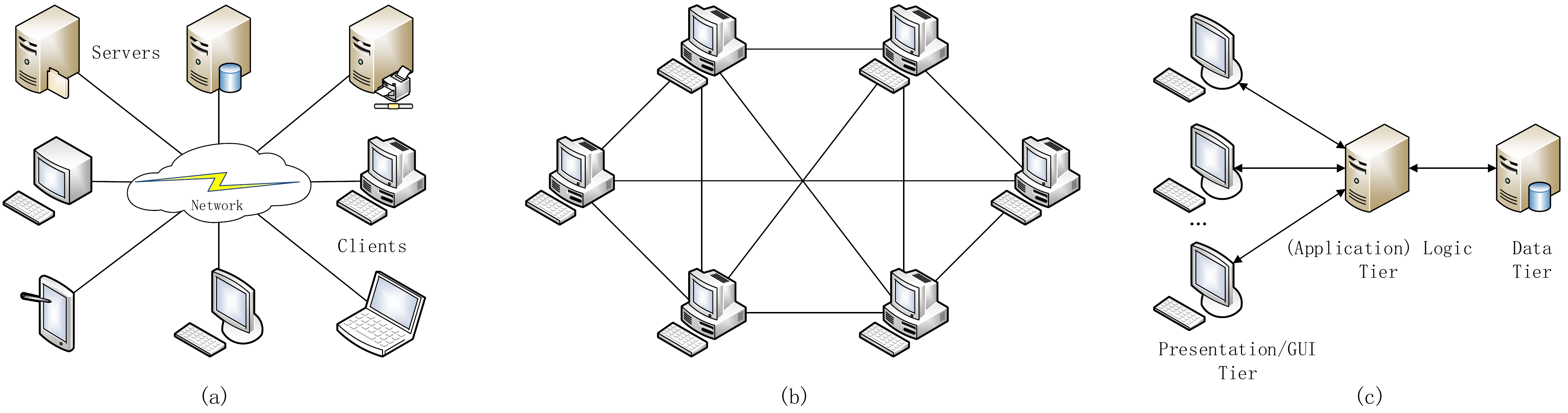}
	\label{fig:distarc}
\end{figure}

There are typically three types of distributed system architectures: {\em client-server}, {\em peer-to-peer}, and {\em n-tier}, as shown in Figure~\ref{fig:distarc}.

{\em Client-server} (CS) is a type of network architecture in which each process/node on the network is either a server or a client.
Servers are powerful for controlling and managing relevant resources (e.g., disk drives, database, printers, network traffic),
while clients rely on servers for those resources~\cite{maglogiannis2006enabling}.
Client-server architecture is simple to implement, without peer-discovery~\cite{hanafiah2015web}.
For example, NioEcho is a {\em client-server} distributed program including a client and a server~\cite{nioecho}.

{\em Peer-to-peer} (P2P) is a type of network architecture in which each process/node has equivalent responsibilities and abilities.
P2P differs from CS, in which some processes/nodes are dedicated to serving other processes/nodes~\cite{maglogiannis2006enabling}.
For instance, OpenChord is a {\em peer-to-peer} distributed system providing network services through a distributed hash table~\cite{openchord}.

The {\em n-tier} architecture breaks up an application into tiers, providing flexibility and reusability for developers who only need to modify or add a specific tier(layer), rather than to rewrite the whole application when they decide to change the application.
In the term {\em n-tier}, "n" can be any number (larger than 1) of distinct tiers used in a specific architecture, such as 2-tier, 3-tier, or 4-tier, etc~\cite{maglogiannis2006enabling}.
For example, Microsoft Azure is a typical {\em n-tier} distributed system that provides cloud computing services~\cite{wilder2012cloud,copeland2015microsoft,azure2022}.

\section{Logic Clocks in Distributed Systems}

A distributed system is based on a computer network where different computers are connected via passing messages or other types of middleware.
This feature helps users share various resources via network communication~\cite{mullender1993distributed}. But there is no global logic or physical clock for concurrent execution of distributed components.


The Lamport timestamp (LTS) algorithm is used to generate a partial ordering of events in a distributed system, maintaining a logical clock for all processes.
In the LTS algorithm, each process maintains an integer value, initially zero, which periodically increments, once after every atomic event; the value is attached to the record of the execution of each event as its timestamp centrally or separately; the traces are mostly maintained by each process~\cite{fidge1987timestamps}. 
The LTS algorithm has asynchronous and synchronous communication methods, with respect to the following rules:
\begin{enumerate}
\item A process increments its counter for each event in it.
\item When sending a message, a process includes its counter value with the message.
\item When receiving a message, the counter of the recipient is updated (adding 1).
\end{enumerate}

\begin{figure}
    \centering
  \caption{The Lamport timestamp (LTS) algorithm used in multiple processes of a distributed program}
  \includegraphics[width=0.75\textwidth]{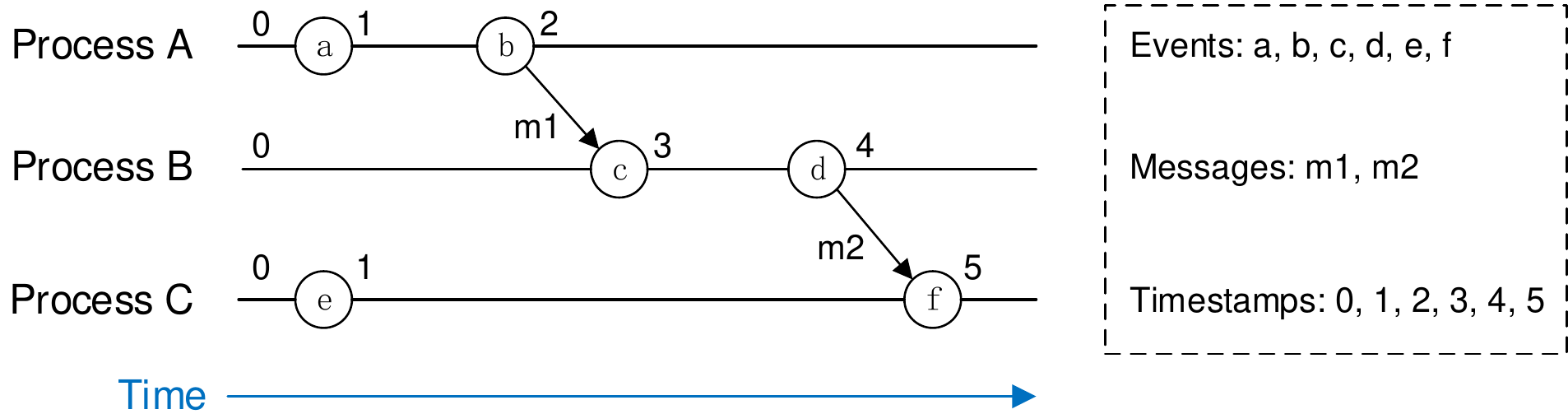}
  \label{fig:lamport}
\end{figure}

\indent
I use an example to explain the LTS algorithm in detail.
I suppose that there are three processes (i.e., {\em Process A}, {\em Process B}, {\em Process C}) in a distributed program, as shown in Figure~\ref{fig:lamport}. %
Each process has its logical clock initialized to zero and the clock value (i.e., timestamp) immediately increments after an event, such as 1 for the event {\em a}, 2 for the event {\em b}, etc.
When the message {\em m1} was sent from {\em Process A} to {\em Process B}, the timestamp 2 was piggybacked and then the event {\em c} (of {\em Process B}) gets its timestamp 3,
which is the maximum value between piggybacked timestamp 2 and local timestamp 0 (initial value), added 1, as (max(0, 2) + 1 = 3).
Then, the timestamp of the event {\em d} is 4 (= 3 + 1).
Next, the message {\em m2} is sent from {\em Process B} to {\em Process C} with the clock value 4 piggybacked.
Finally, the event {\em f} (of {\em Process C}) gets its timestamp 5 (= max(1, 4) + 1), where 1 is the timestamp of the previous event ({\em e}) in the same process ({\em Process C})~\cite{fidge1987timestamps,lamport2019time}.

Like the LTS algorithm, vector logical clocks are used to determine the partial ordering of events and to detect causality violations in distributed systems,
by comparing event timestamps~\cite{baldoni2002fundamentals}.
For a distributed system with $N$ processes, there is a vector (i.e., an array) of $N$ logical clocks, one clock per process;
and each process remains a local vector (i.e., array) that includes the largest possible values of the global clock vector \cite{zheng2019optimal,Vectorclock}.

However, vector logical clocks have a major fault: If a distributed system has too many processes whose count $N$ is very large,
the timestamp data (i.e., communication overhead) attached to each message would be too large to be acceptable~\cite{sun2002capturing}.

Besides the vector clock and LTS timestamps algorithms, matrix clocks are also used to capture causal and chronological relationships in distributed programs .
As a generalization of the vector clock notion, a matrix clock maintains a vector of the vector clocks for each process (communicating host)~\cite{drummond2003reducing,Matrixclock}.

\section{Dependence Analysis}
In program execution, there are two types of explicit dependencies: traditional dependencies and thread dependencies.
\emph{Control dependency} and \emph{data dependency} are traditional dependencies.
In a concurrent program, there may be three types of thread dependency: synchronization, ready, and interference dependencies.
In a common distributed software system, besides explicit dependencies (via references and/or invocations) among code entities, dependencies across distributed components/processes (referred to as {\em inter-component/interprocess} dependencies) are {\em implicit} because these components/processes are decoupled by networking facilities.

Analyzing dependencies among program entities of a software system can help developers better understand the structure and behaviors of the system.
Thus, dependence analyses are very useful for users to develop, test, and maintain the system, because these tasks rely on the understanding of system structure and behaviors.

Program dependencies can be deduced by both static and dynamic analyses.
Static dependence analysis computes dependencies via analyzing the program code without executing the software.
By contrast, a dynamic dependence analysis infers dependencies from the data gathered during the execution(s).
In particular, as a special type of dynamic analysis, hybrid dependence analysis combines static and dynamic analysis modalities~\cite{shar2013mining}.
Hybrid dependence analysis approaches can integrate run-time information extracted from dynamic analysis techniques into static analysis algorithms to precisely compute program dependencies.
Hybrid dependence analyses are becoming popular because they generate more accurate results than other analysis modalities.
However, a hybrid dependence approach is often complex, and its analysis process may be time-consuming \cite{rao2017comparative}.

\subsection{Dependence analysis for single-process programs}
As a dynamic analysis approach, {\diver}~\cite{cai14diver} computes dependence sets as impact sets using dependence analysis techniques.
As a recent advance in (offline) dynamic analysis, it achieves higher precision and provides a more cost-effective option 
over EAS-based approaches (which derive dynamic dependencies based on execution orders), such as {\pieas}~\cite{apiwattanapong2005efficient}.
{\diver} utilizes a static dependency analysis to significantly decrease the size of the dependence
set produced by {\pieas}.

With significantly smaller resulting dependence sets, 
the cost-effectiveness tradeoff of {\diver} is much higher even with the additional static dependence analysis cost.
{\diver} works in three technical phases: static analysis (Phase 1), runtime tracing (Phase 2), and post-processing analysis (Phase 3).
{\diver} first computes traditional control/data dependencies~\cite{horwitz1990interprocedural} and instruments the input program in Phase 1.
In Phase 2, the instrumented version of the program is executed for tracing {\em entry} (i.e., program control entering a method) and {\em returned-into} (i.e., program control returning from a callee into a caller) events. 
In Phase 3, 
the technique computes the dependence set from the trace for any query given by the user.

An online version of Diver, {\diveronline}~\cite{cai2017hybrid},
avoids execution tracing costs (e.g., space and I/O costs) that are ineluctable in offline analyses,
via computing dependence sets during the execution of
the program under analysis. 
%
%
In addition, 
{\diveronline} provides an {\em All-in-One} analysis, which
computes the dependence sets for all possible queries (methods), and then corresponding dependence sets are directly delivered to the user within a short response time.
As such, an {\em All-in-One} online dynamic dependence analysis may be suitable for large-scale software systems.


\subsection{Dependence analysis for multi-threaded programs}

The increasing use of multi-threaded (concurrent) programs invokes challenges for dependence analyses, and it is also relatively difficult for users to understand multi-threaded systems.
For instance, in a shared memory model, one thread accessing a memory location may potentially interfere with another thread to access the same location,
leading to dependencies between these two threads.

Generally, among threads, there are two particular types of control dependencies---{\em ready dependencies} and {\em synchronization dependencies},
and a type of data dependencies---{\em interference dependencies}~\cite{Nanda2006ISM}.
The main task of a dynamic analysis of threading-induced dependencies is thus to infer {\em ready dependencies}, {\em synchronization dependencies}, and {\em interference dependencies} across multiple threads that occurred during program executions.
For multi-threaded programs,
several dependence analysis algorithms~\cite{xu2005brief,xiao2005improved,Nanda2006ISM,giffhorn2009precise} have been developed.

Indus is a sound framework to analyze and slice multiple-threaded (current) Java programs. In Indus, Java source code is first transferred to Jimple code as an intermediate representation.
However, since it is purely static, those dependencies are only approximated by Soot with very little cost~~\cite{corbett2000bandera,ranganath2007slicing}. 
Indus gives developers the most common dependence analyses for intra-thread control and data dependencies, inter-thread {\em ready}, {\em synchronization}, and {\em interference} dependencies.



\subsection{Dependence analysis for distributed programs}

For a complex distributed system with multiple processes, the developer needs to understand various (explicit and implicit) dependencies both within a single process and across multiple processes.
Krinke proposed a slicing algorithm incorporating dependencies across distributed components induced by socket-based message passing~\cite{krinke2003context},
but the dependencies were approximated over-conservatively because they are computed through purely static analysis.
Another approach~\cite{barpanda2011dynamic} infers various kinds of dependencies due to interprocess communications.
However, the approach potentially suffers a scalability problem due to its heavyweight nature.

To overcome the scalability challenges, a lightweight dynamic analysis for distributed programs, {\distia}~\cite{cai2016distea}, was developed.
The analysis monitors and records method events and their timestamps during the system execution, and then
approximates run-time dependencies among relevant methods, either within or across processes, based on the happens-before relations among method execution events.

For example, if a method $A$ has the last returned-into event which was executed before the first entry event of another method $B$, the partial-order is $A$ {\em before} $B$, and {\distia} approximately supposes that $B$ is dependent on $A$.
Similarly, if one method $D$ is dependent on another method $C$, $C$ must execute before $D$;
otherwise, $C$ cannot affect $D$.
Thus, dependencies computed by {\distia} are safe for executed methods,
but not for all methods since some methods were not covered during the execution.
On the other hand, if $B$ is executed after $A$, $B$ may or may not be actually dependent on $A$.
Therefore, we know that {\distia} is not precise.

\section{Dynamic Information Flow Analysis and Dynamic Taint Analysis}
Tracking/checking 
dynamic information flow 
underlies various security applications (e.g.,~\cite{suh2004secure,masri2008application,masri2004detecting,schwartz2010all,goli2019security}),
It addresses a general source-sink problem for a program execution, in which a {\em source} is where confidential or untrusted (i.e., {\em sensitive}) information is produced and flows into the program,
while a {\em sink} consumes the information and makes it flow out of the program execution~\cite{terminology2018sourcesink,alashjaee2019dynamic}.
Due to its focused reasoning about actual executions, this approach has
precision merits over statically inferring information flow.

One 
technique realizing the approach is to
compute the chains of dynamic control and data dependencies 
hence to infer full information flow paths between given sources and 
sinks
during the execution (e.g.,~\cite{shroff2007dynamic,masri2009algorithms,masri2008application,masri2004detecting}),
called dynamic information flow analysis (DIFA).
An alternative technique is dynamic taint analysis (DTA), which applies tags to (i.e., {\em taint}) the data 
entering the program from the sources,
propagates the taint tags during the execution,
and checks the data at the sinks against the presence of the tags
(e.g., ~\cite{clause2007dytan,zavou2011taint,newsome2005dynamic,shankar2017androtaint,ming2015taintpipe,she2020neutaint,ashouri2019hybrid,banerjee2019iodine,zhu2011tainteraser,karim2018platform,jiang2017kakute,sun2016pileus,zavou2015information,you2017taintman,enck2014taintdroid,yin2007panorama,hauser2013intrusion,chow2004understanding,kang2011dta++}).
Unlike DIFA, DTA does not compute full information flow paths.
DIFA thus provides better support in
usage scenarios that require more detailed flow information (e.g., diagnosing data leaks in the paths).

DIFA and DTA can be differentiated
(1) by their inner workings as mentioned earlier (i.e., DIFA works by computing dynamic dependencies, while DTA works via data tainting and taint propagation)
and (2) by their results---DIFA provides full information flow paths while DTA just tells which data is tainted.
On the other hand, both DIFA and DTA concern information flow paths between given sources and sinks.

\section{Reinforcement Learning}
Reinforcement learning (RL) is an area of machine learning, beyond supervised learning,
to suggest software agents taking actions in an environment to maximize the total reward for all possible successive actions~\cite{littman1996generalized}.
RL is based on the information measured from the environment, and thus it can be called action-based learning.
RL refers to a learning approach whose agent or actor modifies its actions according to the response to its interaction with the environment.
And RL hardly requires a priori knowledge, so that it can be applied to varying and uncertain environments where standard supervised/unsupervised learning approaches are not applicable~\cite{lewis2010reinforcement}.

Unlike supervised learning and unsupervised learning, RL does not need training data.
Described by Markov Decision Processes, whose states are decided by the previous states~\cite{puterman2014markov},
the output of RL depends on the corresponding input,
and the next input depends on the current output~\cite{ackley1990generalization}.

\begin{wrapfigure}{r}{0.3\textwidth}
\vspace{-10pt}
\caption{Q-learning workflow}
\includegraphics[width=0.3\textwidth]{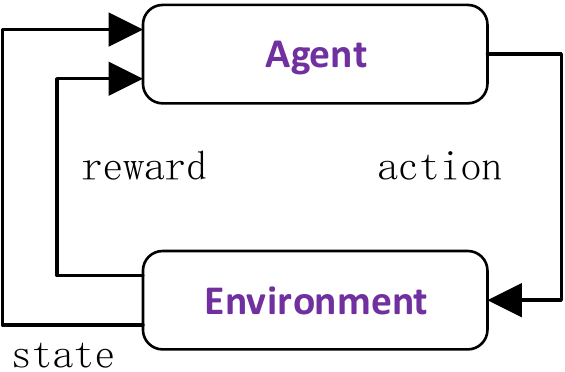}
\label{fig:ql}
\vspace{-10pt}
\end{wrapfigure}
In particular, as a particular type of RL, Q-learning uses a Bellman equation to minimize its cumulative cost~\cite{barron1989bellman}.
It does not require explicit or exact descriptions of software systems and only needs state measurements in its feedback control loop~\cite{lewis2010reinforcement}.
Q-learning also has an exact capability of learning the next state according to only previous states.
Starting from an initial state, Q-learning tries to find a way to maximize cumulative reward values by selecting an action after measuring how good the action is in a particular state~\cite{watkins1992q}.
It is an off-policy and model-free algorithm, as it does not require an existing policy or model~\cite{wawrzynski2004model}.
The overall workflow of a Q-learning algorithm is depicted in Figure~\ref{fig:ql}, including the following steps:

(1) The algorithm initializes Q-learning components and a Q-table (i.e., a lookup table for storing expected rewards calculated).

(2) At the current state, the agent selects an action referencing the maximal value in the Q-table or by random.

(3) The agent receives a state and a reward from the environment.

(4) The algorithm updates the Q-table using the Bellman equation~\cite{dolcetta1984approximate}.

(5) The algorithm repeats steps (2), (3), and (4) until the learning meets predefined conditions (e.g., when the agent finishes its ultimate tasks).

\section{Analysis with Variable Cost-effectiveness}
Most existing analysis approaches commonly suffer from challenges of balancing the analysis cost and effectiveness. 
They are either precise but too expensive or efficient but too imprecise. 
To deal with these challenges, one existing solution is to offer variable cost-effectiveness balances to satisfy varying user needs.
For example, the {\diapro} framework provides flexible cost-effectiveness choices for a variety of levels of cost-effectiveness tradeoffs with the best options for variable user requirements and budgets~\cite{cai2016diapro}.
By combining the static and dynamic data, {\diapro} unifies {\pieas}~\cite{apiwattanapong2005efficient}, {\diver}~\cite{cai14diver}, and three dependence-based dynamic dependence analysis techniques: one using coverage and trace, one using aliasing and trace, and the other using all these dynamic data (aliasing, coverage, and trace).

Another example is {\textsc{D$^2$Abs}}~\cite{cai2021d2abs},
which aims at practical scalability and offers various levels of cost-effectiveness tradeoffs in the dynamic
dependence analysis for distributed programs with four versions. 

%% file: chapters/chapter3.tex
\chapter{{\distmeasure}: A Framework for Measuring and Understanding Distributed Software Systems via the Lens of Interprocess Communication} \label{ch:distmeasure}
The goal of my doctoral research is to address scalability, cost-effectiveness, and other (applicability/portability) challenges in distributed software analysis and its applications to other quality problems.
And to start with, I looked at the security problems of distributed systems.
There is a communication mechanism that basically connects the different processes in a distributed system.
Interprocess communication (IPC) may be one of the most important features of a distributed system.
Thus, I developed {\distmeasure} (the expansion of \cite{fu2019measuring,fu2019towards}), including a set of metrics for common distributed systems, with a focus on
their IPC, a vital aspect of their run-time behaviors. 
I also implemented a toolkit {\distfax}~\cite{fu2022distfax} for measuring IPCs and quality of distributed software.
And I demonstrated the practicality of characterizing IPC dynamics and complexity via the proposed IPC metrics, by computing the measures against the executions of 11 real-world distributed systems, with a large number of test inputs.
To show the practical usefulness of IPC measurements, I extensively investigated how the proposed metrics may help predict and understand various quality metrics of distributed systems, via classifying system quality levels based on the IPC measurements, with respect to several different quality aspects.

\section{Motivation}
In general, a distributed system consists of multiple collaborating components each running in a process typically
located at a separate computing node.
Since these components interact primarily through IPC~\cite{sharifi2012platform}, measuring IPC is essential for understanding the behaviors of distributed systems.

Apparently, IPC dominates the activities of this system, excluding which each component alone would be largely trivial.
Thus, the complexity of this system's execution essentially lies in the complexity of its IPC.
First, I wanted to diagnose the communication security issues of this system reported by users with the inputs that can reproduce the issues.

A rewarding first step would be to understand how communication currently works in terms of the IPC with respect to the user input.
Measuring IPC would help in this scenario.
Moreover, the IPC measurements may help the developer assess further understand quality aspects of an IPC-induced system.
For instance, if the IPC coupling is very high, intuitively the system might be difficult to update.
Thus, its maintenance cost may be quite high also.

Unfortunately, there is a lack of tool support for measuring distributed systems executions with respect to IPC,
and it is unknown which quality factors might be analyzable through the IPC measures.
There, I addressed these questions by developing new IPC metrics and applying them to distributed software quality assessment.

Meanwhile, it may be difficult or even impossible to directly measure some quality metrics of distributed systems.
For example, the quality metric {\em code churn size} is commonly used  to quantify the varying parts of an evolving program.
However, computing this quality metric typically requires the data of program releases,
and it is impossible to directly measure the first release of the program.
The correlations between the IPC metrics and the quality (metrics) of a distributed system can be leveraged by developers to indirectly assess correlated quality metrics of the system and further to understand the system and its quality aspects.

\section{Approach}
I first give an overview of {\distmeasure} high-level workflow and architecture.
Then, I describe IPC metrics and quality metrics used in {\distmeasure}.

\subsection{Overall workflow}
\begin{figure*}[tp]
	\centering
	\caption{An overview of the {\distmeasure} workflow}
	\includegraphics[width=1\textwidth]{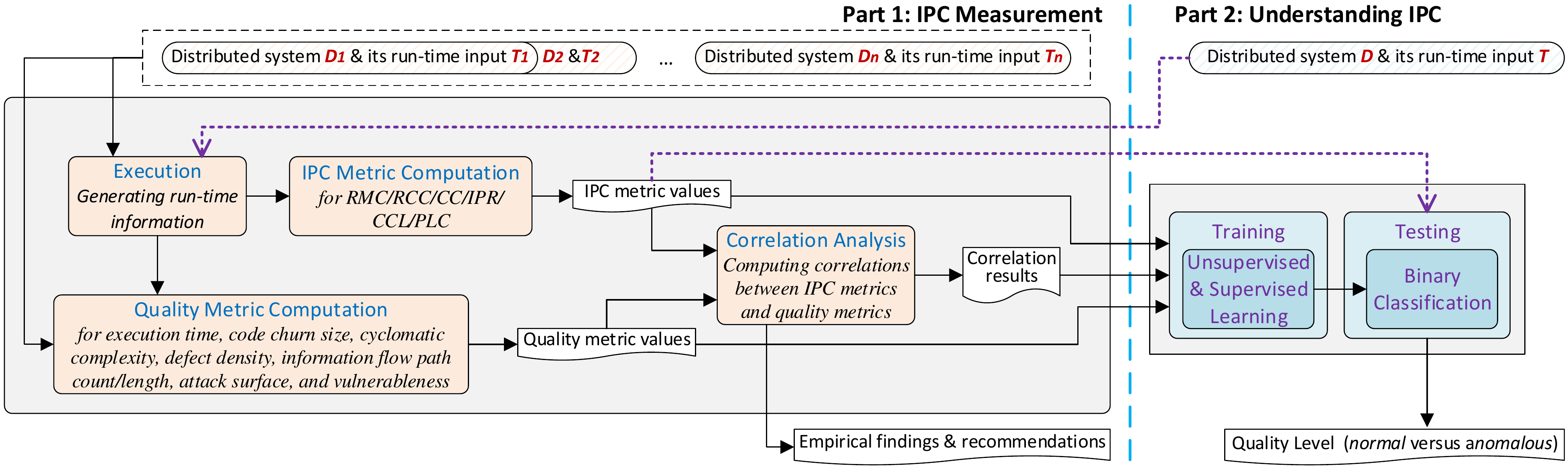}
	\label{fig:distmoverview}
\end{figure*}

Figure~\ref{fig:distmoverview} depicts the overview of {\distmeasure}, including two closely connected parts:
{\em IPC Measurement (Part 1)} and {\em Understanding IPC (Part 2)}.
In {\em Part 1},
{\distmeasure} takes two {\bf inputs} from the user:
multiple ($N$) distributed systems $D_1$, $D_2$, ..., $D_n$ and their run-time inputs $T_1$, $T_2$, ..., $T_n$ including system commands, SQL statements, text messages, etc.
{\distmeasure} first executes the systems against corresponding inputs, 
to produce the system run-time data.
Then, {\distmeasure} computes six IPC metrics (i.e., RMC/RCC/CCC/IPR/CCL/PLC) from the data.
\vspace{15pt}

Meanwhile, eight quality metrics (i.e., {\em execution time}, {\em code churn size}, {\em cyclomatic complexity}, {\em defect density}, {\em information flow path count}, {\em information flow path length}, {\em attack surface}, and {\em vulnerableness}) of the systems are measured directly.
From the values of these IPC metrics and quality metrics, {\distmeasure} computes the correlations between them.
These correlation results induce to empirical findings and recommendations about the system quality,
as the {\bf output} of {\distmeasure} Part 1.

In {\em Part 2},
{\distmeasure} takes the IPC and quality metric values, distributed systems $D$, and $D$'s input $T$, as {\bf inputs} to classify the same quality metrics of  $D$.
It first trains unsupervised learning and supervised learning models from the metric values,
and then uses these models to classify each of such quality metrics as {\tt anomalous} or {\tt normal}.
{\tt Anomalous} indicates a warning of low (than average level) quality, while {\tt normal} means no such a warning (i.e., the corresponding quality metric value is higher than or equals or the average level). 
In particular, models are trained and used for classification of quality metrics, each of which is significantly (and negatively or positively) correlated to one or more of the six IPC metrics (correlation absolute value $>=$ 0.4).
These classification results provide developers' guides for the quality assurance tasks in software maintenance, such as software debugging, software testing, and so on.

\subsection{IPC metrics}

As shown in Table~\ref{tab:ipcs}, I defined five IPC coupling metrics (i.e., {\em run-time message coupling (RMC)}, {\em run-time class coupling (RCC)}, {\em class central coupling (CCC)}, {\em inter-process reuse (IPR)}, and {\em class communication load (CCL)}) and one IPC coupling metric
(i.e., {\em process-level cohesion (PLC)}~\cite{fu2022distfax}), and explained their computation (in the second column) and justification (in the third column).
Six IPC metrics cover four levels of measurement granularity: method ($IPR$), class ($RCC$, $CCC$, and $CCL$), process/component ($RMC$, $RCC$, and $PLC$), and system ($RMC$, $RCC$, $CCC$, $IPR$, $CCL$, and $PLC$), listed in the last column ({\bf NS. L.}, short for non-system level) of Table~\ref{tab:ipcs}.
System level is not listed in Table~\ref{tab:ipcs} because it covers all six IPC metrics.

{\em RMC} is the extent of run-time interactions among processes.
A process-level RMC is the number of messages sent from one process to another.
And a system-level RMC is the average of the process-level RMC on all communicating process pairs.

{\em RCC} concerns methods from a class in one process to access methods in other processes.
A class-level RCC is the ratio of the total number of methods in the first class in the first process that is dependent on any method in the second class in the second process, to the total number of methods in any process, other than the first process; 
while a process-level RCC metric is then set to the size of the union set of entire dependence sets of all methods in all classes of the process, where the numerator is the size of the union set of remote dependence sets of those methods.
Moreover, a system-level RCC metric is the average of all process-level RCC measures.

{\em CCC} concerns the importance of a class affecting other classes in remote processes.
A class-level CCC is the sum of all RCC measures between the class and other classes in all remote processes.
A system-level CCC is the mean of process-level CCC measures.

\begin{table*}[tp]
  \centering
  \caption{Summary of IPC Metrics}
\resizebox{1\columnwidth}{!}{%
    \begin{tabular}{l|p{33em}|p{13em}|l}
    \hline
     & \multicolumn{1}{c|}{\textbf{Type \& Definition \& computation}} & \multicolumn{1}{c|}{\textbf{Justification}} & \multicolumn{1}{c}{\textbf{NS. L.}} \\
    \hline
RMC   & Coupling \& the interprocess message coupling \& the number of messages sent from a process to another one & the extent of run-time interactions among processes & process \\
    \hline
RCC   & Coupling \&  the class coupling between two processes and further at system-level \& the ratio of the total number of methods in the second class that are dependent on any method in the first class, to the total number of methods, which are dependent on any method in the second class and in all processes but the first process & how methods from a class in one process access methods in other processes & \makecell[l]{class,\\process}  \\
    \hline
CCC   & Coupling \& the aggregate coupling as regards an individual class executed in a local process with respect to classes in all remote processes/ the aggregate RCC metrics between the class and other classes in all remote processes & the importance of a class with respect to its coupling strength & class \\
    \hline
IPR   & Coupling \& interprocess coupling at method levels \& the intersection of the local and remote dependence sets divided by the size of the union set of methods executed & code overlapping and reuse across processes & method \\
    \hline
CCL   & Coupling \& the communication loads of an individual class communicating with others in all remote processes \& the sum of the sizes of remote dependence sets of the class's methods divided by the size of the set of all executed methods in the class & how much a class contributes to communication loads among processes & class \\
    \hline
    \hline
PLC   &  Cohesion \& internal connections within an individual process \& the sum of the sizes of local dependence sets of the process' methods divided by the size of the set of all executed methods in the process & the degree to which the methods of a process belong together & process \\
    \hline
    \end{tabular}%
}
   \label{tab:ipcs}%
\end{table*}

For inter-process code overlapping and reuse, a method-level {\em IPR} is the ratio of the intersection of the local dependence set (set of methods in the process that depends on the given method), and remote dependence set (set of methods that depend on the given method but are in any process other than the first process), to the size of the entire set of methods covered in the execution.
A system-level IPR metric is then set to the ratio of the sum of method-level IPR measures on all methods in the execution, to the size of the entire set of those methods.

{\em CCL} measures how much a class contributes to the communications loads between the process executing the class and other processes.
A higher CCL means that a class has more communications among system components.
A class-level CCL is computed as the sum of the sizes of remote dependence sets of the class's methods divided by the size of the set of all executed methods in the class.
The system-level CCL is the mean of class-level CCL measures over all classes executed (across all processes).

Differing from previous metrics all about coupling, {\em PLC} measures cohesion.
It is the degree to which all of the methods executed in a process belong together.
A process-level PLC is computed as the sum of the sizes of local dependence sets of all methods in the process,
The system-level PLC is the mean of process-level PLC measures over all processes.

Some of these metrics (i.e., $RMC$, $RCC$) explicitly measure interprocess coupling, while some metrics (i.e., $CCC$, $IPR$, and $CCL$) are derivative coupling metrics.
And $IPR$ measures the common dependencies of two processes, differing from four other coupling metrics (i.e., $RMC$, $RCC$, $CCC$, and $CCL$) that measure mutual dependencies of one process on the other.

Differing from measuring process cohesion,
measuring interprocess coupling in distributed systems, however, 
is different and challenging~\cite{cai2016distea}.
Coupling metrics are often defined on the basis of certain relationships (e.g., dependency and inheritance)~\cite{tahir2012systematic}.
However, deriving the interprocess dependencies from which the metrics are computed,
is not trivial in the context of distributed system executions.
The main reason lies in the lack of global timing across the system 
together with the lack of explicit references/invocations across distributed components. 
To overcome this challenge, I leverage a framework for dynamic dependence analysis of distributed programs~\cite{cai2021d2abs}.
Using this framework, I reason about interprocess dependencies through the happens-before relation between
executing methods across processes, derived from a global partial ordering of method execution events.
I further exploit the semantics of message passing to improve the precision of such derived dependencies, so as to
enhance the validity of IPC coupling and cohesion metrics.

\subsection{Quality metrics correlated to IPC metrics}

\begin{figure*}[tp]
	\centering
	\caption{The reference-quality model underlying {\distmeasure}, adopted from and compliant with the ISO/IEC 25010~\cite{iso25010} standard}
	\includegraphics[width=1\textwidth]{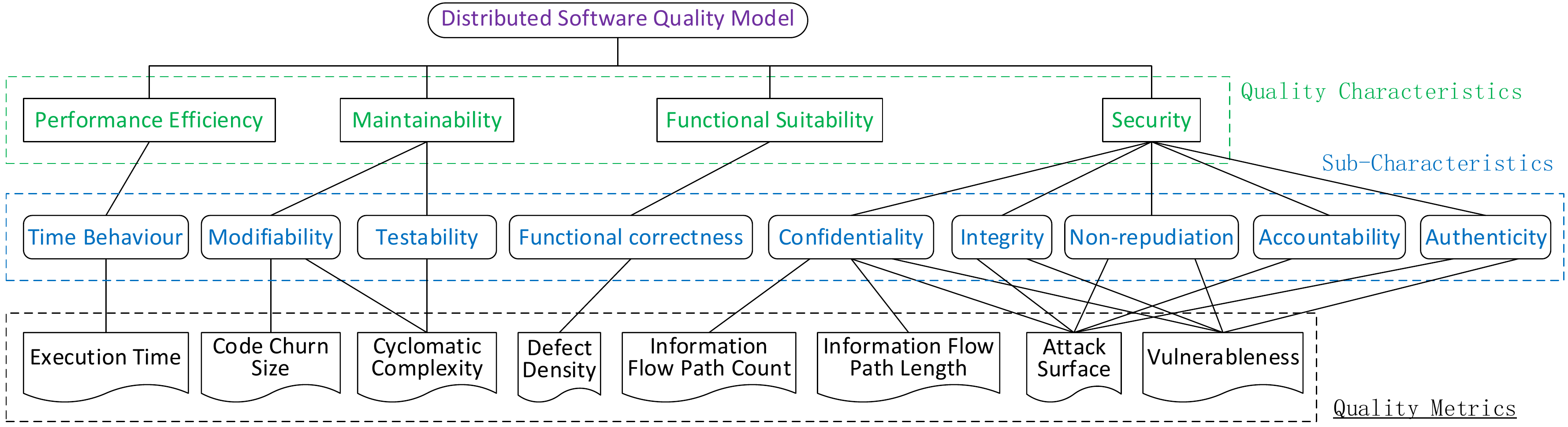}
	\label{fig:model}
\end{figure*}

To demonstrate the usefulness of the coupling and cohesion measurement in aiding the analysis of distributed systems quality,
I adopted the standard quality model ISO/IEC 25010~\cite{iso25010} with necessary customization selection, as depicted in Figure~\ref{fig:model}.
The model has three layers: quality characteristics (i.e., factors),  sub-characteristics (i.e., sub-factors), and quality metrics.
The top and second layers are specified by the standard quality characteristic/sub-characteristic names (in ISO/IEC 25010~\cite{iso25010}),
but the bottom layer is customized by quality metric names used in {\distmeasure}.

There are four quality characteristics (i.e., {\em performance efficiency}, {\em maintainability}, {\em functional suitability}, and {\em security}), nine sub-characteristics (i.e., {\em time behaviour}, {\em modifiability}, {\em testability}, {\em functional correctness}, {\em confidentiality}, {\em integrity}, {\em non-repudiation}, {\em accountability}, {\em authenticity}), and eight quality metrics
(i.e., {\em execution time}, {\em code churn size}, {\em cyclomatic complexity}, {\em defect density}, {\em information flow path count}, {\em information flow path length}, {\em attack surface}, and {\em vulnerableness}), some of which for measuring multiple sub-characteristics.

{\distmeasure} uses a direct metric {\em execution time} to measure the time behaviour of a distributed system,
and then normalized it by the logical source lines of code (SLOC) of the system under analysis.

A version-level {\em code churn size} is the source line number of code changed (i.e., added, deleted, or updated),
between two adjacent versions of a program~\cite{zeiss2007applying}.
Then, it is normalized by the logical SLOC of the latter version.
Finally, the system-level code churn size is computed as the mean of all version-level code churn sizes of all the program's releases.

The {\em cyclomatic complexity} of a specific system is a direct quality metric~\cite{mccabe1976complexity},
normalized by the logical SLOC of the system.

The {\em defect density} of a specific system is the number of defects related to the system, normalized by the system's logical SLOC.

The {\em information flow path count} is the total number of dynamic information flow paths, while {\em information flow path length} is computed as the average length of the dynamic information flow paths.

The {\em attack surface} is defined as a quality metric indicating the relative security of a system in
three dimensions (methods, channels, and data) to determine if the system is more secure than others~\cite{manadhata2010attack}.
I first consider a triple $<$\textbf{N1}, \textbf{N2}, \textbf{N3}$>$,
where \textbf{N1} is the number of methods including sources/sinks (i.e., entry/exit points), \textbf{N2} is the number of network ports (i.e., channel) used, and \textbf{N3} is the number of files read/written by the subject during the execution).
Next, I calculate the Euclidean distance between it and the origin (0,0,0) as $\sqrt{N1^{2}+N2^{2}+N3^{2}}$.
Finally, the distance is normalized by the subject's logical SLOC, as the system-level attack surface.

The vulnerableness of most subjects is based on CVSS 2.0 score of public security databases (e.g., National Vulnerability Database (NVD)~\cite{nvd}) and the time of a vulnerability in CVSS.
If the vulnerability was discovered recently, it should be concerned more than another discovered long ago.
However, NVD does not contains the defects of some subjects (e.g., Grizzly, OpenChord, Voldemort, XNIO, xSocket).
Thus, other online bug resources (e.g., Jira, SourceForge, GitHub) should be considered too.
The vulnerableness of a subject is computed as
\vspace{-4pt}\begin{equation}\label{eq_evulnerableness}
	\begin{split}
		& N_{nn} + \sum (CVSS_v *(100 - Diff_v / 100))
	\end{split}
\end{equation}
where $N_{nn}$ is the number of vulnerabilities found in non\_NVD bug sources (without CVSS scores), $CVSS_{v}$ is the CVSS score of a vulnerability $v$, and $Diff_v$ is the difference between the current year and the $v$ found year,

Then, I explored the relationships between IPC measurements and eight quality metrics quantified through respective direct measurements, to show the usefulness of IPC measurements and their statistical relationships with distributed software quality measurements and
to demonstrate the capabilities of {\distmeasure} to help understanding the software quality via deeply characterizing IPC-related behaviors of distributed software.

For the goal, I performed extensive statistical analyses to examine the correlation of each pair between the six IPC metrics and the eight quality metrics.
Spearman`s rank correlation analysis~\cite{sedgwick2014spearman} rather than alternatives (e.g., Kendall's~\cite{abdi2007kendall}, Pearson's~\cite{benesty2009pearson} was selected.
The reason is that the former does not assume any relationship between the two involved variables.

Furthermore, I considered correlation strength according to the value of Spearman`s rank coefficient $r$ in~\cite{schober2018correlation}:
If $|r|$ $>=$ 0.4, the correlation is significant;
otherwise (i.e., $|r|$ $<$ 0.4), the correlation is weak.
In addition, $p$ values were also computed, indicating the statistical significance of correlation coefficients.

\subsection{Classification for understanding and predicting quality}

In {\em Understanding IPC (Part 2)},
{\distmeasure} first trains unsupervised learning and supervised learning models from IPC/quality measurements,
and then leverages these models to classify the quality metrics significantly correlated to IPC metrics.

\vspace{2pt}
\noindent
\textbf{Learning features.}
The six proposed IPC metrics (i.e., RMC, RCC, CCC, IPR, CCL, and PLC) are {\em potential} features for unsupervised and supervised classifiers for direct quality metrics
significantly correlated to one or more IPC metrics.
Both positive and negative correlations were considered.
Only the system-level metric values were used as feature values in {\bf Part 2} of {\distmeasure}.

\vspace{2pt}
\noindent
\textbf{Unsupervised learning.}
Grouping together given data points with similar characteristics, partition the data points into $k$ clusters by assigning them to the nearest cluster centers~\cite{pham2005selection},
$k$-means clustering algorithm (implemented in the Python Scikit-learn library~\cite{pedregosa2011scikit}) was selected for unsupervised learning quality classifiers.
I set $k$ $=$ 2 since {\distmeasure} aims to classify quality metrics as {\tt normal} versus {\em anomalous} for detecting quality anomalies and further understanding the system quality.

\vspace{2pt}
\noindent
\textbf{Supervised learning.}
After comparatively exploring some alternatives, I selected bootstrap aggregation (or bagging for short) to construct supervised learning quality classifiers.
Bagging generates multiple versions of a model via bootstrap the training data set replication used as new training data sets,
to provide high prediction accuracy, especially for unstable data---the accuracy can be retained with large variations
across the training data~\cite{breiman1996bagging}.
Distributed systems cover various application domains, execution scenarios, and scales.
Thus, in system executions, there are expected large variations in different training samples that are suitable for bagging.
As binary classifiers, the supervised ones in {\distmeasure} are also utilized to discover quality anomalies for predicting and understanding the quality of distributed systems, like those unsupervised classifiers.

\section{Tool Implementation}
\begin{figure*}[tp]
	\centering
	\caption{An overview of {\distfax} architecture}
	\includegraphics[width=1\textwidth]{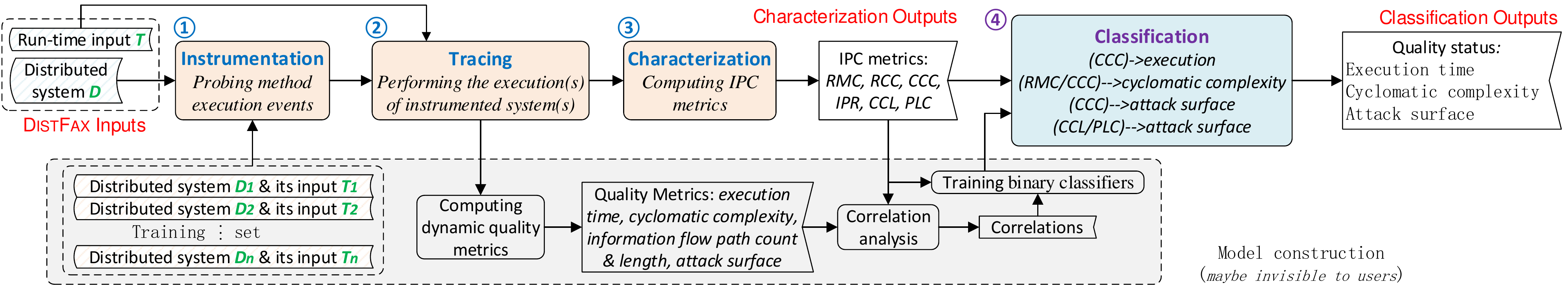}
	\label{fig:distfoverview}
\end{figure*}

To characterize common distributed systems, I implemented {\distfax} (short for \underline{Dist}ributed software systems facts (sounding \underline{Fax}))~\cite{fu2022distfax},
a toolkit for characterizing common distributed systems,
concerning their interprocess communications (IPCs), an important aspect of the run-time behaviors of the systems.
As shown in Figure~\ref{fig:distfoverview},
{\distfax} measures the system coupling/cohesion via IPC metrics defined and characterizes the system run-time quality via dynamic quality metrics referring to the standard quality model ISO/IEC 25010~\cite{iso25010}.
Then, {\distfax} analyzes statistical correlations between the IPC metrics and quality metrics.
Furthermore, it leverages the correlations to build learning models for classifying the system quality status with respect to the IPC metrics of the systems.

\section{Evaluation}         \label{sec:distmeasureeval}

\subsection{Experiment Setup} \label{ss:distmeasurees}

I measured the IPC traits of nine real-world distributed software projects, mostly
enterprise-scale systems.
Table~\ref{tab:subjects} lists subject names/versions ({\bf the first column}),
the logical source lines of code ({\bf the second column}), test types ({\bf the third column}), and test input counts ({\bf the last column}).

For all subjects, their integration tests were created via putting together steps in the quick start guides from their official websites.
In particular, for the integration test of six frameworks/libraries (XNIO, xSocket, QuickServer, Thrift, Grizzly, and Netty), I developed applications to cover their major functions and then performed all of the applications.
In addition, for ZooKeeper, its load and system tests came with the project packages.
I now introduce all evaluation subjects, their original integration test operations, and corresponding expansions of their run-time inputs.

\setlength{\tabcolsep}{3.5pt}
\begin{table*}[tp]
	\centering
	\caption{Statistics of experimental subjects in {\distmeasure}}
	\begin{tabular}{|l||r|l|r|}
		\hline
		\multicolumn{1}{|c||}{\textbf{Subject (version)}} & \multicolumn{1}{c|}{\textbf{\#Logical SLOC}} & \multicolumn{1}{c|}{\textbf{Test Type}} & \multicolumn{1}{c|}{\textbf{\# Test Input}} \\
		\hline
		XNIO (2.0.0)  & 3,963 & Integration & 247 \\
		\hline
		OpenChord (1.0.5) & 6,391 & Integration & 1,999 \\
		\hline
		xSocket (2.8.15) & 11,628 & Integration & 1,698 \\
		\hline
		QuickServer (1.4.6) & 13,369 & Integration & 34 \\
		\hline
		Thrift (0.11.0) & 13,543 & Integration & 525 \\
		\hline
		Grizzly (2.4.0) & 22,725 & Integration & 2,000 \\
		\hline
		Karaf (2.4.4) & 46,810 & Integration & 26 \\
		\hline
		\multirow{3}[1]{*}{ZooKeeper (3.4.11)} & \multirow{3}[3]{*}{50,577} & Integration & 2,000 \\
		\cline{3-4}          &       &  Load & 1,409 \\
		\cline{3-4}          &       & System & 2,000 \\
		\hline
		Voldemort (1.9.6) & 66,754 & Integration & 1,631 \\
		\hline
		Netty (4.1.19) & 109,450 & Integration & 1,919 \\
		\hline
		Derby (13.1.1) & 423,662 & Integration & 1,392 \\
		\hline
	\end{tabular}%
	\label{tab:subjects}%
\end{table*}

\begin{enumerate}
\item XNIO is a non-blocking I/O library used to build efficient networking applications~\cite{vsimanskyintegration}.
In its original integration test, I started one server and one client and then sent arbitrary text messages from the client to the server.
To augment the test suite, I created 2,000 files, each of which includes different text contents randomly generated.
During each execution, the XNIO client read one file and sent the full content to the server.
\item OpenChord is a peer-to-peer network service~\cite{dabek2001building}.
In its original integration test, I first started three nodes A, B, and C. Then,
I performed the following operations: On node A, create
an overlay network; on the other nodes B and C, join the
network; on the node C, insert a new data entry to the
network; on the node A, search and then remove the data
entry; Lastly, on the node B, list all data entries.
To augment the test suite, I created 2,000 files, each of which includes a set of OpenChord commands (e.g., {\tt retrieveN -key test}).
Some of these files include invalid commands or command combinations to construct malformed inputs.
During each execution, the nodes read the commands from one file and then perform them in order.
\item xSocket is a NIO-based library for building high-performance computing (HPC) software~\cite{hammerton2014evaluation}.
In its original integration test, after one server and one client were started, the client sent a sequence of manually composed text messages to the server.
I expanded the test suite using the same way that I used for XINO.
\item QuickServer is an open-source library for users to quickly develop multi-client TCP applications~\cite{aydin2015architecture}.
In its original integration test, after its server was started, the client connected to it and then sent a set of text messages to the server.
I expanded the original test suite like that for XINO.
\item Thrift is a framework for developing scalable cross-language services~\cite{rakowski2015learning}.
In its original integration test, I used its libraries to develop a calculator consisting of a server and a client component.
I ran the calculator (from its client) against basic arithmetic operations (i.e., addition, subtraction, multiplication, and division).
To expand the test suite, I created 2,000 files, each of which includes a distinct arithmetic expression (e.g., {\tt 1 add 2 minus 3 multiply 4}),
with some invalid ones to represent malformed inputs.
During each execution, the Thrift client read the expression from one file, sent it to the server, and then took the computation result back from the server.
\item Grizzly is an NIO-based server framework from the GlassFish community~\cite{pelegri2007glassfish}.
In its original integration test, I started a server and a client, and then sent random text messages from the client to the server, and finally waited for the echo of each message.
I expanded the original test inputs almost like for XNIO, except for the addition of a command for awaiting each message's echo sent by the client.
\item Karaf is a modular container as an open-source runtime environment supporting the standard OSGi~\cite{nierbeck2014apache}.
In its original integration test, I created a container hosted by the server and then executed two commands: listing all packages ({\tt la}) and listing OSGi bundles ({\tt list}).
To expand the original input set, I created 2,000 files, each of which includes various Karaf commands (e.g., {\tt config:proplist}).
I purposely included invalid ones in some cases to construct abnormal inputs.
During each execution, the Karaf client read the commands from one file and performed them in order.
\item ZooKeeper achieves consistency and synchronization in distributed systems~\cite{hunt2010zookeeper}.

In its original integration test, the operations were: create two nodes, search them, look up their attributes, update their data association,
and remove these two nodes.
To expand the integration test inputs, I created 2,000 files each of which includes various ZooKeeper commands (e.g., {\tt ls /zk-temp}), including invalid ones.
During each integration test execution, the Zookeeper client read the commands from one file and then performed them in order.

In the original load test, after the server was started, I started a container instance and then generated workloads.
To expand the load test suite, I generated the workloads with 2,000 different sets of configuration parameters (e.g., {\tt the number of clients involved in the load}, {\tt the request size}).
During each load test execution, I ran the load test with one parameter set.

In the original system test, I started a server instance and a system test container, and then launched the system test.
To expand the system test suite, I created 2,000 configuration files each with different configuration parameters (e.g., {\tt initLimit}, {\tt tickTime}, {\tt syncLimit}).
During each system test execution, I ran the original system test code but with one configuration file.
\item Voldemort is a distributed key-value store underlying LinkedIn's service~\cite{sumbaly2012serving}.
In its original integration test, I performed the following operations in order:
add a key-value pair, find the key for its value, remove the key, and retrieve the pair.
To expand the original test suite, I created 2,000 files, each of which includes various Voldemort commands (e.g., {\tt getmetadata}).
Invalid commands or command sequences were included for invalid inputs.
During each execution, the Voldemort client read the commands from one file and then performed each of the commands.
\item Netty is a framework for rapid HPC application development~\cite{maurer2016netty}.
In its original integration test, after starting a server and a client, I sent text messages from the client to the server.
I expanded the original input set using the same way for XINO.
\item Derby is an open-source relational database~\cite{christudas2019derby}.
In its original integration test, I searched all the data records ({\tt SELECT *}) from a relational database (including one table) created beforehand.
To expand the original test, I created 2,000 files, each of which includes a distinct set of SQL statements that are compatible with Derby (e.g., {\tt show settable\_roles}).
Invalid SQL statements or invalid statement sequences were included to construct invalid inputs.
For each execution, the Derby client reads SQL statements from one file and then executed them in sequence.
\end{enumerate}

There are 26 to 2,000 test inputs ({\bf the last column}) for each subject and test type, for a total of 16,880 test inputs for these evaluation subjects.
These 16,880 subject executions formed the basis of all the experiments of {\distmeasure}.

Two dynamic analyzers~\cite{cai2021d2abs,fu2021flowdist} were used to detect information flow paths, 
hence the two direct quality metrics (i.e., information flow path count and information flow path length) based on such paths.
The sources and sinks involved were generated according to JDK security/cryptography APIs.
Source code measurement tool LocMetrics~\cite{aswini2017assessment,kaur2016comparative} was utilized to calculate the scales (i.e., logical SLOC) of distributed systems.
To compute the dynamic (run-time) cyclomatic complexity of each subject execution, I instrumented the subject, executed the instrumented version of the subject, and counted the number of simple conditional decisions exercised in the execution.
Comparison tool {\tt diff} ~\cite{baer2009measuring} was used to calculate the code churn sizes between the adjacent versions of each subject.
All evaluation experiments were performed in Ubuntu 18.04.1 workstations, 
each of them with 256GB DRAM and 2.27GHz CPU.

\subsection{IPC measurements}

My results on IPC measurements are summarized in Table~\ref{tab:rmcrccipr}.
Each number represents one of the six proposed IPC metrics computed for one subject test.

\setlength{\tabcolsep}{9.5pt}
\begin{table}[htbp]
  \centering
  \caption{System-level IPC measurement results} 
    \begin{tabular}{|l||r|r|r|r|r|r|r|}
    \hline
 \multicolumn{1}{|c||}{\textbf{Subject Executions}} & \multicolumn{1}{c|}{\textbf{RMC}} & \multicolumn{1}{c|}{\textbf{RCC}} & \multicolumn{1}{c|}{\textbf{CCC}} & \multicolumn{1}{c|}{\textbf{IPR}} & \multicolumn{1}{c|}{\textbf{CCL}} & \multicolumn{1}{c|}{\textbf{PLC}} \\
    \hline
    XNIO  & 16.82 & 58.96 & 2.76  & 0.51  & 74.43 & 41.10 \\
    \hline
    OpenChord & 3.14  & 67.15 & 5.29  & 0.78  & 267.36 & 255.78 \\
    \hline
    xSocket & 20.74 & 76.77 & 3.63  & 0.36  & 257.98 & 208.19 \\
    \hline
    QuickServer & 2.12  & 36.21 & 2.42  & 0.40  & 113.27 & 70.85 \\
    \hline
    Thrift & 11.59 & 25.27 & 3.17  & 0.56  & 23.40 & 41.59 \\
    \hline
    Grizzly  & 39.68 & 152.09 & 2.16  & 0.67  & 665.13 & 673.34 \\
    \hline
    Karaf & 2.85  & 22.31 & 1.06  & 0.45  & 70.16 & 78.32 \\
    \hline
    ZooKeeper & 6.26  & 191.83 & 3.04  & 0.42  & 506.32 & 461.36 \\
    \hline
    ZooKeeper Load & 4.01  & 90.02 & 1.19  & 0.37  & 391.41 & 369.29 \\
    \hline
    ZooKeeper System & 4.84  & 131.32 & 2.62  & 0.39  & 382.56 & 332.70 \\
    \hline
    Voldemort & 40.17 & 301.54 & 5.02  & 0.54  & 528.32 & 569.79 \\
    \hline
    Netty & 1.00  & 129.00 & 2.38  & 0.54  & 863.39 & 765.36 \\
    \hline
    Derby & 3.31  & 29.45 & 2.22  & 0.72  & 717.70 & 734.12 \\
    \hline
    \end{tabular}%
  \label{tab:rmcrccipr}%
\end{table}%

Two real-world distributed systems, Grizzly and Voldemort, had the largest RMC values.
The reason is that two systems exhibited the highest levels of inter-component dependencies and a lot of message exchange (and corresponding network communications) among their processes during their integration-test executions that needed closely collaborating processes, leading to the highest RMC levels.
And RMC values did not seem to be typically related to the system scales (in terms of logical SLOCs).
For example, the RMC value in the largest system, Derby, is smaller than that in the smallest system, XNIO, with respect to the same test type (i.e., integration test).

Voldemort had the largest RCC value and hence the highest process coupling at class level, followed by Zookeeper;
while Karaf had the small one, meaning that in a Karaf process, a class hardly affects the class in other processes.
It seemed that RCC values did not have significant relationships with RMC values and system scales in terms of logical SLOCs.

In terms of the numbers, the mean CCC was mostly between one and five, meaning that every class collaborated
with about 1 - 5 other classes in remote processes.

OpenChord had the largest IPR value (0.78), while xSocket had the smaller one (0.36).
This meant a lot of functional sharing among the processes of OpenChord.
Conversely, xSockets processes had very low levels of common dependencies.

CCL values were between 23 to 863,
meaning that each class had 23 - 863 times the communication loads of other classes in remote processes.
The largest CCL value 863 was for Netty, while the smallest one was for Thrift.
This suggests the highest level of interprocess communication complexity for Netty and further the most difficult debugging for Netty IPC correctness/performance.
Instead, Thrift had the lowest interprocess communication complexity and hence could be debugged easily.

I saw that PLC values ranged from 41.10 to 765.36,
meaning a process' methods depending on about 41 to 765 others in the same process.
I also observed that high/low CCL values came along with high/low PLC values in general.
The reason is that the evaluation executions had overall closeness between process-level coupling and cohesion,
in terms of CCL for coupling and PLC for cohesion.

In sum, I found that four IPC metrics (i.e., RMC/RCC/CCC/IPR) were related to neither subject scales nor with any other IPC metrics.
This implies that the associations among these four IPC metrics were few and each of the RMC/RCC/CCC/IPR hardly influenced others.
In addition, I observed an obvious difference between RMC/RCC/CCL/PLC and CCC/IPR, that RMC/RCC/CCL/PLC varied sharply with very small variations in CCC/IPR.
This difference suggests that RMC/RCC/CCL/PLC changed in relatively wide ranges across systems, but CCC/IPR changed little.

\subsection{Quality measurements correlated to IPC measurements} \label{sss:qmcim}

Information flow path count, which is the total number of such paths, and information flow path length is the average length of the paths.
Both of then are normalized by the subject logic thousand SLOC (KSLOC), as listed in Table~\ref{tab:paths}.
\setlength{\tabcolsep}{1.5pt}
\begin{table*}[tp]
	\centering  
	\caption{Measurement results for two direct quality metrics: {\em information flow path count} and {\em information flow path length}}
\resizebox{1\textwidth}{!}{%
		\begin{tabular}{l||r|r|r|r|r}
			\hline
			Subject Execution & \multicolumn{1}{c|}{Thrift} & \multicolumn{1}{c|}{xSocket} & \multicolumn{1}{c|}{Voldemort} & \multicolumn{1}{c|}{ZooKeeper\_Load} & \multicolumn{1}{c}{Netty} \\
			\hline
			\hline			
			Information flow path count & 2.22E-04 &  1.72E-04 &  6.28E-04 &  1.27E-03 &  1.83E-05 \\
			\hline
			Information flow path length & 1.26E-02 &  4.69E-03 &  5.43E-03 &  9.22E-03 &  2.26E-05 \\
			\hline
		\end{tabular}%
}
	\label{tab:paths}%
\end{table*}%

The direct measures of each of the six quality metrics (except for information flow path count and length) over all subjects are summarized in Table~\ref{tab:otherfactors}.
When quantifying a quality factor, 
I selected logic SLOC/KSLOC, 
as the normalizing unit according to the value range of the quality factor's measure,
to avoid the metric value being too small to be ignored.

Then, I conducted extensive statistical analysis to discover significant associations between the coupling metrics and
the quality metrics, via a non-parametric correlation analysis using Spearman's method~\cite{sedgwick2014spearman}.
I showed, through empirically validated correlations, promising applications of IPC metrics
in predicting and understanding the quality/behaviors of complicated distributed systems.
The results showed widely varying correlations between six IPC metrics and eight quality metrics for 6 x 8 = 48 pairs.
I marked significant correlations in {\bf boldface}, as shown in Table~\ref{tab:correlations}.

\begin{table*}[htbp]
	\centering
	\caption{Measurements of other (six) direct quality metrics, normalized by logical SLOC or KSLOC}
	\resizebox{1\textwidth}{!}{%
		\begin{tabular}{|l||r|r|r|r|r|r|}
			\hline
			\multicolumn{1}{|c||}{\multirow{2}[1]{*}{\textbf{Subject}}} & \multicolumn{1}{c|}{\textbf{Execution}} & \multicolumn{1}{c|}{\textbf{Code Churn}} & \multicolumn{1}{c|}{\textbf{Cyclomatic}} & \multicolumn{1}{c|}{\textbf{Defect}} & \multicolumn{1}{c|}{\textbf{Attack}} & \multicolumn{1}{c|}{\multirow{2}[1]{*}{\textbf{Vulnerableness}}} \\
			& \multicolumn{1}{c|}{\textbf{Time (seconds) }} & \multicolumn{1}{c|}{\textbf{Size}} & \multicolumn{1}{c|}{\textbf{Complexity}} & \multicolumn{1}{c|}{\textbf{Density}} & \multicolumn{1}{c|}{\textbf{Surface}} &  \\
			\hline
			XNIO  & 1.51E-03 & 9.28E-02 & 2.03E-01 & 9.34E-03 & 4.55E-03 & 1.01E+00 \\
			\hline
			OpenChord & 8.45E-03 & 8.87E-02 & 1.72E-01 & 2.35E-03 & 1.13E-02 & 6.26E-01 \\
			\hline
			xSocket & 9.46E-04 & 4.83E-02 & 2.38E-01 & 5.16E-04 & 1.98E-03 & 2.58E-01 \\
			\hline
			QuickServer & 8.23E-04 & 4.73E-02 & 2.45E-01 & 2.99E-04 & 7.10E-04 & 4.38E-01 \\
			\hline
			Thrift & 5.91E-04 & 6.57E-02 & 1.45E-01 & 4.28E-03 & 2.51E-03 & 8.14E+00 \\
			\hline
			Grizzly  & 3.08E-04 & 7.81E-02 & 2.21E-01 & 1.98E-03 & 6.30E-04 & 8.80E-01 \\
			\hline
			Karaf & 5.13E-04 & 1.17E-02 & 1.66E-01 & 9.08E-03 & 3.01E-04 & 1.96E+00 \\
			\hline
			ZooKeeper & 4.60E-04 & 1.37E-02 & 5.17E-02 & 7.41E-03 & 6.16E-03 & 9.00E-01 \\
			\hline
			Voldemort & 4.19E-04 & 3.97E-03 & 2.42E-01 & 4.93E-04 & 4.61E-03 & 1.80E-01 \\
			\hline
			Netty & 1.10E-04 & 1.11E-02 & 2.33E-01 & 5.03E-03 & 7.13E-04 & 1.80E-01 \\
			\hline
			Derby & 5.90E-05 & 2.70E-02 & 1.55E-01 & 9.70E-03 & 4.91E-04 & 5.12E-01 \\
			\hline
		\end{tabular}%
	}
	\label{tab:otherfactors}%
\end{table*}


From the results, five (out of six total) IPC metrics (i.e., RMC, RCC, CCC, CCL, and PLC) were significantly correlated to and informative of one or more of six (out of eight total) quality metrics (i.e., Execution Time, Code Churn Size, Cyclomatic Complexity, Defect Density, Attack Surface, and Vulnerableness),
despite varying correlation strengths.
High IPC coupling based on the dependencies between a process and others (i.e., high RMC/RCC/CCC) was significantly
correlated to long execution time, high complexity, low defect density, large attack surface, and few vulnerabilities reported.
Also, a large number of communication loads of a class with others in remote process (i.e., high CCL) suggest small attack surface and low vulnerableness.
Meanwhile, high cohesion in an individual process (i.e., high PLC) implies small code churn size, small attack surface, and few vulnerabilities.

According to Table~\ref{tab:correlations}, from IPC metric(s) for predicting dynamic quality metrics, I constructed four classifiers (IPC metric(s) -$>$ corresponding quality metric):
(CCC) -$>$ execution time, (RMC,CCC) -$>$ (run-time) cyclomatic complexity, (CCC) -$>$ attack surface, and (CCL, PLC) -$>$ attack surface.
For static quality metrics, in a similar way, I built one classifier (CCL, PLC) -$>$ vulnerableness.

\begin{table*}[tp]
  \centering
  \caption{Spearman`s correlation coefficients (p-values) between IPC and quality metrics}
	\resizebox{1\textwidth}{!}{%
    \begin{tabular}{l|l|l|r|r|r|r|r|r}
    \hline
    \multicolumn{3}{c|}{\textbf{Quality metrics}} & \multicolumn{6}{c}{\textbf{IPC metrics}} \\
    \hline
    \multicolumn{1}{c|}{\textbf{Level}} & \multicolumn{1}{c|}{\textbf{Type}} & \multicolumn{1}{c|}{\textbf{Name}} & \multicolumn{1}{c|}{\textbf{RMC}} & \multicolumn{1}{c|}{\textbf{RCC}} & \multicolumn{1}{c|}{\textbf{CCC}} &  \multicolumn{1}{c|}{\textbf{IPR}}     & \multicolumn{1}{c|}{\textbf{CCL}} & \multicolumn{1}{c}{\textbf{PLC}} \\
    \hline
    \multirow{3}[3]{*}{System} & \multirow{3}[3]{*}{Static} & Code Churn Size & {\makecell[l]{0.2320\\(4.46E-01)}}& {\makecell[l]{-0.3260\\(2.77E-01)}}& {\makecell[l]{0.2541\\(4.02E-01)}}& {\makecell[l]{0.2707\\(3.71E-01)}}& {\makecell[l]{-0.3978\\(1.78E-01)}}& {\makecell[l]{\textbf{-0.4972}\\(8.38E-02)}}\\
\cline{3-9}          &       & Defect Density &
{\makecell[l]{-0.2210\\(4.68E-01)}}& {\makecell[l]{-0.3149\\(2.95E-01)}}& {\makecell[l]{\textbf{-0.4475}\\(1.25E-01)}}& {\makecell[l]{0.0497\\(8.72E-01)}}& {\makecell[l]{0.0166\\(9.57E-01)}}& {\makecell[l]{0.0331\\(9.14E-01)}}\\
\cline{3-9}          &       & Vulnerableness &
{\makecell[l]{0.0553\\(8.58E-01)}}& {\makecell[l]{\textbf{-0.4205}\\(1.53E-01)}}& {\makecell[l]{-0.2407\\(4.28E-01)}}& {\makecell[l]{-0.0470\\(8.79E-01)}}& {\makecell[l]{\textbf{-0.6224}\\(2.31E-02)}}& {\makecell[l]{\textbf{-0.5616}\\(4.58E-02)}}\\
    \hline
    \hline

\multirow{5}[4]{*}{Execution} & \multirow{5}[4]{*}{Dynamic} & Execution Time &
{\makecell[l]{-0.2845\\(0.00E+00)}}& {\makecell[l]{-0.0864\\(2.39E-29)}}&
{\makecell[l]{\textbf{0.4735}\\(0.00E+00)}}& {\makecell[l]{0.0461\\(2.11E-09)}}&
{\makecell[l]{-0.1554\\(1.10E-91)}}& {\makecell[l]{-0.2471\\(3.32E-233)}}\\
\cline{3-9}          &       & {\makecell[l]{Cyclomatic\\Complexity}} &
{\makecell[l]{\textbf{0.4239}\\(0.00E+00)}}& {\makecell[l]{0.0573\\(9.16E-14)}}&
{\makecell[l]{\textbf{0.4831}\\(0.00E+00)}}& {\makecell[l]{0.1988\\(5.77E-150)}}&
{\makecell[l]{-0.3644\\(0.00E+00)}}& {\makecell[l]{-0.3941\\(0.00E+00)}}\\
\cline{3-9}          &       & {\makecell[l]{Information Flow\\Path Count}} &
{\makecell[l]{0.1295\\(4.99E-64)}}& {\makecell[l]{0.0777\\(4.97E-24)}}&
{\makecell[l]{-0.0091\\(2.38E-01)}}& {\makecell[l]{-0.3771\\(0.00E+00)}}&
{\makecell[l]{-0.1327\\(3.31E-67)}}& {\makecell[l]{-0.1570\\(1.15E-93)}}\\
\cline{3-9}          &       & {\makecell[l]{Information Flow\\Path Length}} &
{\makecell[l]{-0.1616\\(3.45E-99)}}& {\makecell[l]{0.0349\\(5.77E-06)}}&
{\makecell[l]{-0.1159\\(1.50E-51)}}& {\makecell[l]{-0.2494\\(1.24E-237)}}&
{\makecell[l]{0.1134\\(2.07E-49)}}& {\makecell[l]{0.0757\\(6.59E-23)}}\\
\cline{3-9}          &       & Attack Surface &
{\makecell[l]{0.1556\\(5.35E-92)}}& {\makecell[l]{-0.2132\\(9.49E-173)}}&
{\makecell[l]{\textbf{0.4534}\\(0.00E+00)}}& {\makecell[l]{-0.0868\\(1.42E-29)}}&
{\makecell[l]{\textbf{-0.7424}\\(0.00E+00)}}& {\makecell[l]{\textbf{-0.7968}\\(0.00E+00)}}\\
    \hline
    \end{tabular}%
}
  \label{tab:correlations}%
\end{table*}%

In addition, the p-value of a correlation coefficient is the probability that the coefficient is not significant.
If the p-value is less than 5\%, the correlation coefficient is called statistically significant.
As shown in Table ~\ref{tab:correlations}, zero (0.00E+00) p-values for corresponding coefficients signal significant correlations between the five (out of six total) IPC metrics (i.e., RMC/CCC/CCL/IPR/PLC) and one or more of four (out of five total) dynamic quality metrics (i.e., Execution Time, Cyclomatic Complexity, Information Flow Path Length, and Attack Surface) considered.
However, for static quality metrics (versus IPC metrics), corresponding p-values (most
of them $>$ 0.05) could not indicate significant coefficients due to too few (only 11) static data points.

\subsection{Classification for predicting and understanding quality}
In this section, I present my experimental results and conclusions, focusing on the results of classifying dynamic quality metrics with respect to a large number of (16,880) subject executions as samples.
However, for classifying static quality metrics, only eleven available samples (i.e., subjects) were too few for training practical models, and hence possibly leading to a solid conclusion.

For unsupervised classifiers for predictable dynamic quality metrics, Table~\ref{tab:kmeanss} shows their precision, recall, and F1 accuracy.
For the classifier, the first and second columns show IPC metric(s) -$>$ corresponding quality metric.
The 3--5th columns and 7--8th columns show hold-out validation results and 10-fold cross-validation (CV) results, respectively.
Average 74\% F1 demonstrated generally useful accuracy for classifying the distributed system quality with respect to dynamic quality metrics (i.e., {\em execution time}, {\em run-time cyclomatic complexity}, and {\em attack surface}), achieved by the unsupervised learning (k-means) algorithm.
In addition, two validation results showed very similar precision, recall, and F1 accuracy (average 90.96\% versus 88.20\%, 66.86\% versus 66.71\%, and 74.12\% versus 74.51\%), both on overall average and for each individual classifier.
This meant that these evaluation results were consistent and hence developers could consolidate the result reliability and validity.

\setlength{\tabcolsep}{3.5pt}
\begin{table}[tp]
  \centering
  \caption{The effectiveness of unsupervised learning ($k$-means) classification for 
  	dynamic predictable quality metrics}
	\resizebox{\columnwidth}{!}{%
    \begin{tabular}{|lr||r|r|r||r|r|r|}
    \hline
    \multicolumn{2}{|c||}{\textbf{Model}} & \multicolumn{3}{c||}{\textbf{Hold-out Validation}} & \multicolumn{3}{c|}{\textbf{10-fold Cross-validation}} \\
    \hline
    \multicolumn{1}{|c|}{\textbf{IPC Metric}} & \multicolumn{1}{c||}{\textbf{Quality Metric}} & \multicolumn{1}{c|}{\textbf{Precision}} & \multicolumn{1}{c|}{\textbf{Recall}} & \multicolumn{1}{c||}{\textbf{F1}} & \multicolumn{1}{c|}{\textbf{Precision}} & \multicolumn{1}{c|}{\textbf{Recall}} & \multicolumn{1}{c|}{\textbf{F1}} \\
    \hline
    \multicolumn{1}{|l|}{CCC} & \multicolumn{1}{l||}{Execution Time} & 100.00\% & 37.84\% & 54.90\% & 83.33\% & 41.27\% & 55.20\% \\
    \hline
    \multicolumn{1}{|l|}{RMC, CCC} & \multicolumn{1}{l||}{Cyclomatic Complexity} & 73.23\% & 84.91\% & 78.64\% & 99.68\% & 66.85\% & 80.03\% \\
    \hline
    \multicolumn{1}{|l|}{CCC} & \multicolumn{1}{l||}{Attack Surface} & 90.87\% & 83.22\% & 86.88\% & 98.10\% & 77.73\% & 86.73\% \\
    \hline
    \multicolumn{1}{|l|}{CCL, PLC} & \multicolumn{1}{l||}{Attack Surface} & 99.73\% & 61.47\% & 76.06\% & 71.70\% & 81.01\% & 76.07\% \\
    \hline
    \multicolumn{2}{|l||}{\textbf{Average:}} & \textbf{90.96\%} & \textbf{66.86\%} & \textbf{74.12\%} & \textbf{88.20\%} & \textbf{66.71\%} & \textbf{74.51\%} \\
    \hline
    \end{tabular}%
	}
  \label{tab:kmeanss}%
\end{table}%

However, the unsupervised learning ($k$-means) algorithm did not achieve the high levels of effectiveness in classifying {\em execution time} (The recall is lower than 42\% and F1 accuracy is lower than 56\%).
This suggests that the clustering algorithm
was not very effective to capture the {\em time behavior} status of a distributed system, according to the system IPC measurement result (CCC).
The main reason is that the ($k$-means) algorithm is based on the similarity of data points, but system executions, which have near IPC measurements (e.g., CCC), might not have close execution time.
Therefore, the unsupervised learning (e.g., $k$-means) classification might not be the best option for assessing the quality metric {\em execution time} through CCC.

\setlength{\tabcolsep}{3.5pt}
\begin{table}[htbp]
  \centering
  \caption{The effectiveness of supervised learning (bagging) classification for dynamic predictable quality metrics}
	\resizebox{\columnwidth}{!}{%
    \begin{tabular}{|lr||r|r|r||r|r|r|}
    \hline
    \multicolumn{2}{|c||}{\textbf{Model}} & \multicolumn{3}{c||}{\textbf{Hold-out Validation}} & \multicolumn{3}{c|}{\textbf{10-fold Cross-validation}} \\
    \hline
    \multicolumn{1}{|c|}{\textbf{IPC Metric}} & \multicolumn{1}{c||}{\textbf{Quality Metric}} & \multicolumn{1}{c|}{\textbf{Precision}} & \multicolumn{1}{c|}{\textbf{Recall}} & \multicolumn{1}{c||}{\textbf{F1}} & \multicolumn{1}{c|}{\textbf{Precision}} & \multicolumn{1}{c|}{\textbf{Recall}} & \multicolumn{1}{c|}{\textbf{F1}} \\
    \hline
    \multicolumn{1}{|l|}{CCC} & \multicolumn{1}{l||}{Execution Time} & 99.80\% & 99.80\% & 99.80\% & 99.70\% & 99.70\% & 99.70\% \\
    \hline
    \multicolumn{1}{|l|}{RMC, CCC} & \multicolumn{1}{l||}{Cyclomatic Complexity} & 99.10\% & 99.10\% & 99.10\% & 99.70\% & 99.70\% & 99.70\% \\
    \hline
    \multicolumn{1}{|l|}{CCC} & \multicolumn{1}{l||}{Attack Surface} & 96.50\% & 96.50\% & 96.50\% & 95.20\% & 95.20\% & 95.20\% \\
    \hline
    \multicolumn{1}{|l|}{CCL, PLC} & \multicolumn{1}{l||}{Attack Surface} & 97.50\% & 97.50\% & 97.50\% & 96.80\% & 96.80\% & 96.80\% \\
    \hline
    \multicolumn{2}{|l||}{\textbf{Average:}} & \textbf{98.23\%} & \textbf{98.23\%} & \textbf{98.23\%} & \textbf{97.85\%} & \textbf{97.85\%} & \textbf{97.85\%} \\
    \hline
    \end{tabular}%
	}
  \label{tab:baggings}%
\end{table}%

On the contrary, supervised learning classifications between IPC metrics and dynamic predictable quality metrics were more effective for predicting these quality metrics, as shown in Table~\ref{tab:baggings}.
For the same classification tasks between the same IPC and quality metrics, related to supervised classifiers,
both the hold-out validation and 10-fold CV revealed high levels of precision, recall, and F1 accuracy, which all were above 95\%.
In particular, the overall average F1 accuracy was about 98\%.
Compared with unsupervised classification results, these supervised classification results also had consistency in all of these three effectiveness metrics,
not only for individual classifiers but for overall.
There, with respect to available labeled training samples, supervised classifiers were clearly more suitable for accurate classifications in {\distmeasure} than unsupervised classifiers.

\begin{figure}[tpbh]
  \centering
	\caption{Ranking of features (IPC metrics) by importance score (shown on the $x$ axis) for supervised (bagging) classifiers for quality metrics. The left figure shows the importance scores of RMC versus CCC in classifying {\em run-time cyclomatic complexity}, while the right one shows the scores of CCL versus PLC in classifying {\em attack surface}}
  \includegraphics[width=0.49\textwidth]{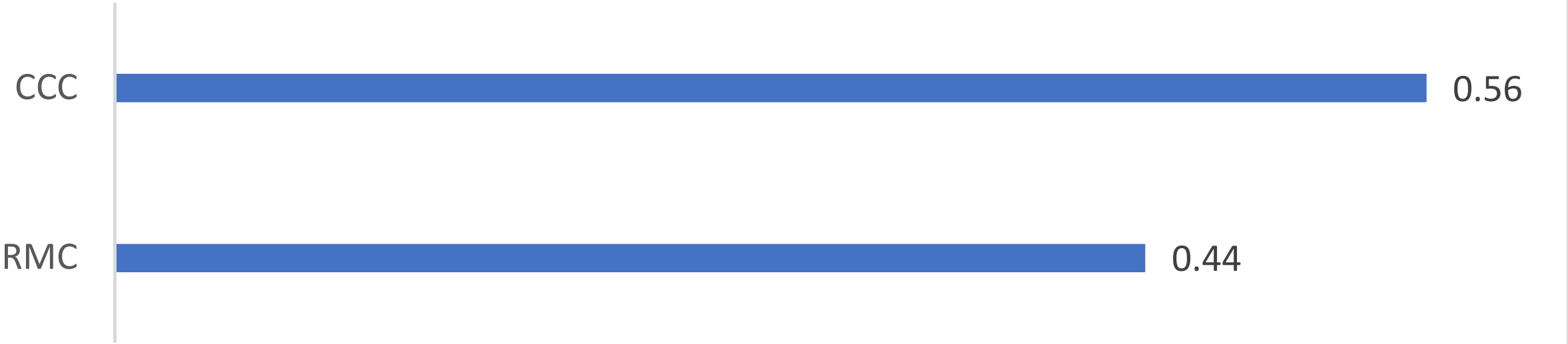}
  \includegraphics[width=0.49\textwidth]{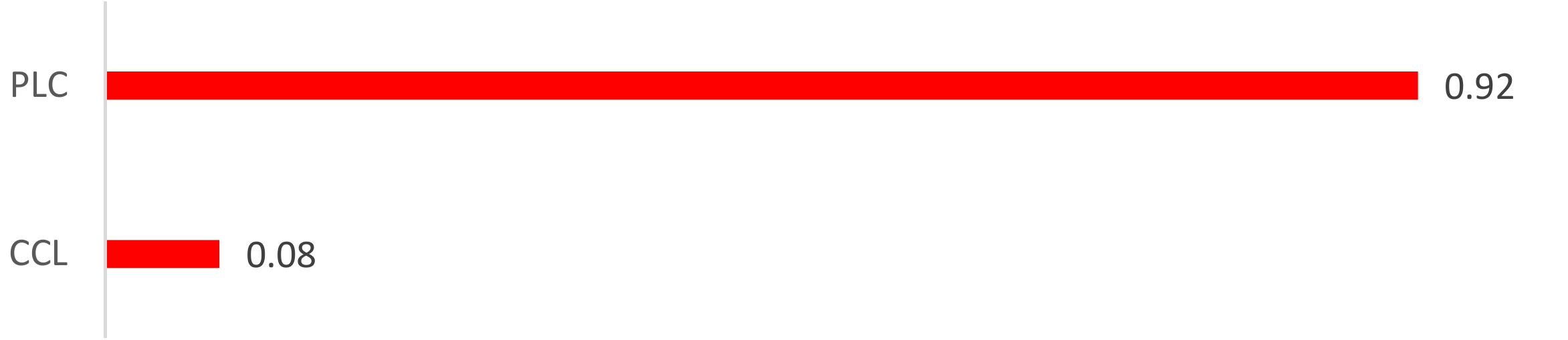}
  \vspace{-10pt}
  
  \label{fig:importances}
\end{figure}
I used Python Scikit-learn library~\cite{pedregosa2011scikit} to compute feature importance scores as the standard deviation and mean of the impurity decrease accumulation \cite{cassidy2014calculating} (i.e., feature\_importances in Python) for the supervised classifiers in {\distmeasure}.
The results (i.e., features ranking by importance scores) are depicted in Figure~\ref{fig:importances}.
Yet two classifiers "(CCC) -$>$ execution" and "time (CCC) -$>$ attack surface" have one feature (CCC) only, without the requirement of feature selection.
The results show that CCC, the strongest IPC coupling metric as discussed earlier,
was clearly more important than RMC for classifying the quality metric {\em run-time cyclomatic complexity} (with the importance score of 0.56 versus 0.44).
And PLC, the only IPC cohesion metric in {\distmeasure},
was much more important than CCL for classifying the quality metric {\em attack surface} (with the importance score of 0.92 versus 0.08).
Note that feature importance and ranking are not relevant in $k$-means clustering, thus this study was only conducted on my supervised classifiers.

\begin{figure*}[tp]
  \centering
  \caption{The average accuracy of alternative supervised learning algorithms compared with the default algorithm (bagging)}
  \includegraphics[width=1\textwidth]{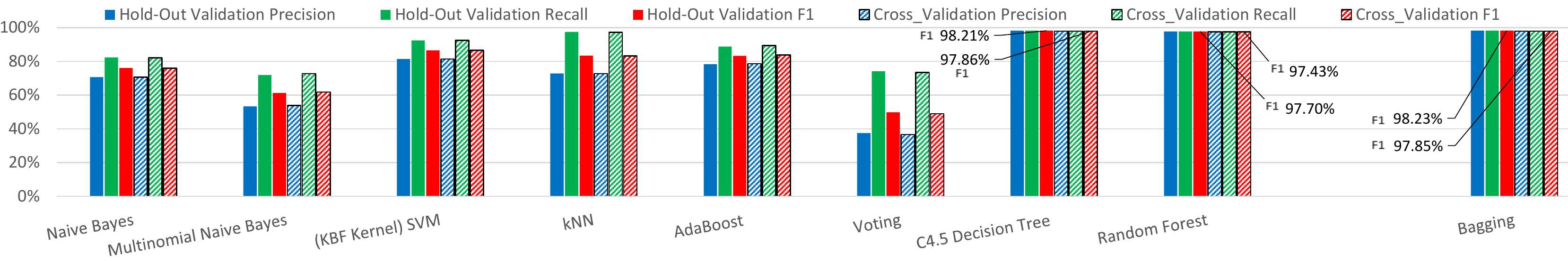}
  \label{fig:accuracy}
\end{figure*}

I compared the default supervised classification (bagging) algorithm with eight alternative supervised learning algorithms when classifying dynamic, predictable quality metrics
based on the IPC metrics.
Figure~\ref{fig:accuracy} shows the overall average precision, recall, or F1 accuracy across, with both hold-out validation and 10-fold cross-validation.
From the figure, the {\em bagging} algorithm demonstrates the best performance among all these supervised learning algorithms considered.
Specially,
{\em C4.5 Decision Tree} algorithm achieved the second highest F1 accuracy, followed by {\em Random Forest},
while two algorithms, {\em Voting} and {\em Multinomial Naive Bayes}, had the worst F1 accuracy.
Other algorithms had similar classification effectiveness in terms of precision, recall, or F1 accuracy, in either hold-out validations or 10-fold cross-validations.

\subsection{Threats to Validity}\label{s:distmeasurethreats}
The empirical results are subject to various common kinds of validity threats according to~\cite{wohlin2000experimentation}.
I describe below each major kind and then discuss how I control or mitigate relevant validity threats.

\vspace{2pt}
\noindent
\textbf{Internal validity.}
The main threat to \emph{internal} validity lies
in possible errors in the implementation in measuring/computing IPC/quality metrics,
analyzing correlations, and classifying quality.
To reduce this threat, I carefully reviewed the code and manually validated the analysis/prediction results of two smallest subjects.
In particular, I confirmed the correctness of these subject's executions' dynamic dependencies used in {\distmeasure} computations.
As a result, I did not find such issues.

In addition, {\distmeasure} results might have been influenced by various factors not considered but have affected
the process for collecting the data needed for directly quantifying the studied quality metrics.
For example, in order to measure the quality metric code churn size,
I needed to gather the data on system release/version histories, from various sources.
However, some system releases might not be available online.
To mitigate this threat, I carefully searched/reviewed all possible subject release data
and contracted relevant developers with requests for the missing releases.

The measurement results of some particular quality metrics may not be valid with respect to other possible relevant quality measures that I did not consider.
For example, run-time cyclomatic complexity is currently used to the quality sub-characteristic testability, without considered other metrics that may affect testability.
Systems with similar (run-time) cyclomatic complexity may not always be similarly testable.
For instance, two programs have similar code but read/write various external files.
The former handles large files with complex data structures, while the latter operates simple files (e.g., text files).
Thus, they have close (run-time) cyclomatic complexity but different testability.
Some other quality metrics (e.g., availability, network throughput, response time, scalability) were not considered
since I only concerned the quality metrics related to IPC-related system behaviors justifiably.

\vspace{2pt}
\noindent
\textbf{External validity.}
The main threat to \emph{external} validity is that my study results may not generalize to
other distributed programs and executions.
To reduce this threat, I selected subjects covering various architectures, domains, and scales.

With limited run-time inputs, a relevant validity threat arises from classifying the software quality in {\bf Part 2}, with respect to {\em static} quality metrics.
With different run-time inputs, the correlations between the IPC metrics and static quality metrics, and
the performance of the classifiers based on IPC metrics for classifying quality metrics, may be different from those reported.

Another external validity threat is that it is hard to collect all three test types (e.g., integration, system, and load tests) that are only for ZooKeeper.
Fortunately, integration tests were available for each subject and exercised {\em whole-system} behaviors.
Furthermore, I have manually constructed lots of unique run-time inputs for subjects, to significantly increase the total run-time coverage for all subjects, and hence to reduce the threat.

\vspace{2pt}
\noindent
\textbf{Construct validity.}
The main threat to \emph{construct} validity is for the use of statistical analysis to cause my conclusions.
In computing the system-level IPC metrics, I took the means of lower-level metrics without concerning the variations (e.g., standard deviations).
To reduce the threat as regards the correlation analysis, I purposely chose Spearman`s method over alternatives as it is a non-parametric method that does not assume normality of underlying data distribution
or relationships between the data groups.

On the other hand, although I applied largely augment the run-time inputs to the subjects,
this augmentation did not help get more samples for training and testing the quality classifiers with respect to static quality metrics.
As a result, the evaluation for these classifiers might be quite premature.
I would need thousands of different Java distributed systems as subjects for a much stronger evaluation in this regard.

\vspace{2pt}
\noindent
\textbf{Conclusion validity.}
The lack of enough distributed system subjects (only 11) causes a threat to \emph{conclusion} validity regarding static quality metric classifiers.
In building these classifiers, I measured the IPC metrics during subject executions and aggregated corresponding results.
This process ignored the different characteristics of dynamic behaviors across various executions of each subject,
and essentially treated all executions of each subject as representations of its holistic behaviors.

This aggregation ignored the varying characteristics of dynamic behavioral profiles across different executions of each subject, and essentially treated all of the executions of each subject as being collectively representative of that subject's holistic behaviors.
Such a treatment may lead to biases about the performance of these classifiers.

Eventually, the quality classifiers only estimate whether the quality metrics are anomalous or not (whether they are worth warning or not) only at high levels.
Currently, {\distmeasure} is not sufficient for pinpointing such anomalies when warned.

\section{Lessons Learned and Takeaways}
My exploration of IPC metrics not only demonstrated the practicality of
measuring IPC in large, real-world distributed systems,
but also revealed the substantial presence (albeit with varying degrees) of {\em implicit} coupling among distributed components (generally decoupled in architectural design).
And I showed that one way to reveal such implicit coupling is through measuring interprocess coupling.
My results on IPC measurements revealed that higher coupling in terms of inter-process dependencies
is generally bad for quality (by being significantly indicative of lower quality with respect to five out of the
six factors considered).

In particular,
in distributed program executions, high IPC coupling was typically associated with low quality in some characteristics/sub-characteristics
(e.g., bad time behaviour and further low performance efficiency, low modifiability/testability and further low maintainability, low security)
and corresponding quality metrics (e.g., long execution time, high complexity, large attack surface).
This largely consolidates the drawbacks of high coupling, generally leading to low quality~\cite{kan2002metrics}.

Conversely, higher process-level cohesion benefited the system quality via high security in multiple sub-characteristics
(i.e., high confidentiality, integrity, non-repudiation, accountability, and authenticity)
and relevant quality metrics (e.g., low attack surface and vulnerableness).
This confirmed the previous finding that high cohesion (in an individual software process/component) is typically related to high quality~\cite{gelinas2006cohesion}.

In summary, distributed system developers are recommended to achieve and maintain low overall (implicit) coupling among system components for
high system quality, especially when they concern quality characteristics including performance efficiency, functional suitability, and security.
On the other hand, developers prefer to program high cohesive components with respect to the merits of higher process-level cohesion, when concerning better maintainability and security~\cite{iso25010}.
In short, low coupling and high cohesion should be the targets of distributed system developers.

From the statistical results of {\distmeasure} {\bf Part 1},
there was only one classifier for one {\em static} predictable quality metric: (CCL, PLC) -$>$ vulnerableness, discussed in $\S$\ref{sss:qmcim}.
However, the corresponding dataset was not sufficient (only 11 data points, as shown in Tables~\ref{tab:otherfactors}) for unsupervised learning and supervised learning in {\distmeasure} {\bf Part 2}.
In particular, 10-fold cross-validation could not be performed due to too few samples.
Much more data points and samples are required for training/testing a sound classifier for the static predictable quality metric (i.e., vulnerableness).

As one of the six IPC metrics in {\distmeasure}, IPR did not significantly correlate to any direct quality metric considered.
However, it is still useful for understanding run-time code reuse in distributed systems.
Meanwhile, the other five IPC metrics have shown to be indicative of distributed software quality with respect to at least one of the quality metrics, static and/or dynamic.

In particular, the strongest indicator of interprocess coupling, CCC, which captures the coupling {\em between all} of the distributed processes (as opposed to measuring that {\em between two} processes as RCC does) at a fine-grained granularity level (as opposed to measuring that at a coarse level as RMC does) tended to be the most indicative of distributed software quality. Indeed, CCC was found to be significantly correlated with more (four) quality metrics than any other IPC metrics in the experiments. As a learning feature, CCC also contributed to the quality classifications more than other features wherever more than one feature was used in the classifiers.
Meanwhile, PLC, the only metric in the framework that captures process cohesion, exhibited similar merits to CCC---it was found to be significantly correlated with three quality metrics, more than other IPC metrics except for CCC, while having contributed to relevant classifiers much more than other features.
These results demonstrated the great usefulness of measuring both coupling and cohesion
at process level for assessing distributed software quality.

The results also show that supervised classifiers had clear advantages over unsupervised classifiers, at least for dynamic, predictable quality metrics of distributed systems.
This comparison suggests that predicting those quality metrics purely based on the similarity of corresponding IPC metrics across different system executions might not be effective.
The main reason is that varied system executions may have very different quality measurements even if they are close to enough IPC measurements.
Instead, a better way would be based on supervised classifications with respect to labeled data.

On the other hand, the advantages of supervised classifications over unsupervised classifications in {\distmeasure} also carry a price that handling relevant quality metric values might be expensive, especially because a large number of such values were computed and labeled as training samples
in the evaluation of {\distmeasure}.

In addition, for supervised (bagging) classifiers, computed Spearman's rank correlation coefficients and feature importance scores
were consistent, as shown in Table~\ref{tab:correlations} and Figure~\ref{fig:importances}.
For the quality metric {\em run-time cyclomatic complexity}, the IPC metric CCC had larger correlation coefficient than RMC did (0.448 versus 0.42);
while CCC was more important than RMC in the classification for {\em run-time cyclomatic complexity} (importance scores of 0.56 versus 0.44).
A similar consistency was observed in the classifier "(CCL, PLC) -$>$ attack surface"---The IPC metric CCL was more strongly correlated with {\em attack surface}, and also more important in the classifier, than PLC.

\section{Related Work}
In this section, I discuss other previous works in three categories: dynamic coupling metrics ($\S$\ref{sec:dcm}), run-time cohesion ($\S$\ref{sec:cohesion}), and 
predictive software quality assessment based on machine learning ($\S$\ref{sec:mlfpsq}).


\subsection{Dynamic coupling metrics}\label{sec:dcm}
Jin et al.~\cite{jin2018dynamic} defined a dynamic component coupling metric ($CPC$) directly based on inter-component
dependencies derived from method executions with timing information. Conceptually, the $CPC$ metric is closely related to
our $IPR$ metric, in that both are based on approximated dynamic dependencies across components. However, the interprocess
dependencies on which our $IPR$ computation is based are significantly more precise than the
purely control-flow-based dependencies approximated in~\cite{jin2018dynamic}, according to ~\cite{cai2016distea}.
In addition, $CPC$ was defined for measuring structural complexity, while $IPR$ is proposed primarily as a
reusability metric.
Previous reuse metrics mainly concern reusing library code and connectivity between server and
client nodes as a whole~\cite{frakes1996software}. Instead, I measure interprocess reusability at code level in terms of metrics defined based on code dependencies.

\subsection{Run-time cohesion}\label{sec:cohesion}
Jin et al.~\cite{jin2016dynamic} also proposed a dynamic cohesion measurement approach for distributed software, which includes two component-level cohesion metrics (i.e., {\em CC} and {\em CCW}) by extending the metric {\em lack of cohesion in methods (LCOM)}, a classical cohesion metric for single-process programs.
A structural quality attribute {\em cohesion factor of component (CHC)} was later introduced also for distributed software~\cite{jin2018dynamic}.
These cohesion metrics were 
only evaluated against
{\em specialized} distributed programs (e.g., Netflix RSS Reader, RSS Reader Recipes, and/or the distributed version of
iBATIS JPetStore)~\cite{jin2016dynamic,jin2018dynamic}.
The underlying measurement tool used was also designed for these specialized systems only.
In comparison, the cohesion metric PLC is defined based on method-level dynamic dependencies and
evaluated against real-world {\em common} distributed systems.

\subsection{Software quality assessment based on machine learning}\label{sec:mlfpsq}
Software quality prediction helps developers with better utilization of resources, effort estimates, and making testing plans for components that may have defects, hence reducing development costs and mitigating risks at initial stages~\cite{rana2007survey,sheoran2016software}.
Various machine learning techniques, using unsupervised or supervised learning algorithms (e.g., logistic regression, support vector machine, neural networks, and $k$-means clustering) have been used for software quality assessment.

For instance,
Khoshgoftaar and Allen proposed a software quality assessment model using logistic regression~\cite{khoshgoftaar1999logistic}.
Xing et al. employed a support vector machine technique for the classification of software modules based on
a complexity metric to predict software quality in early development stages~\cite{rashid2014machine}.
Abe et al. used a Bayesian classifier to estimate the success or failure of a software project~\cite{abe2006estimation}.
In addition, neural network techniques were also applied to the prediction of software quality~\cite{khoshgoftaar1996using,thwin2002application,wang2004extract,sheoran2016software}.
Furthermore, Zhang et al.~\cite{zhang2015smplearner} presented SMPLearner 
that trained a software maintainability prediction model by gathering the real maintenance efforts computed from code change histories; it experimented with 24 common machine learning techniques, including SVM regression, random forest, $K$-star, etc.

As a type of software quality assessment model, 
numerous software defect prediction models have been built from single or multiple software projects for within- or cross-project  prediction~\cite{zhang2014towards,zhang2016cross,zhou2018far}, respectively.
For example,
Zhang et al. built a universal defect prediction model with a large number of software projects from various contexts by clustering projects
based on the similarity of distribution across multiple predictors,
deriving the rank transformations,
and fitting the model on the data transformed beforehand~\cite{zhang2014towards}.

An unsupervised learning model does not require any training data and thus avoids any homogeneity requirement (e.g., a similar distribution of metrics) among projects~\cite{zhang2016cross}.
Moreover, some simple unsupervised models outperformed supervised models for a special type of software defect prediction (i.e., effort-aware just-in-time defect prediction) for open-source software systems~\cite{yang2016effort,fu2017revisiting}.
In particular, unlike distance-based unsupervised learning (e.g., $k$-means clustering) models,
connectivity-based unsupervised defect classifiers are based on the assumption of a similar intuition that defective entities are likely to cluster around the same area~\cite{zhang2016cross}.

In contrast to these prior defect prediction models that are commonly {\em static}---they are based on project features rather than system execution traits, the quality classifiers in {\distmeasure} are dynamic as they are based on IPC measurements in specific executions.
I also differ from prior defect prediction works in that I leverage IPC characteristics as a particular aspect of system behaviors, which have not been exploited before.
{\distmeasure} also focuses on addressing the quality of distributed software systems, which were not particularly addressed in earlier works.
Finally, unlike prior works on defect prediction that mainly aim at predicting whether a software unit (e.g., a file) contains functional defects, my quality classifiers address a variety of (both functional and non-functional) quality characteristics.
Meanwhile, I recall that the main goal of {\distmeasure} is to enable and explore measuring IPC-induced behaviors and understanding the measurement results from a perspective of their quality correlations, rather than developing a defect prediction model, for distributed software systems.

%% file: chapters/chapter4.tex
\chapter{{\flowdist}: Multi-Staged Refinement-Based Dynamic Information Flow Analysis for Distributed Software Systems} \label{ch:flowdist}
Through working on {\distmeasure}, I found that interprocess communication is very important for understanding the information flow security problems in distributed systems.
{\distmeasure} could help us indirectly understand potential information problems.
However, it could not directly detect information flow paths that might leak sensitive data.
This motivated me to consider interprocess dependencies and information flow paths in overall data flow analyses for distributed systems.
As the next step in my research, I developed a dynamic information flow analysis (DIFA) framework for distributed systems, which could detect the concrete sensitive information flow paths.
\section{Motivation}
\begin{figure}[tp]

\caption{A case of sensitive information flow (marked by arrowed lines) in Apache ZooKeeper across three components (processes)}
\vspace{-5pt}
\centerline{\includegraphics[width=0.95\columnwidth]{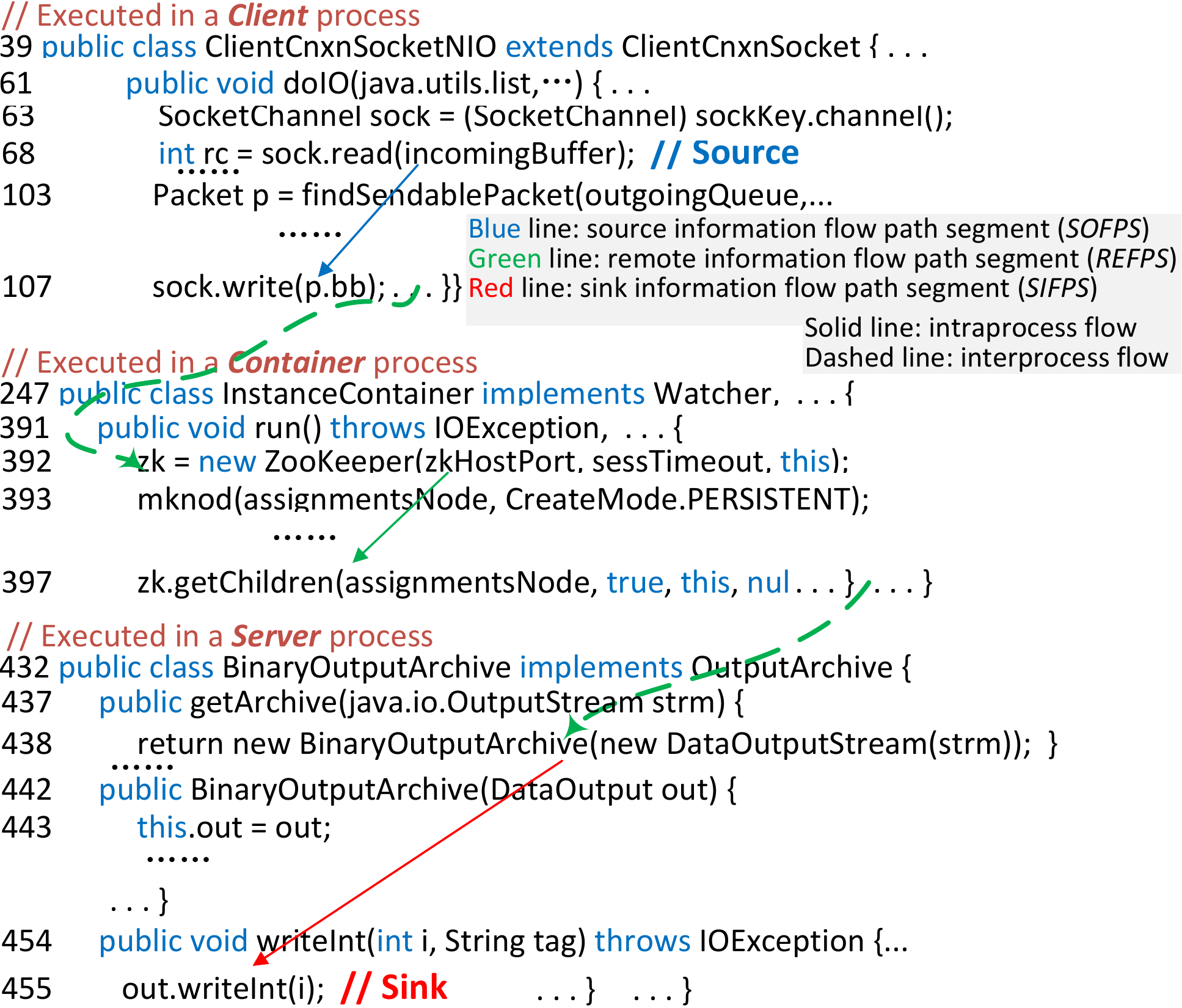}}

 \label{fig:motivatingexample}
\end{figure}

With increasing demands for computation at large scale, distributed software has been increasingly developed.
As other domains of software applications, distributed software also suffers from varied security vulnerabilities.
For example, a real-world distributed system, Apache Zookeeper~\cite{zookeeper}, had a security vulnerability as reported in CVE-2018-8012~\cite{cve20188012}: \emph{"There is no enforced authentication or authorization when a ZooKeeper server attempts to join a quorum ......"}. Relevant information flow is shown in Figure~\ref{fig:motivatingexample}.

With this vulnerability, attackers might easily gain access to a Zookeeper server and might lead to severe damage or losses.
Especially, compared to centralized programs, distributed systems usually have larger code sizes, with their decoupled components
running on physically separated machines without a global timing mechanism. These characteristics, among others, contribute to
the greater complexity of distributed software, making it even more difficult to defend the code security for these systems.

As popular alternative techniques~\cite{kang2011dta++},
taint analysis approaches have been used for defending against these vulnerabilities.
They help users identify where the sensitive data may be leaked to untrustworthy parties as revealed by tainted information flow paths (i.e., {\em taint paths}).
However, as conservative taint analyzers, static taint analysis tools often suffer from possible unsoundness due to the use of dynamic language constructs (e.g., reflective calls and dynamic code loading) in modern software.

In contrast, dynamic taint analysis (DTA) has been regarded as a powerful technique for software security that is more precise than static approaches, since it monitors and/or computes information flows that are actually exercised during the program executions~\cite{russo2010dynamic}.
Unfortunately, they were mostly developed for single-process programs and cannot be immediately applied/adapted to common distributed systems.
A major reason for the {\em applicability} issue lies in that these tools compute information flows based on explicit dependencies, which do not exist among distributed (decoupled) components in common distributed software.

Other existing dynamic taint analyzers, such as Panorama~\cite{yin2007panorama} and Taintdroid~\cite{enck2014taintdroid}, may not be subject to the {\em applicability} barrier, yet typically rely on the underlying runtime platform (e.g., operating system or JVM) being customized or modified (e.g., instrumented).
These tools thus suffer from {\em portability} problems---for each of their updated versions,
and thus the runtime platform may need to be modified again, which may require substantial effort and not always be possible.

In a distributed system, for an information flow path segment, if its statements or methods are all in one component/process,
it should be called the \emph{local information flow path segment}.
Relatively, all statements or methods in the \emph{remote information flow path segment} (\emph{REFPS}) are in different components/processes.
A \emph{local information flow path segment} is called a \emph{source information flow path segment} (\emph{SOFPS}) if it contains one or more source statements or methods.
Oppositely, the \emph{local information flow path segment}, including sink statements or methods, is called a \emph{sink information flow path segment} (\emph{SIFPS}).

An information flow in a distributed program may go across decoupled components via message passing.
Figure~\ref{fig:motivatingexample} shows a code excerpt of Apache ZooKeeper (version 3.4), a widely used distributed coordination service, 
including a sensitive flow that is responsible for a vulnerability case CVE-2018-8012~\cite{cve20188012}.

The data-leaking flow crossed three processes:
(1) The sensitive data (a security key) was read into \emph{incomingBuff} in class {\scriptsize\tt ClientCnxnSocketNIO} of a {\em Client} process (at the {\em {\color{blue}Source}}), (2) passed through class {\scriptsize\tt InstanceContainer} of a {\em Container} process, and (3) reached class {\scriptsize\tt BinaryOutputArchive} of a {\em Server} process where the data leaked out of the system (at the {\em {\color{red}Sink}}).
This leakage caused an authentication/authorization failure when a ZooKeeper server tries to join a quorum, which thus may propagate fake changes to the ZooKeeper leader node.
In addition, there are no explicit dependencies among processes: {\em Client}, {\em Container}, and {\em Server} in the information flow. However, there may be implicit dependencies among them in the interprocess information flows shown as broken lines in Figure~\ref{fig:motivatingexample}.

\section{Approach}
I developed {\flowdist}~\cite{fu2021flowdist}, a DIFA for common distributed software, working at purely application level to avoid platform customization, hence achieving high portability.
It infers implicit, interprocess dependencies from global partially ordered execution events for applicability to distributed software.
Most of all, it achieves high scalability while remaining effective, via introducing a multi-staged refinement-based scheme for application-level DIFA, where
an otherwise expensive data flow analysis is reduced by method-level results from a cheap pre-analysis.

\subsection{Overview}

\begin{figure*}[tp]
  \centering
  \caption{An overview of {\flowdist} architecture}
  \includegraphics[width=1\textwidth]{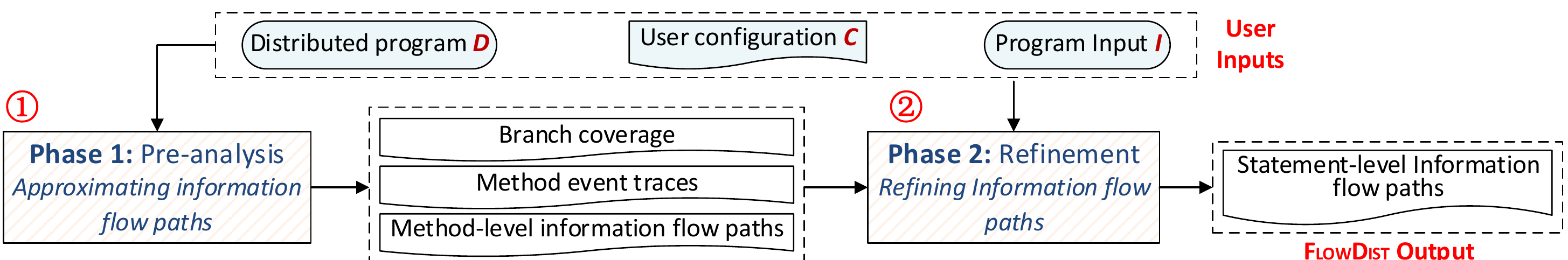}
  \label{fig:levels}
\end{figure*}

An overview of {\flowdist} architecture is shown in Figure~\ref{fig:levels}.
{\flowdist} needs three {\bf user inputs}: the distributed program \emph{D} , the run-time input set \emph{I} to drive \emph{D}, and the user configuration \emph{C}. This configuration specifies the sources and sinks of user interest and a list of message-passing APIs that {\flowdist} probes for monitoring and profiling inter-process message communication events.
With these user inputs, {\flowdist} computes the information flow paths between any source and any sink of {\em C} with respect to {\em I}, in {\em two phases}.

In the first phase ({\bf Pre-analysis}), {\flowdist} computes method-level flow paths and the statement-level coverage for methods on the paths to avoid otherwise expensive computation (overcoming the scalability challenge) of the next phase by narrowing down their analysis scopes.
Then, in the second phase ({\bf Refinement}), {\flowdist} infers the statement-level information flow paths as the final {\flowdist} {\bf output} (i.e., fine-grained {\em sensitive flows}) through a hybrid dependence analysis as guided by the pre-analysis result.

\subsection{Phase 1: Pre-analysis}

\begin{figure*}[tp]
  \centering
  \caption{Phase 1 workflow of {\flowdist}}
  \includegraphics[width=0.95\textwidth]{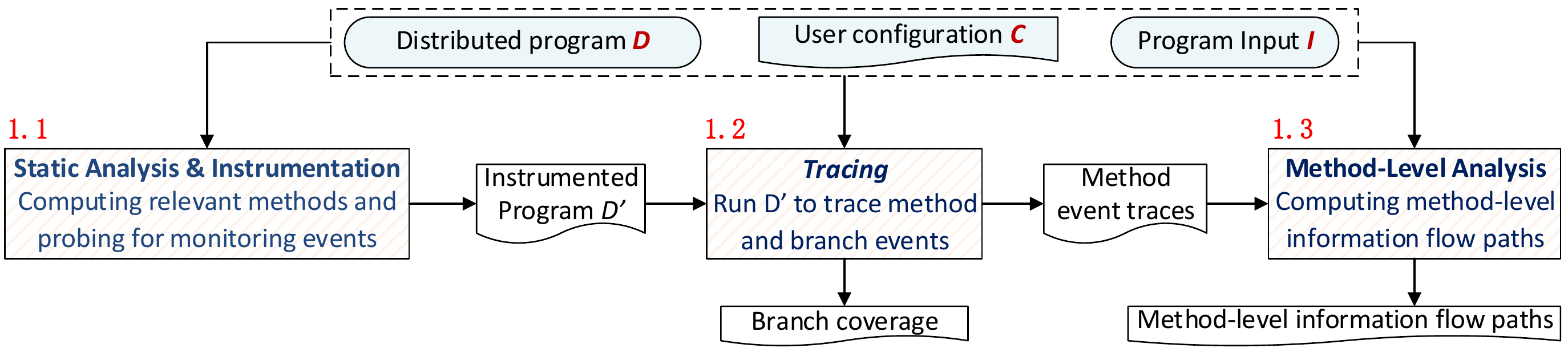}
  \label{fig:phase1}
\end{figure*}

The goal of this phase is to provide a coarse (method) level of analysis result, via a rough (conservative) and rapid analysis,
that will reduce the costs of the next phase, hence enabling the overall scalability. 
In this phase, {\flowdist} computes branch coverage and method-level information flow paths between each source-sink pair (defined in configuration $C$), in 
three major steps, as shown in Figure~\ref{fig:phase1}.

\vspace{2pt}
\noindent
\textbf{Step 1.1.}
In this step, 
{\flowdist} leverages three types of dynamic data in its hybrid dependence analysis to balance the analysis cost effectiveness balance: (1) {\em entry} (i.e., program control entering a method) and {\em returned-into} (i.e., program control returning from a callee into a caller) method events,
(2){\em sending} and {\em receiving}  message events,
and (3) statement branch coverage events.

{\flowdist} produces the instrumented version $D'$ of program $D$, by inserting probes to monitor these events.
To identify where to probe the message-passing events, {\flowdist} refers to the message-passing API list in {\em C}.
If it is not provided by the user, {\flowdist} would use a default list of the most commonly used APIs in the Java SDK.
Only the methods on static control flow paths between source-sink pairs are probed, referred to as {\em relevant methods}.
Also, only statement branches in relevant methods (i.e., called {\em relevant branches}) are probed.

For each distributed component in $D$, an interprocedural control flow graph (ICFG) is constructed.
Each message-sending and message-receiving API callsite in the component are treated as an additional sink and source, respectively.

\vspace{2pt}
\noindent
\textbf{Step 1.2.}
In this step, 
{\flowdist} records the {\em first} entry and {\em last} returned-into events for every method.
The reason that only these events are monitored is that they suffice for inferring the happens-before relations among all method-execution events, hence the approximate (control-flow-based) dependencies among associated methods~\cite{cai2016distea} across all processes.
Meanwhile, message-passing events, albeit themselves not traced, are handled during this step to partial-order method-execution events based on
the Lamport time-stamping algorithm~\cite{Lamport}, so as to determine the happens-before relation between any two method-execution
events later.
The method event trace ($T_i$) is produced for each of the $n$ processes ($P_i$) in the execution.
In addition,
a mapping ($pid2fm[i]$) is produced for $P_i$ to keep the timestamp ($p2fm[i][j]$) of each event of receiving the first message from another process $P_j$ ($i,j\in[1,n]$).

\vspace{2pt}
\noindent
\textbf{Step 1.3.}
In this step, 
{\flowdist} computes method-level information flow paths as the output of this phase,
by identifying the sequence of methods 
between any source and any sink exercised during the execution.

\begin{algorithm}[htbp]
\scriptsize{
    \caption{Computing method-level flow paths}
\begin{flushleft}
    \scriptsize{Let $SO$ and $SI$ be the list of source and sink enclosing methods, respectively} \\
    \scriptsize{Let $T_i$ be time-stamped method execution event trace in process $P_i$, $i$$\in$$[1,n]$}
    \vspace{-8pt}
\end{flushleft}
    \label{algo:phase1}
    \begin{algorithmic}[1]
    \vspace{-0pt}
        \State {\emph{$ps$} = $\emptyset$}\label{algo1:line1} \hfill{\color{green}\scriptsize// initialize the set of all method-level paths between the given pair} 
        \For{$i$=1 to $n$}\label{algo1:line2} \hfill{\color{green}\scriptsize// traverse the $n$ processes of the given execution}
            \State $S_d$ = $\{s|s\in SO \land s\in T_i\}$\label{algo1:line3}
            \If{$S_d$==$\emptyset$}~~\textbf{continue}\label{algo1:line4}
            \EndIf
            \For{{\bf each} method $q\in S_d$}\label{algo1:line5} \hfill{\color{green}\scriptsize// first compute intraprocess dependencies}
                \State $DS(q)$ = $\{m|m\in T_i~\land~fe(q)\le lr(m)\}$\label{algo1:line6}
                \For{$j$=1 to $n$}\label{algo1:line7} \hfill {\color{green}\scriptsize// then compute interprocess dependencies}
                    \If{$i$==$j$ $\lor$ $p2fm[i][j]$==$null$}~~\textbf{continue}\label{algo1:line8}
                    \EndIf
                    \State $DS(q)$ $\cup$$=$ $\{m|m\in T_j\land fe(q)\le p2fm[i][j]\le lr(m)\}$\label{algo1:line9}
                \EndFor
                \If{$DS(q)\cap SI$==$\emptyset$}~~\textbf{continue}\label{algo1:line10}
                \EndIf
                \State $ps$ $\cup$$=$ $\{<m1,...,m_k>|m1==q\land m_k\in SI \land \forall_{i<j,~i,j\in[1,k]} fe(m_i)\le lr(m_j) \land \forall_{i\in[1,k]} m_i\in DS(q)\}$\label{algo1:line11}
            \EndFor
        \EndFor
        \State\Return $ps$\label{algo1:line12}
    \end{algorithmic}
}
\end{algorithm}
With the event traces and mapping from Step 1.2, {\flowdist} 
computes the method-level information flow
paths according to Algorithm~\ref{algo:phase1}.
The core idea is that {\flowdist} combines method-level control flows and process-level data flows to approximate dynamic method-level dependencies.

The algorithm traverses the $n$ per-process traces to search paths $ps$ by  (lines~\ref{algo1:line2}-\ref{algo1:line11}).
In each trace $T_i$, the set $S_d$ of covered source-enclosing methods is obtained (line~\ref{algo1:line3}).
If there is no source executed in $P_i$, no path would start in $P_i$ (corresponding to $T_i$)  (line~\ref{algo1:line4}).
Otherwise, the algorithm identifies paths starting at $q$ by computing its dynamic dependence set $DS(q)$, for each method $q$ in $S_d$. (lines~\ref{algo1:line5}-\ref{algo1:line10}).

I let $fe(m)$ and $lr(m)$ be the timestamp of the first entry event and last returned-into event of a method $m$, respectively.
The {\em local} (intraprocess) dependencies are first identified (line~\ref{algo1:line6}) according to the happens-before relation between $q$ and each other method $m$ executed in $P_i$ (treated as a {\em local} process).
The design reason is that method $m2$ is not dependent on method $m1$ if $m2$ is not executed after $m1$~\cite{cai14diver}.

Next, dependencies in each other ({\em remote}) process $P_j$ are identified (lines~\ref{algo1:line7}-\ref{algo1:line9}).
If $P_j$ never sent a message (line~\ref{algo1:line8}) or the timing of message passing implies no dependence, 
relevant methods in $P_j$ are not considered. 
Otherwise, they are added to $DS(q)$ (line~\ref{algo1:line9}).
The rationale is that for two methods $m1$ and $m2$ executed in two processes, $p1$ and $p2$, respectively, $m2$ depends on $m1$ only if $p2$ receives at least one message before $lr(m2)$ that is sent (directly or transitively) from $p1$ after $fe(m1)$.

When $DS(q)$ is computed, if it includes a sink-enclosing method $m_k$ (line~\ref{algo1:line10}), these partially ordered methods in $DS(q)$ form an information flow path from $q$ to $m_k$. All such paths are gathered into $ps$ (line~\ref{algo1:line11}) and then returned (line~\ref{algo1:line12}).

\subsection{Phase 2: Refinement}
\begin{figure*}[tp]
  \centering
  \caption{Phase 2 workflow of {\flowdist}}
  \includegraphics[width=0.95\textwidth]{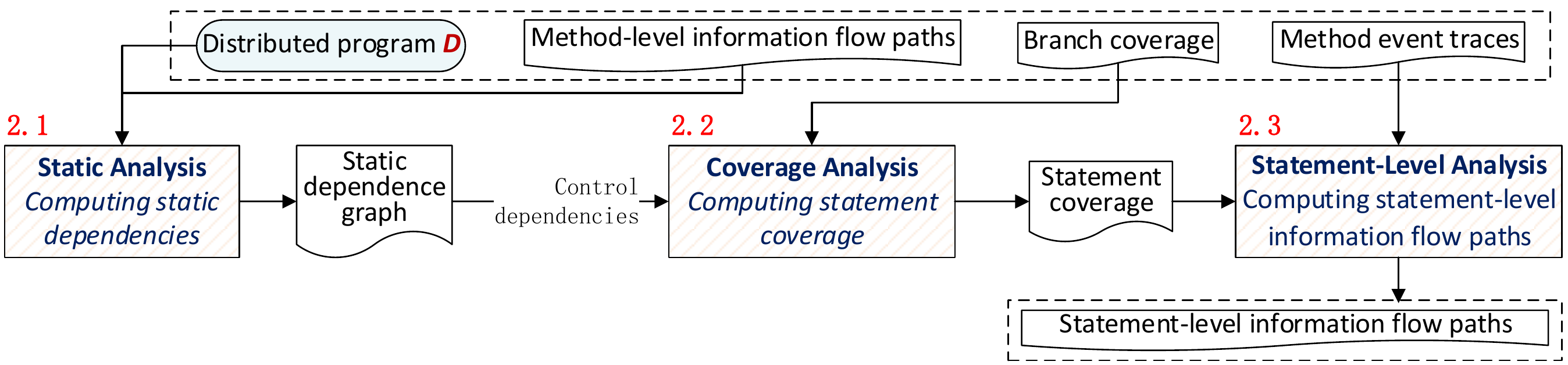}
  \label{fig:phase2}
\end{figure*}

In this phase,
{\flowdist} aims to computed fine-grained (statement-level) information flow paths by refining the coarse results (i.e., method-level information flow paths) inferred in
{\em Phase 1}, leveraging program data of two modalities (i.e., {\em static} and {\em dynamic}), in three steps, as shown in Figure~\ref{fig:phase2}.
The primary motivation of this hybrid analysis design is to balance the total analysis cost and precision~\cite{cai2016diapro}.

\vspace{2pt}
\noindent
\textbf{Step 2.1.}
This step is mainly a static analysis that builds a static dependence graph of the program {\em D}.
Differing from a whole-system dependence graph, the graph is partial, only involving methods on the (method-level) flow paths computed in the first phase
(pre-analysis).
The rationale is that {\em one method on a statement-level information flow path from a source to a sink must be on the statement-level information flow path between two methods enclose the source and sink}.
{\flowdist} stops interprocedural propagation of relevant flow facts, when encountering methods that are not on the (method-level) information flow paths.

The static dependencies are computed at a statement level, to be used as an essential type of information by the hybrid data flow analysis in the last step.
Specifically, {\flowdist} computes data/control dependencies~\cite{horwitz1990interprocedural} within and across threads.
On the other hand, when computing such static dependencies, in order to cover all of the components in {\em D}, the static analysis searches for all possible entry points (i.e., all classes containing the {\tt main} method) of {\em  D} and starts the data flow computation from each of the entry points found. All the control dependencies are also computed.

The static dependence analysis here is chosen to be context-insensitive because its results are only used in Step 2.3, which will use method-execution events to provide the necessary
context.
Furthermore, its interprocedural analysis part is flow-insensitive because those events are ordered by their timestamps, while its intraprocedural analysis part remains flow-sensitive.
These selections reduce the total analysis cost of {\flowdist}.


\vspace{2pt}
\noindent
\textbf{Step 2.2.}
This step aims to produce the statement coverage for each process, the fine-grained dynamic data only for refining the hybrid analysis in Step 2.3, referring to 
the static dependencies from Step 2.1 and branch coverage from Step 1.2.
Importantly, only methods on method-level paths computed in Phase 1 are considered.
The relational is that {\em statements in other methods will not appear on the statement-level information flow paths}.
%

\vspace{2pt}
\noindent
\textbf{Step 2.3.}
With the statement coverage ($SC$), method event traces ($T_i$) in process $P_i$, and partial static dependence graph ($sDG$) of $D$,
{\flowdist} now infers statement-level information flow paths between each source-sink callsite pair ($<$$s$,$t$$>$).
{\flowdist} exploits message sending and receiving APIs to identify the callsites within the methods on the method-level flow paths,
indicating where information flows out from and into each process, referred to as {\em outlets} and {\em inlets}, respectively.

In Algorithm~\ref{algo:phase3}, per-process method event sequences are first merged as a global event sequence $ES$ (line~\ref{algo2:line2}) ordered by event timestamps. Then, the function {\tt build-dDG} constructs a dynamic dependence graph $dDG$ (line~\ref{algo2:line3}) by referring to the static dependencies in $sDG$ while traversing $ES$, using a hybrid dependence analysis inspired by {\diver}~\cite{cai14diver}.
The insights include:
(1) 
interprocedural dependencies in $sDG$ are categorized into {\em adjacent} (due to parameter or return-value passing)
and {\em posterior}(due to the def-use associations);
and (2) when scanning $ES$, if that dependence is adjacent and $m2$ happens immediately after $m1$ in $ES$,
or the dependence is posterior and $m2$ happens anywhere after $m1$ in $ES$, a static dependence of method $m2$ on another method $m1$ is activated and hence added to $dDG$. 

\begin{algorithm}[tp]
\scriptsize{
    \caption{Computing statement-level flow paths}
\begin{flushleft}
    \scriptsize{let $S_1, S_2, ..., S_n$  be the $n$ per-process instance-level event traces} \\
    \scriptsize{let $M_1, M_2,...,M_k$ be the $k$ methods on the method-level path} \\
    \scriptsize{Let $sDG$ be the partial static dependence graph} \\
    \scriptsize{Let $<$$s$,$t$$>$ be a source-sink callsite 
                pair between which paths are computed} \\
    \scriptsize{Let \emph{outlets} be the list of all outlets} \\
    \scriptsize{Let \emph{inlets} be the list of all inlets} \\
\end{flushleft}
    \label{algo:phase3}
    \begin{algorithmic}[1]
        \State \emph{SOFPS}=$\emptyset$, \emph{REFPS}=$\emptyset$, \emph{SIFPS}=$\emptyset$, \emph{intraFP}=$\emptyset$ \label{algo2:line1}
        \State merge $T_i$, $i$$\in$$[1,n]$ into a global partially ordered sequence $ES$ \label{algo2:line2}
        \State $dDG$ = {\tt build-dDG}($sDG$, $<$$s$,$t$$>$, \emph{$ES$})\label{algo2:line3}\hfill\textcolor{green}{\scriptsize// hybrid analysis}
        \State $dDG'$ = {\tt prune-dDG}($dDG$, \emph{$SC$})\label{algo2:line4}\hfill\textcolor{green}{\scriptsize// statement-level pruning}
        \State \emph{SOFPS} = {\tt findPaths}($dDG'$, \{$s$\}, \emph{outlets}, $tr(s)$)\label{algo2:line5}
        \For{$i$= to $n$}\label{algo2:line6}\hfill\textcolor{green}{\scriptsize// compute remote segments of interprocess paths}
            \State \emph{intraFP}~$\cup$=~{\tt findPaths}($dDG'$,\{$s$\},\{$t$\},$S_j$)\label{algo2:line7}\hfill\textcolor{green}{\scriptsize// intraprocess paths}
            \If{$tr(s)==T_i$~ $\lor$ ~$tr(t)==T_i$}~~\textbf{continue}\label{algo2:line8}
            \EndIf
            \State \emph{REFPS}~$\cup$=~{\tt discoverPaths}($dDG'$, \emph{inlets}, \emph{outlets}, $T_i$)\label{algo2:line9}
        \EndFor
        \State \emph{SIFPS} = {\tt discoverPaths}($dDG'$, \emph{inlets}, \{$t$\}, $tr(t)$)\label{algo2:line10}
        \State\Return [{\tt spliceSegs}(\emph{SOFPS}, \emph{REFPS}, \emph{SIFPS}), \emph{intraFP}]\label{algo2:line11}
    \end{algorithmic}
}
\end{algorithm}

The algorithm 
considers all static {\em intraprocedural} dependencies in a method that is in $ES$ as activated and then adds them to $dDG$.
And it constructs the graph from 
the source $s$, only including dependencies that reach the sink $t$.
The resulting dependencies in $dDG$ would not be precise at statement level.
Therefore, {\flowdist} proceeds with a function {\tt prune-dDG} which prunes spurious dependencies in $dDG$
per the statement coverage $SC$:
Vertices corresponding to uncovered statements and
their associated edges are deleted from the graph, causing the pruned graph  (line~\ref{algo2:line4}).
I made this choice to contain the overall analysis cost of {\flowdist} for scalability. %

With the $dDG'$, the algorithm then deduces both intraprocess and interprocess information flow paths, 
using a function, {\tt discoverPaths(G, In, Out, T)}, which detects paths
from any statement in {\tt In} to any statement in {\tt Out} on a graph {\tt G} while only considering nodes in {\tt T}.
Simply traversing $dDG'$, 
the algorithm computes intraprocess information
paths (\emph{intraFP}) in each process 
(line~\ref{algo2:line7}).

However, for any interprocess information flow path, the sink is not explicitly reachable from the source on $dDG'$ 
due to no connectivity in it.
Thus, {\flowdist} computes the three segments separately: \begin{enumerate}
\item The segment within the source ($s$)'s process (\emph{SOFPS}) is computed via a traversal on $dDG'$ (line~\ref{algo2:line5}) that retrieves paths from $s$ to a relevant outlet within that process's trace---$tr(x)$ denotes the trace that includes an event of
the method that encloses a statement $x$. 
\item The segment within the sink ($t$)'s process (\emph{SIFPS}) is computed similarly (line~\ref{algo2:line10}), 
but by searching for paths from any inlet to $t$.
\item The remaining segment (\emph{REFPS}) is searched within each process (lines~\ref{algo2:line6}-\ref{algo2:line9}) except the one that encloses the one that encloses $s$ or $t$ (line~\ref{algo2:line8}), 
 by traversing $dDG'$ and looking for paths from all inlets to all outlets in the process.
\end{enumerate}

Eventually, these segments are spliced into interprocess information flow paths with the function {\tt spliceSegs}, according to the timestamps
of relevant inlets and outlets. 
This splicing works so that there are no events between the end of an \emph{SOFPS} and the start of an \emph{REFPS},
nor between the end of the \emph{REFPS} and the start of an \emph{SIFPS} as per the global partially ordered sequence $ES$.
With the intraprocess paths, these spliced interprocess paths are then returned as the algorithm output
(line~\ref{algo2:line11}).

\subsection{Alternative designs}
The default design of {\flowdist} as presented above targets common distributed software systems in general.
To more systematically explore the multi-staged refinement-based methodology for dynamic information flow analysis,
I developed two alternative designs: {\flowdistsim} and {\flowdistmul}.
They might offer even greater cost-effectiveness and scalability for systems that meet certain conditions,
by further reducing analysis costs while without compromising precision and soundness.

\vspace{2pt}
\noindent
\textbf{{\flowdistsim}.}
In Step 1.1 of {\flowdist}, the goal of the static analysis is to reduce
the instrumentation scope, hence the costs of tracing method and statement branch events.
However, with certain systems, probing for and tracing all such events is cheap, and
the cost incurred by this static analysis itself may be larger than the cost reduced.
Optimized for systems that meet these conditions, {\flowdistsim} skips the static analysis and simply instruments all methods and statement branches.

\vspace{2pt}
\noindent
\textbf{{\flowdistmul}.}
With some systems, the {\flowdistsim} design is well justified.
But probing for and then tracing all method and branch events in $D$ incurs substantial costs.
To reduce these costs, I introduce an intermediate phase to {\flowdistsim}, with a multi-staged and refinement-based design in Phase 1.

First, 
the new Phase 1 only probes for and traces the first entry and last returned-into events of each method, and then
computes method-level flow paths from those events.
The intermediate phase then probes for and traces the coverage of statement branches, and all instances of both kinds of events of methods on such paths.
Finally, Step 2.2 is removed from Phase 2.

Since {\flowdistmul} requires multiple executions of the same system against the same input (in the first and intermediate phases),
this design is optimized for systems with deterministic executions.
The first condition is that the inconsistencies between the two executions could compromise the soundness of the DIFA as a whole.
Another condition is that the cost reduction should outweigh the costs incurred by the intermediate phase.

\vspace{2pt}
\noindent
\textbf{{\flowdistsim} versus {\flowdistmul}.}
According to the rationale of each alternative design,
{\flowdistmul} is the best for such systems without non-deterministic executions, 
while {\flowdistsim} is expected to perform the best for small/simple systems with non-deterministic executions. 
For large/complex distributed systems, {\flowdist} (default design) would perform the best.
These contrasts are justified by the {\em conditions} (as described above).
If the system does not meet any of those conditions, the default design of {\flowdist} is superior in general.

\subsection{Limitations}
{\flowdist} does not address the problem of identifying the sources/sinks of interest, which are assumed in default list or to be given by users.
Also, 
the analyses in {\flowdist} are limited to the program parts executed.
Thus, the capabilities of discovering bugs rely on that
(1) the relevant source and sink are specified and (2) the source and sink are covered by the run-time inputs considered.
Moreover, considering the security context in specific usage scenarios (e.g., external protection mechanisms applied to the source or sink),
{\flowdist} may suffer from false positives as they do not analyze, nor have access to, those external/context factors.

Also, {\flowdist} requires static instrumentation and thus does not suit systems that cannot be modified.
Additional limitations of {\flowdistsim} and {\flowdistmul} are those implied by the respective system
conditions discussed earlier.

\section{Tool Implementation}
\begin{figure*}[tp]
  \centering
  \caption{An overview of {\disttaint} architecture}
  \includegraphics[width=1\textwidth]{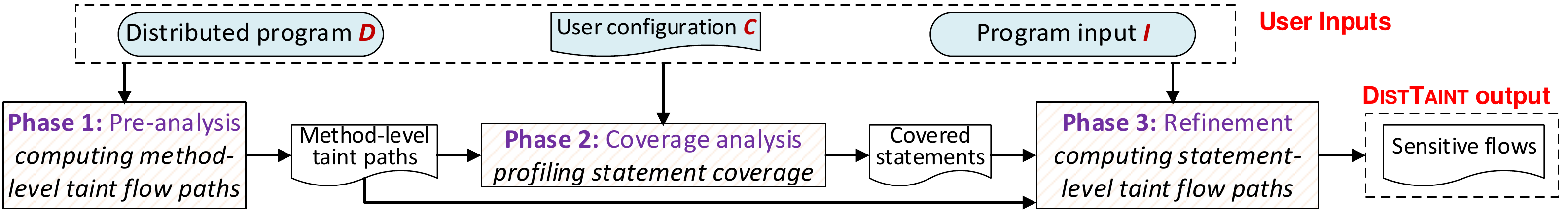}
  \label{fig:disttaintoverview}
\end{figure*}

For distributed systems, information flow security solutions face multiple challenges, including technique {\em applicability}, tool {\em portability},
and analysis {\em scalability}. 
Due to these challenges, I implemented {\disttaint}~\cite{fu2019dynamic,fu2020scaling,fu2019scalable,fu2019towards}, a dynamic information flow (taint) analyzer for distributed systems, 
as shown in Figure~\ref{fig:disttaintoverview}.

By partial-ordering method events during the execution, {\disttaint} computes implicit dependencies in distributed programs to resolve the applicability challenge.
It overcomes the portability challenge by working fully at application level, without customizing the run-time platform.
To achieve scalability, it reduces analysis costs using a multi-phase analysis, where the method-level results (produced by pre-analysis phase) are used to narrow down the scope of the following statement-level analysis.
Applied it to different large-scale distributed software against diverse executions,
{\disttaint} demonstrated its applicability to, portability with, and scalability for industry-scale distributed systems, along
with its capability of finding existing and new vulnerabilities.

\section{Evaluation}         \label{sec:flowdisteval}

\subsection{Experiment setup} \label{ss:flowdistes}
\begin{table}[htbp]
  \centering
   \caption{Subject distributed programs and test inputs used}
    \begin{tabular}{|l|r|r|l|l|}
    \hline
    \multicolumn{1}{|c|}{\textbf{Subject}} & \multicolumn{1}{c|}{\textbf{\#SLOC}} & \multicolumn{1}{c|}{\textbf{\#Method}} & \multicolumn{1}{c|}{\textbf{Scenario}} & \multicolumn{1}{c|}{\textbf{Tests}} \\
    \hline
    NIOEcho & 412   & 27    & Client-server & Integration \\
    \hline
    MultiChat & 470   & 37    & Peer-to-peer & Integration \\
    \hline
    ADEN & 4,385 & 260   & Peer-to-peer & Integration \\
    \hline
    Raining Sockets & 6,711 & 319   & Client-server & Integration \\
    \hline
    OpenChord & 9,244 & 736   & Peer-to-peer & Integration \\
    \hline
    Thrift  & 14,510 & 1,941 & Client-server & Integration \\
    \hline
    xSocket & 15,760 & 2,209 & Peer-to-peer & Integration \\
    \hline
          &       &       & Client-server & Integration \\
\cline{4-5}    ZooKeeper & 62,194 & 5,383 & N-tier & Load \\
\cline{4-5}          &       &       & N-tier & System \\
    \hline
    \multirow{2}[4]{*}{RocketMQ} & \multirow{2}[4]{*}{105,444} & \multirow{2}[4]{*}{6,198} & N-tier & Integration \\
\cline{4-5}          &       &       & N-tier & System \\
    \hline
          &       &       & Client-server & Integration \\
\cline{4-5}    Voldemort & 115,310 & 20,406 & N-tier & Load \\
\cline{4-5}          &       &       & N-tier & System \\
    \hline
    Netty & 167,961 & 12,389 & N-tier & Integration \\
    \hline
    \multirow{2}[4]{*}{HSQLDB} & \multirow{2}[4]{*}{326,678} & \multirow{2}[4]{*}{10,095} & Client-server & Integration \\
\cline{4-5}          &       &       & N-tier & System \\
    \hline
    \end{tabular}%
  \label{tab:flowdistsubjects}%
\end{table}%

As shown in Table~\ref{tab:flowdistsubjects}, I used 12
Java distributed systems as subjects. 
The subject sizes are measured by numbers of non-blank non-comment Java source code lines (\emph{\#SLOC}),
numbers of methods defined in the subject (\emph{\#Method}),
and execution scenarios/architectures (\emph{Scenario}) including client-server, peer-to-peer, and n-tier.

I now describe each subject and its integration test operations, except for those that were already introduced ($\S$\ref{ss:distmeasurees}).
\begin{enumerate}
\item NioEcho~\cite{nioecho} provides an echoing service for any message sent by clients.
In its integration test,
I started a server and a client, sent random text messages from the client to the server, and then waited for the echo of each message.
\item MultiChat~\cite{multichat} is a chat service broadcasting client messages. 
In its integration test, I started a server and three clients. From one client, I sent random text messages to the server, which broadcast them to all other clients.
\item ADEN~\cite{ADEN} offers a UDP-based alternative to TCP sockets.
In its integration test, I started two nodes, each of which sends messages to and receives messages from the other node.
\item Raining Sockets~\cite{RainingSockets} is a non-blocking and sockets-based framework.
In its integration test, I started a server and a client, and then the client sent text messages to the server.
\item RocketMQ~\cite{RocketMQ} is a distributed messaging platform.
In its integration test, there are four components: a name server, a broker, a producer, and a consumer. The name server provided reading and writing services and records full routing information, the broker stored messages, the producer sent messages to the broker, and the customer received messages from the broker.
\item Netty:
In its integration test, I developed a n-tier (3-tier) application with three nodes.
The first node read an email list from a file and then sent relevant emails to the second node.
Then, the second node encrypted the emails using the RSA algorithm and then sent them to the third node. Finally, the third node used Postfix to send emails received.
\item HSQLDB (HyperSQL DataBase)~\cite{Hsqldb} is an SQL relational database system.
I started a database server and a client. Then, the client sent a SQL query to the server and then received the SQL result from the server.
\end{enumerate}

For all subjects, their integration tests were created according to the guides at their official websites.
Particularly for the integration test of five frameworks/libraries (ADEN, Raining Sockets, Thrift, xSocket, and Netty), I developed applications to cover their major functions and then performed all of the applications.

I selected a state-of-the-art dynamic taint analyzer Phosphor~\cite{bell2014phosphor} and a static taint analyzer JOANA~\cite{tooljoana2013atps}, as two baselines compared with {\flowdist}.
For the experiments, I used Ubuntu 16.04.3 LTS workstations, each of them equipped with an Intel E7-4860 2.27GHz CPU and 32GB DMI RAM.

\subsection{Results and analysis}

\vspace{2pt}
\noindent
\textbf{Effectiveness.}
\setlength{\tabcolsep}{1pt}
\begin{table}[tp]
	\small
	\centering
	\caption{Numbers of intraprocess ({\em Ir}) source/sink pairs ({\em Pr}) and information flow paths ({\em Ps}), versus interprocess ({\em Int}) ones}
	\vspace{-0pt}
	\begin{tabular}{|l|r|r|r|r|r|}
		\hline
		\textbf{Execution} & \multicolumn{1}{l|}{\textbf{\#IrPr}} & \multicolumn{1}{l|}{\textbf{\#IrPs}} & \multicolumn{1}{l|}{\textbf{\#IntPr}} & \multicolumn{1}{l|}{\textbf{\#IntPs}} & \multicolumn{1}{l|}{\textbf{IntPs/AllPs}} \\
		\hline
		NioEcho & 66    & 21    & 12    & 6     & 22.22\% \\
		\hline
		\textcolor[rgb]{ .502,  .502,  .502}{MultiChat} & \textcolor[rgb]{ .502,  .502,  .502}{42} & \textcolor[rgb]{ .502,  .502,  .502}{0} & \textcolor[rgb]{ .502,  .502,  .502}{12} & \textcolor[rgb]{ .502,  .502,  .502}{0} & \textcolor[rgb]{ .502,  .502,  .502}{0.00\%} \\
		\hline
		\textcolor[rgb]{ .502,  .502,  .502}{ADEN} & \textcolor[rgb]{ .502,  .502,  .502}{0} & \textcolor[rgb]{ .502,  .502,  .502}{0} & \textcolor[rgb]{ .502,  .502,  .502}{5} & \textcolor[rgb]{ .502,  .502,  .502}{0} & \textcolor[rgb]{ .502,  .502,  .502}{0.00\%} \\
		\hline
		Raining Sockets & 12    & 3     & 0     & 0     & 0.00\% \\
		\hline
		\textcolor[rgb]{ .502,  .502,  .502}{OpenChord} & \textcolor[rgb]{ .502,  .502,  .502}{14} & \textcolor[rgb]{ .502,  .502,  .502}{0} & \textcolor[rgb]{ .502,  .502,  .502}{24} & \textcolor[rgb]{ .502,  .502,  .502}{0} & \textcolor[rgb]{ .502,  .502,  .502}{0.00\%} \\
		\hline
		Thrift & 4     & 0     & 4     & 3     & 100.00\% \\
		\hline
		xSocket  & 10    & 8     & 26    & 2     & 20.00\% \\
		\hline
		\textcolor[rgb]{ .502,  .502,  .502}{Zookeeper Integration} & \textcolor[rgb]{ .502,  .502,  .502}{9} & \textcolor[rgb]{ .502,  .502,  .502}{0} & \textcolor[rgb]{ .502,  .502,  .502}{33} & \textcolor[rgb]{ .502,  .502,  .502}{0} & \textcolor[rgb]{ .502,  .502,  .502}{0.00\%} \\
		\hline
		Zookeeper Load & 1086  & 1     & 6522  & 64    & 98.46\% \\
		\hline
		Zookeeper System & 124   & 0     & 1116  & 46    & 100.00\% \\
		\hline
		RocketMQ Integration & 19    & 23    & 46    & 17    & 42.50\% \\
		\hline
		RocketMQ System & 24    & 0     & 187   & 50    & 100.00\% \\
		\hline
		Voldemort Integration & 198   & 30    & 193   & 138   & 82.14\% \\
		\hline
		\textcolor[rgb]{ .502,  .502,  .502}{Voldemort Load} & \textcolor[rgb]{ .502,  .502,  .502}{6} & \textcolor[rgb]{ .502,  .502,  .502}{0} & \textcolor[rgb]{ .502,  .502,  .502}{6} & \textcolor[rgb]{ .502,  .502,  .502}{0} & \textcolor[rgb]{ .502,  .502,  .502}{0.00\%} \\
		\hline
		Voldemort System & 80    & 30    & 77    & 42    & 58.33\% \\
		\hline
		Netty & 9     & 3     & 7     & 2     & 40.00\% \\
		\hline
		HSQLDB Integration & 140   & 10    & 668   & 0     & 0.00\% \\
		\hline
		HSQLDB System & 7     & 2     & 11    & 4     & 66.67\% \\
		\hline
	\end{tabular}%
	\label{tab:localremotepaths}%
\end{table}%
Table~\ref{tab:localremotepaths} shows the number of
source-sink pairs covered in each execution (i.e., subject-test type) and that of information flow paths between the pairs,
separately for intraprocess and interprocess paths.
For each 
source/sink given in the configuration $C$, {\flowdist} treated each of its exercised callsites as a separate source/sink in counting the pairs and computing the paths.
The last column shows the percentage of interprocess information flow paths over all paths per execution.
The greyed rows are for executions that did not have any information flow paths.

\begin{figure}[htbp]
  \centering
\caption{The accuracy of {\flowdist} versus the baselines Phosphor and JOANA}
  \includegraphics[width=0.6\textwidth]{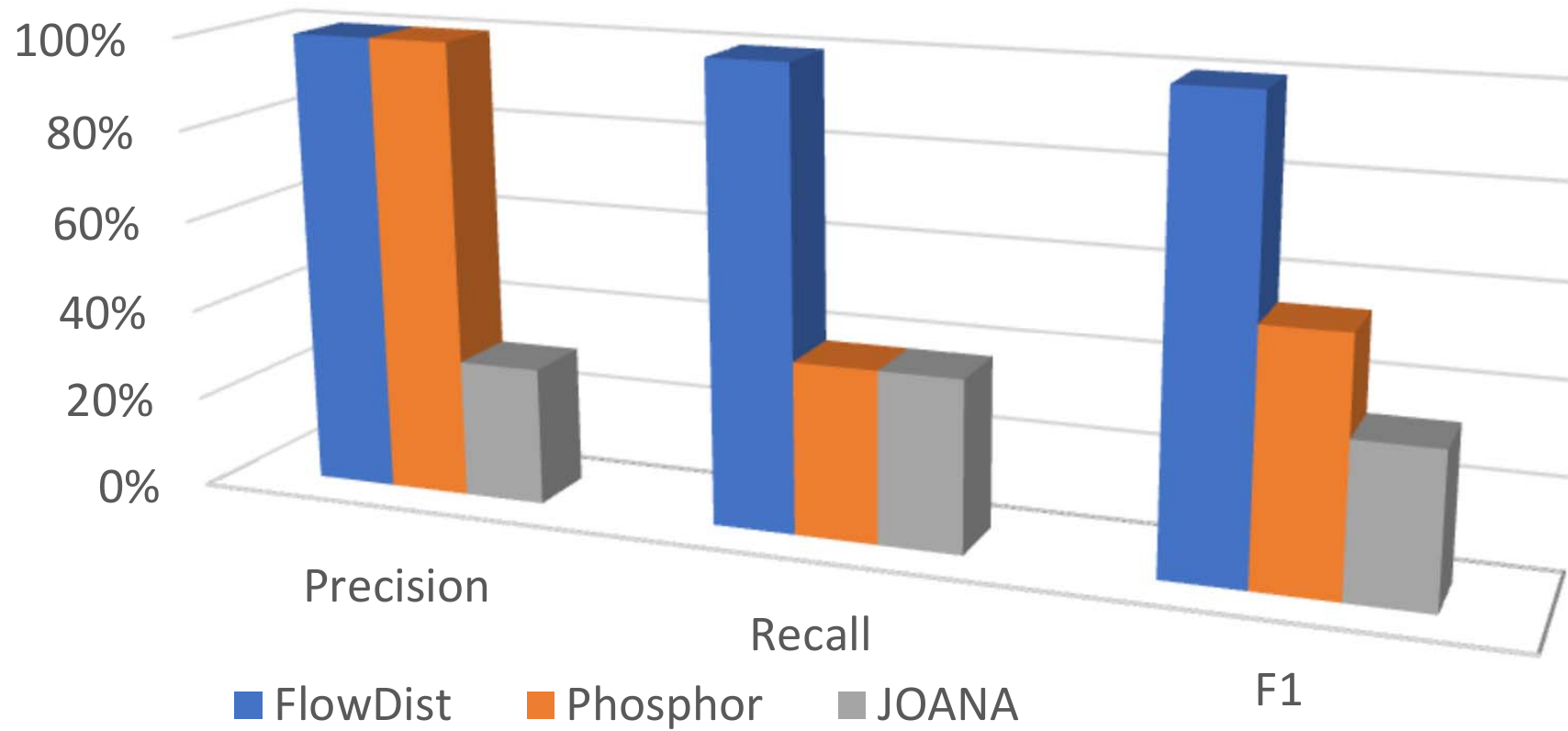}
  \label{fig:fdbaselines}
\end{figure}
Both baselines captured all {\em intraprocess} paths as {\flowdist} did but missed all interprocess ones.
Thus, they had the same but low recall (37.5\%), as shown in  Figure~\ref{fig:fdbaselines}.
For the same reason, {\em none of them discover any known or unknown
vulnerability}.
In addition, JOANA reported many additional paths that were not covered in the executions considered, as false positives, leading
to the lowest precision (30\%).
Therefore, {\flowdist} achieved the highest F1 (100\%) (with Phosphor F1 (54.6\%) and JOANA F1 (33.3\%)).

\setlength{\tabcolsep}{0.5pt}
\begin{table}[htbp]
\captionsetup{font=small}
  \centering
  \caption{Time (in seconds) and storage (in MB) costs of {\flowdist}}	
    \begin{tabular}{|l|r|r|r|r|r|r|r|r|r|}
    \hline
    \multicolumn{1}{|c|}{\multirow{2}[2]{*}{\textbf{Executions}}} & \multicolumn{1}{c|}{Norm} & \multicolumn{4}{c|}{Phase 1 Time} & \multicolumn{3}{c|}{Phase 2 Time} & \multicolumn{1}{r|}{\multirow{2}[2]{*}{Stor.}} \\
\cline{3-9}          & \multicolumn{1}{c|}{Run} & \multicolumn{1}{c|}{Static} & \multicolumn{1}{c|}{Run} & \multicolumn{1}{c|}{Slowdown} & \multicolumn{1}{c|}{Query} & \multicolumn{1}{c|}{Static} & \multicolumn{1}{c|}{Coverage} & \multicolumn{1}{c|}{Query} &  \\
    \hline
    NioEcho & 39    & 53    & 41    & 5.16\% & 0.2   & 50    & 1     & 1.0   & 1.6 \\
    \hline
    MultiChat & 26    & 55    & 28    & 6.12\% & 0.2   & 50    & 1     & 0.1   & 1.0 \\
    \hline
    ADEN  & 21    & 117   & 23    & 10.23\% & 0.3   & 59    & 3     & 0.3   & 4.0 \\
    \hline
    Raining Sockets. & 6     & 40    & 6     & 7.67\% & 0.3   & 122   & 6     & 0.4   & 14.5 \\
    \hline
    OpenChord & 54    & 177   & 59    & 8.54\% & 0.3   & 740   & 41    & 4.7   & 26.7 \\
    \hline
    Thrift & 8     & 146   & 10    & 24.83\% & 0.5   & 79    & 45    & 0.6   & 26.1 \\
    \hline
    xSocket & 11    & 101   & 19    & 63.99\% & 0.5   & 70    & 14    & 0.1   & 29.3 \\
    \hline
    Zookeeper Integration & 71    & 292   & 121   & 70.16\% & 0.5   & 193   & 108   & 1.8   & 231.2 \\
    \hline
    Zookeeper Load & 99    & 292   & 177   & 78.83\% & 0.6   & 137   & 67    & 2.0   & 404.0 \\
    \hline
    Zookeeper System & 98    & 292   & 178   & 81.87\% & 0.5   & 250   & 93    & 1.1   & 417.5 \\
    \hline
    RocketMQ Integration & 105   & 56    & 196   & 87.05\% & 0.6   & 704   & 49    & 21.5  & 291.0 \\
    \hline
    RocketMQ System & 339   & 156   & 753   & 122.09\% & 0.6   & 727   & 52    & 34.0  & 463.2 \\
    \hline
    Voldemort Integration & 28    & 1206  & 58    & 106.06\% & 0.6   & 566   & 317   & 9.1   & 560.4 \\
    \hline
    Voldemort Load & 11    & 1206  & 23    & 113.37\% & 0.6   & 435   & 260   & 14.4  & 523.1 \\
    \hline
    Voldemort System & 31    & 1206  & 65    & 109.81\% & 0.6   & 618   & 344   & 22.2  & 545.1 \\
    \hline
    Netty & 12    & 1132  & 22    & 81.65\% & 0.6   & 381   & 317   & 30.1  & 417.6 \\
    \hline
    HSQLDB Integration & 9     & 659   & 19    & 107.46\% & 0.7   & 2227  & 96    & 41.5  & 591.1 \\
    \hline
    HSQLDB System & 15    & 684   & 36    & 142.71\% & 0.7   & 2771  & 408   & 49.7  & 733.7 \\
    \hline
    \textbf{Overall Average} & \textbf{55} & \textbf{437} & \textbf{102} & \textbf{68.20\%} & \textbf{0.5} & \textbf{565} & \textbf{124} & \textbf{13.0} & \textbf{293.4} \\
    \hline
    \end{tabular}%
  \label{tab:fdcosts}%
\end{table}%
\vspace{2pt}
\noindent
\textbf{Efficiency.}
Table~\ref{tab:fdcosts} gives the breakdowns of the time and storage costs of {\flowdist} over its two phases and major phase steps.
The time costs include those for static analysis (and instrumentation if any) ({\bf Static}), profiling ({\bf Run}), and
on average for computing the (method-level or statement-level) information flow
paths between each source-sink pair ({\bf Query}).
The second column lists the original run time ({\bf Norm Run}) of each execution, from which profiling overheads were computed as runtime slowdown ratios ({\bf Slowdown}).
The eighth column shows the time costs for statement coverage analysis ({\bf Coverage}).
The last column lists the total storage costs ({\bf Stor.}) for all phases per execution, for storing 
the traces of method and branch events in Phase 1, statement coverage and partial static dependence graph in Phase 2,
and the instrumented program.
The overall averages (across all executions) are given in the last row.

On average over the 18 executions, {\flowdist} took 19 minutes ((437 + 565 + 124) / 60) for all one-off analyses,
including the time for all static analyses, instrumentation, and coverage analysis, as shown in the last row of Table~\ref{tab:fdcosts}.
I considered them {\em one-off} because their results are shared by all queries with respect to a given subject execution and source/sink configuration.
In particular,
the {\em partial} dependence analysis (as guided by the method-level paths from Phase 1 was
significantly more efficient than a whole-system analysis without a pre-analysis phase.
For example, the latter did not even finish in 12 hours with otherwise the same setup against Voldemort.

On the other hand, Phosphor and JOANA took 1.38 and 0.43 seconds on average, respectively, for each source/sink pair,
lower than {\flowdist}'s querying cost (13 seconds on average).
{\flowdist} also incurred a higher average storage cost (293.4MB) than Phosphor (21.2MB)
and JOANA (35.2MB).
The reason is that {\flowdist} performed more, heavier analyses (e.g., probing, building the dynamic dependence graph, profiling instance-level method events) than the baselines.
However, these extra costs of {\flowdist} were moderate and should be paid off by its much higher effectiveness.
Also, it did not incur the substantial manual effort (e.g., test case development or source code annotation).

\vspace{2pt}
\noindent
\textbf{Scalability.}
\begin{figure}[tp]
  \centering
  \caption{The run-time slowdowns (\%, $y$ axis) versus \#method execution event instances ($x$ axis) of all subject executions.} 
  \includegraphics[width=0.8\textwidth]{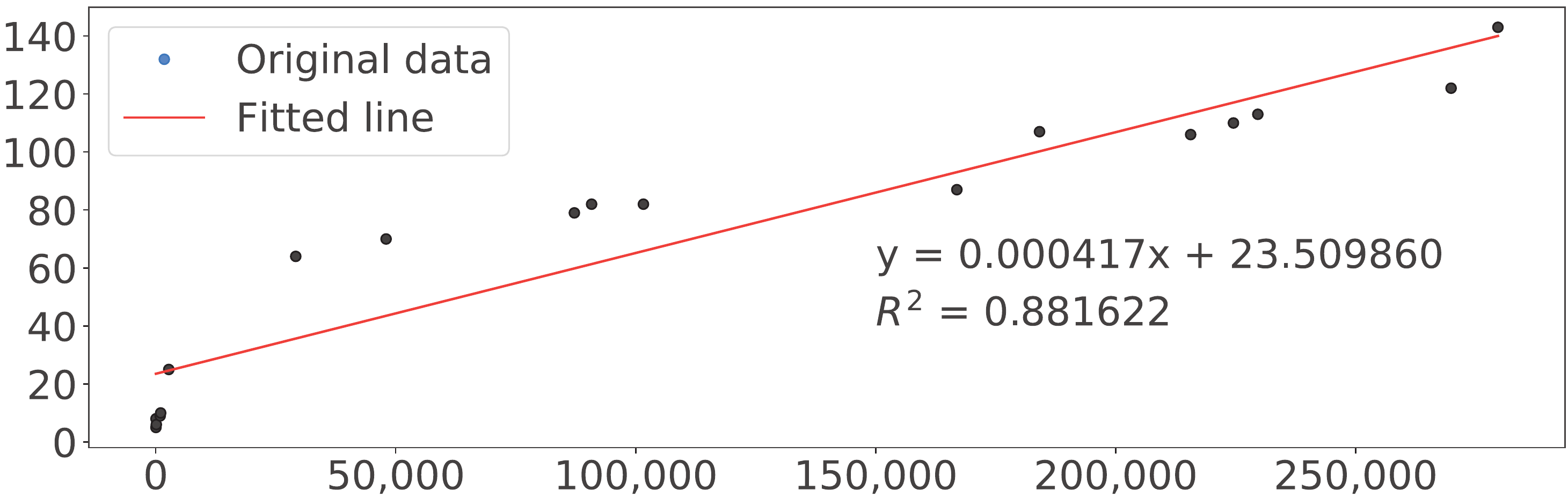}
  \label{fig:scalab}
 \vspace{10pt}
\end{figure}
Figure~\ref{fig:scalab} shows the scalability of {\flowdist} in terms of its runtime slowdown, for all 18 executions, each characterized by the length of the instance-level method execution event sequence in it as a run-time complexity measure.
The fitting curve with $R^2 >$ 0.88, indicates that {\flowdist} scaled gracefully to large-scale systems in terms of the runtime overhead,
where the determination coefficient $R^2\in$[0,1] indicates how close the data are to the curve (the closer $R^2$ is to 1, the better the fitting is).

\begin{figure}[htbp]
 \vspace{10pt}
  \centering
  \caption{The total analysis time (seconds, $y$ axis) versus subject size (\#SLOC, $x$ axis) of all subjects (integration test)}
  \includegraphics[width=0.8\textwidth]{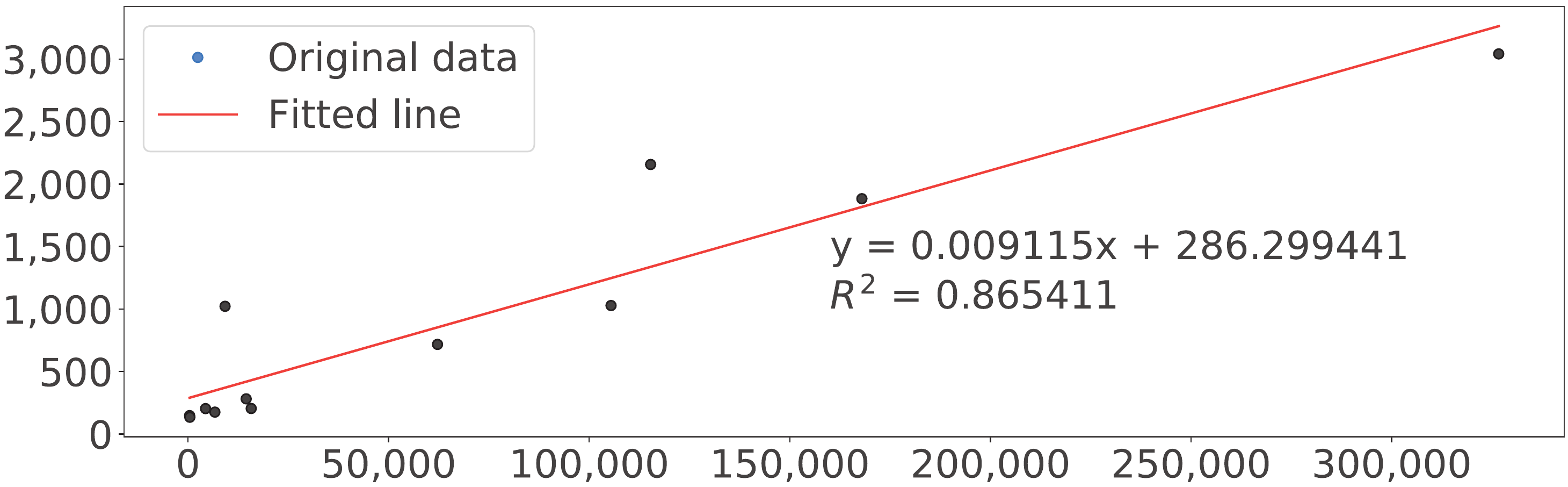}
  \label{fig:totala}
  \vspace{10pt}
\end{figure}

Meanwhile, Figure~\ref{fig:totala} shows how {\flowdist} scaled to subjects of growing sizes in terms of its total time cost (the sum of one-off analysis time, profiling costs, and
the time for querying all possible source/sink pairs), against integration tests because every subject has such a test.
In the same format as Figure~\ref{fig:scalab}, the fitting curve in Figure~\ref{fig:totala} indicates that the total time cost of {\flowdist} grew linearly.

\vspace{2pt}
\noindent
\textbf{Discovering vulnerabilities.}
\begin{table}[tp]
	\scriptsize
	\centering
	\caption{Existing vulnerabilities detected by {\flowdist}}
		\begin{tabular}{|l|l|c|c|c|c|}
			\hline
			\textbf{Subject} & \multicolumn{1}{c|}{\textbf{Vulnerability}} & \multicolumn{1}{c|}{\textbf{Reference}} & \multicolumn{1}{c|}{\textbf{Found}} & \multicolumn{1}{c|}{\textbf{\#Vulnerability}} & \multicolumn{1}{c|}{\textbf{\#False Negative}} \\
			\hline
			\multicolumn{1}{|l|}{HSQLDB} & CVE-2005-3280 & \cite{cve20053280} & \checkmark     & \multicolumn{1}{c|}{1} & \multicolumn{1}{c|}{0} \\
			\hline
			\multicolumn{1}{|l|}{\multirow{10}[2]{*}{Netty}} & CVE-2014-0193 & \cite{cve20140193} & \xmark     & \multicolumn{1}{c|}{\multirow{10}[2]{*}{10}} & \multicolumn{1}{c|}{\multirow{10}[2]{*}{5}} \\
			& CVE-2014-3488 & \cite{cve20143488} & \xmark     &       &  \\
			& CVE-2015-2156 & \cite{cve20152156} & \xmark     &       &  \\
			& CVE-2016-4970 & \cite{cve20164970} & \xmark     &       &  \\
			& Issue 8869 & \cite{netbug8869} & \xmark     &       &  \\
			& Issue 9112 & \cite{netbug9112} & \checkmark     &       &  \\
			& Issue 9229 & \cite{netbug9229} & \checkmark     &       &  \\
			& Issue 9243 & \cite{netbug9243} & \checkmark     &       &  \\
			& Issue 9291 & \cite{netbug9291} & \checkmark     &       &  \\
			& Issue 9362 & \cite{netbug9362} & \checkmark     &       &  \\
			\hline
			\multicolumn{1}{|l|}{RocketMQ} & CVE-2019-17572 & \cite{cve201917572} & \checkmark     & \multicolumn{1}{c|}{1} & \multicolumn{1}{c|}{0} \\
			\hline
			\multicolumn{1}{|l|}{Thrift} & CVE-2015-3254 & \cite{cve20153254} & \checkmark     & \multicolumn{1}{c|}{1} & \multicolumn{1}{c|}{0} \\
			\hline
			\multicolumn{1}{|l|}{\multirow{6}[2]{*}{Voldemort}} & Issue 101 & \cite{voldemortbug101} & \checkmark     & \multicolumn{1}{c|}{\multirow{6}[2]{*}{6}} & \multicolumn{1}{c|}{\multirow{6}[2]{*}{1}} \\
			& Issue 381 & \cite{voldemortbug381} & \checkmark     &       &  \\
			& Issue 387 & \cite{voldemortbug387} & \checkmark     &       &  \\
			& Issue 352 & \cite{voldemortbug352} & \checkmark     &       &  \\
			& Issue 378 & \cite{voldemortbug378} & \checkmark     &       &  \\
			& Issue 377 & \cite{voldemortbug377} & \xmark     &       &  \\
			\hline
			\multicolumn{1}{|l|}{xSocket} & Bug 21 & \cite{x21} & \checkmark    & \multicolumn{1}{c|}{1} & \multicolumn{1}{c|}{0} \\
			\hline
			\multicolumn{1}{|r|}{\multirow{4}[2]{*}{ZooKeeper}} & CVE-2014-0085 & \cite{cve20140085} & \checkmark     & \multicolumn{1}{c|}{\multirow{4}[2]{*}{4}} & \multicolumn{1}{c|}{\multirow{4}[2]{*}{0}} \\
			& Bug 2569 & \cite{zookeeperbug2569} & \checkmark     &       &  \\
			& CVE-2018-8012 & \cite{cve20188012} & \checkmark     &       &  \\
			& CVE-2019-0201 & \cite{cve20190201} & \checkmark     &       &  \\
			\hline
		\end{tabular}%
	\label{tab:fdissues}%
\end{table}%
I searched for real-world vulnerabilities from varied sources (e.g., bug repositories and CVE reports) on our subjects, selected those on information flow security, and then identified one or more vulnerabilities for 7 of the studied subjects, as shown in Table~\ref{tab:fdissues}.
For each of these subjects ({\bf Subject}), vulnerabilities ({\bf Vulnerability}) along with reference links ({\bf Reference}), and vulnerability counts ({\bf \#Vulnerability}) are listed,
with marks ({\bf Found}) indicating whether the vulnerability was found or not.
The last column gives the numbers of missed vulnerabilities ({\bf \#False Negative}).

From the information flow paths inferred (as shown in Table~\ref{tab:localremotepaths}),
{\flowdist} successfully discover most vulnerabilities for all these 7 subjects but Netty.
Five vulnerabilities for Netty and one for Voldemort were missed.
I verified that the reasons for that the missed vulnerabilities were not covered during the executions.
I did not purposely select run-time inputs to cover the vulnerabilities but just used available ones that represented the system operational scenarios.
In addition, for all of the 18 successful cases, corresponding paths were interprocess ones.

\setlength{\tabcolsep}{6.5pt}
\begin{table}[htbp]
  \centering	
  \caption{New vulnerabilities discovered by {\flowdist}}	
    \begin{tabular}{|l||r|r|r|}
    \hline
     {\textbf{Subject}}     & \multicolumn{1}{|l|}{\textbf{\#Fixed}} & \multicolumn{1}{l|}{\textbf{\#Confirmed}} & \multicolumn{1}{l|}{\textbf{\#Pending}} \\
    \hline
    HSQLDB & 0     & 5     & 2 \\
    \hline
    Netty & 1     & 1     & 0 \\
    \hline
    Raining Sockets & 0     & 1     & 0 \\
    \hline
    RocketMQ & 0     & 4     & 0 \\
    \hline
    Thrift & 0     & 5     & 0 \\
    \hline
    Voldemort & 0     & 0     & 4 \\
    \hline
    xSocket & 0     & 0     & 1 \\
    \hline
    Zookeeper & 1     & 1     & 0 \\
    \hline
    \end{tabular}%
  \label{tab:newbugs}%
\end{table}%
Furthermore, from the information flow paths found by {\flowdist},
related to 8 subjects, 24 new vulnerabilities~\cite{flowdistpackage} were identified, as listed in Table~\ref{tab:newbugs}.
After reported by me, 17 have been confirmed and 2 have already been fixed so far.
This suggests that {\flowdist} computes information flow paths in given executions for detecting known or new vulnerabilities/bugs, without the requirements of bug reports;
albeit such reports may benefit the vulnerability detection.

\begin{figure}[htbp]
  \centering
\caption{The total time costs (in seconds) of {\flowdistmul} and {\flowdistsim} against {\flowdist} for all subject executions}
  \includegraphics[width=1\textwidth]{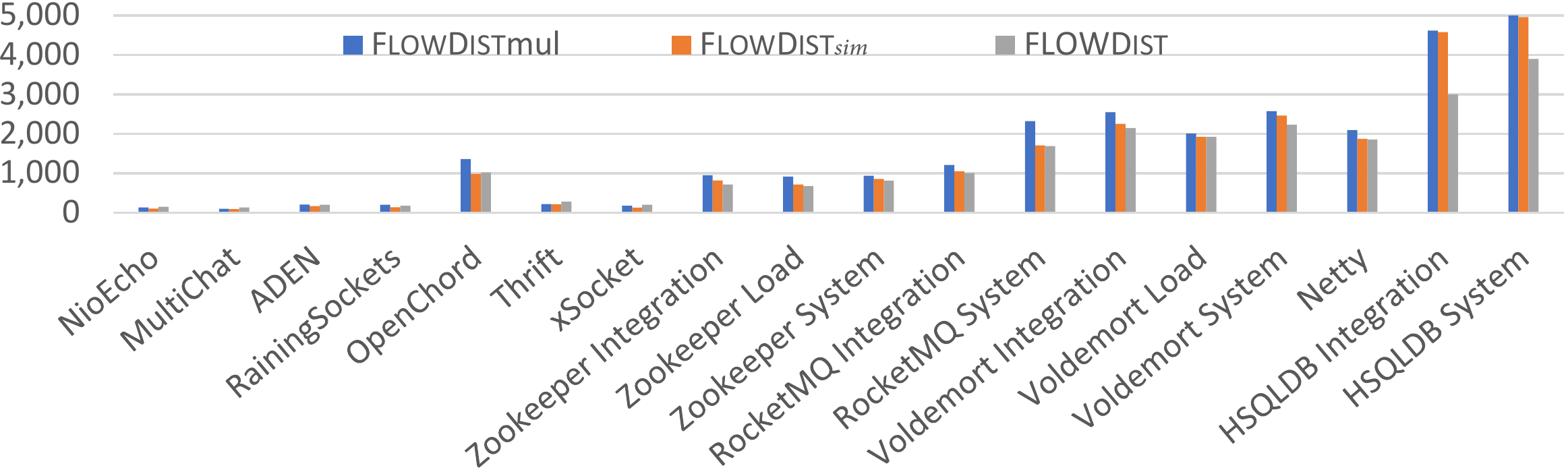}
  \label{fig:flowdistcomp}
\end{figure}

\vspace{2pt}
\noindent
\textbf{Alternative design comparisons.}
As expected, the best performer among the three varied for different systems in terms of efficiency.
Figure~\ref{fig:flowdistcomp} shows the contrasts in the total analysis time of {\flowdist} and its designs for 18 executions studied.
For a relatively large system (ZooKeeper or larger), {\flowdist} was the most efficient.
For the system, the time saved due to the reduced instrumentation and profiling scope in the pre-analysis noticeably outweighed the static analysis time cost, and thus {\flowdist} was better than {\flowdistsim}.
Meanwhile, the time saved due to the reduced scope of profiling instance-level method events was more than the extra time cost of the intermediate phase, and thus {\flowdist} was better than {\flowdistmul}.

These outweighing contrasts were reversed for small systems (those smaller than ZooKeeper), which explains
why the alternative designs were better than {\flowdist}, for those systems.

\begin{figure}[htbp]
	\centering
	\caption{The storage costs (in MB) of {\flowdistmul} and {\flowdistsim} against {\flowdist} for all subject executions}
	\includegraphics[width=1\textwidth]{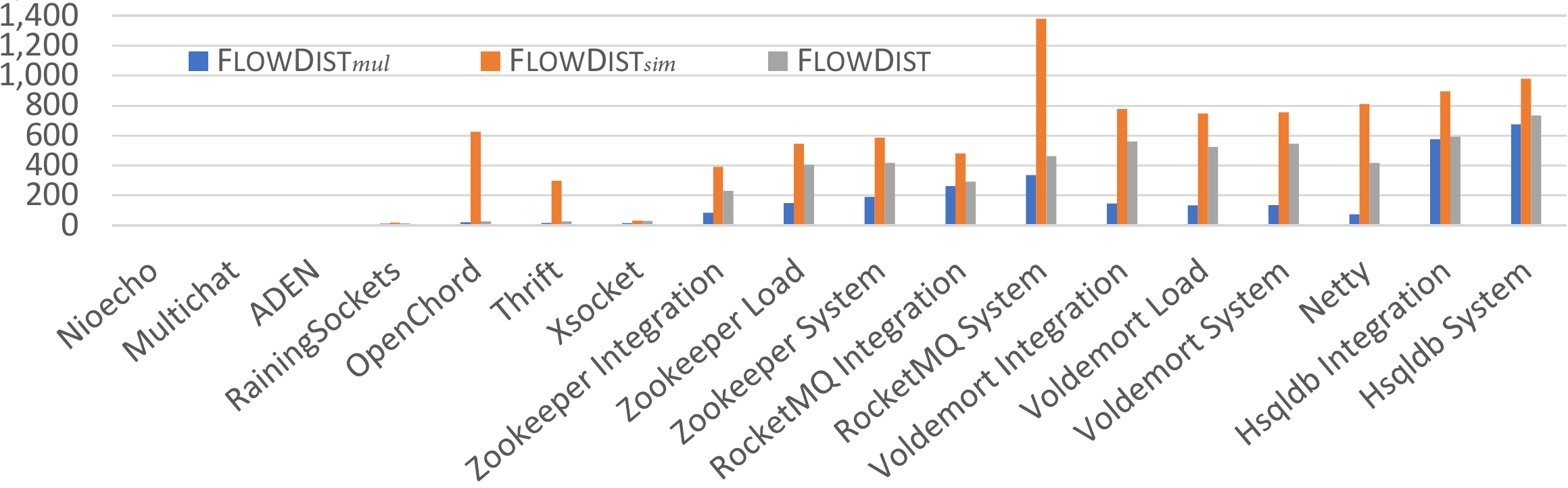}
	\label{fig:flowstorage}
\end{figure}

The comparison of storage costs revealed insignificant differences, as shown in Figure~\ref{fig:flowstorage}.
{\flowdistmul} and {\flowdistsim} had the least and the most storage space requirements, respectively.
And {\flowdist} needed storage spaces between them.
The reason is that {\flowdistmul} only records the first entry and last returned-into events in the pre-analysis phase, and
then only traces methods on the method-level flow paths found in the pre-analysis and branches in those methods.
In contrast, 
{\flowdistsim} traces all instance-level method and statement branch events in the execution during the pre-analysis phase.
On the other hand, {\flowdist} just traces relevant methods and branches only.

\begin{table}[htbp]
  \centering
  \caption{Recommendations on DIFA/DTA tool selection}
    \begin{tabular}{|l|l|l|l|l|}
    \hline
    \multicolumn{3}{|c|}{\multirow{2}[1]{*}{\textbf{System type}}} & \multicolumn{2}{c|}{\textbf{Is the execution non-deterministic?}} \\
\cline{4-5}    \multicolumn{3}{|c|}{} & \multicolumn{1}{c|}{\textbf{Yes}} & \multicolumn{1}{c|}{\textbf{No}} \\
    \hline
    \multirow{5}[2]{*}{\shortstack{Distributed\\(multi-process)}} & \multicolumn{1}{l|}{\multirow{3}[2]{*}{Common}} & \multicolumn{1}{l|}{\multirow{2}[2]{*}{Small}} & \multirow{2}[2]{*}{\flowdistsim} & \multicolumn{1}{c|}{\flowdistsim} \\
\multicolumn{1}{|l|}{} &       &       & \multicolumn{1}{l|}{} & \multicolumn{1}{c|}{or {\flowdistmul}} \\
\cline{3-5}    \multicolumn{1}{|l|}{} &       & \multicolumn{1}{c|}{Large} & {\flowdist} & \multicolumn{1}{c|}{\flowdist} \\
\cline{2-5}    \multicolumn{1}{|l|}{} & \multicolumn{2}{c|}{\multirow{2}[2]{*}{Specialized}} & \multicolumn{2}{c|}{Kakute~\cite{jiang2017kakute} (for Spark~\cite{zaharia2012resilient})} \\
\cline{4-5}    \multicolumn{1}{|l|}{} & \multicolumn{2}{l|}{} & \multicolumn{2}{c|}{Pileus~\cite{sun2016pileus} (for OpenStack~\cite{sefraoui2012openstack}), ...} \\
    \hline
    \multicolumn{3}{|c|}{Single-process} & \multicolumn{2}{c|}{Phosphor~\cite{bell2014phosphor}, JOANA~\cite{tooljoana2013atps}, ...} \\
    \hline
    \end{tabular}%
  \label{tab:recommend}%
\end{table}%

The findings shown in Table~\ref{tab:recommend} led me to recommend how to select the right tool for a particular software system.
Overall, {\flowdist} (default design) best suits large-scale common distributed systems, regardless whether their executions are non-deterministic or not.
For a small common distributed system, if its execution is known to be deterministic, both {\flowdistsim} and {\flowdistmul} might be a great choice; otherwise, only {\flowdistsim} should be considered.
I also list a few peer tools that suite other types of (specialized distributed or single-process) systems in Table~\ref{tab:recommend}.

\section{Related Work}
Two classes of prior works are most relevant to ours: {\em conventional information flow analysis (IFA)} \& {\em taint analysis (TA)} and {\em cross-process information flow analysis (IFA)} \& {\em taint analysis (TA)}.

\subsection{Conventional information flow analysis (IFA) \& taint analysis (TA)}
Most prior analyses of this kind are static~\cite{myers1999jflow,sabelfeld2003language,simonet2003flow,wang2008still,arzt2014flowdroid,li2015iccta,wei2014amandroid,gordon2015information,zhang2019understanding,tooljoana2013atps}.
These approaches suffer from the common imprecision of static analysis, in addition to unsoundness due to dynamic constructs (e.g., dynamic code loading) in
modern languages~\cite{livshits2015defense}.
When applied to distributed software, 
they 
would be subject to even greater inaccuracy due to implicit dependencies among distributed (decoupled) components. 

Among dynamic approaches, TaintDroid~\cite{enck2014taintdroid} customizes the Android OS to track whole-system information flow
at runtime and provides warnings to users when sensitive flows are found.
Panorama~\cite{yin2007panorama} performs system-side
dynamic information flow tracking for Windows malware analysis through dynamic instrumentation based on a processor emulator QEMU.
In~\cite{hauser2013intrusion}, a dynamic taint analysis was used for intrusion detection via a custom Linux security module (i.e., modifying the Linux kernel).
A few others~\cite{chow2004understanding,crandall2004minos,suh2004secure} require specialized hardware to perform taint tracking.
These approaches rely on platform customizations. 
Also, they belong to DTA rather than DIFA since they do not provide full, code-level information flow paths.
Juturna~\cite{loch2020hybrid}, a hybrid approach combining with JOANA, employs bytecode augmentation and modified Java API classes for taint tracking.
Thus, it requires platform customizations too. 

Dytan~\cite{clause2007dytan} provides a generic 
framework for dynamic tainting x86 executables
and an instantiation of the framework for x86 binaries.
Similar to Dytan, 
TaintEraser~\cite{zhu2011tainteraser} leverages a dynamic instrumentation framework Pin~\cite{luk2005pin} to
track run-time information flow in Windows applications.
In~\cite{austin2009efficient,austin2010permissive,karim2018platform}, the authors proposed language semantics for dynamic taint analysis of JavaScript code.
In particular,~\cite{karim2018platform} uses the Jalangi framework~\cite{sen2013jalangi} to instrument the ECMAScript 5~\cite{deitel2011internet} code of target systems.
LabelFlow~\cite{chinis2013practical} is an extension of PHP to simplify security policy implementation in web applications.
TaintMan~\cite{enck2014taintdroid} is an Android RunTime (ART) compatible DTA that statically instruments the code of both the target Android application and libraries to track data and control flows.
TaintTrace~\cite{cheng2006tainttrace} uses DynamoRIO~\cite{brueningbuilding} to perform instrumentation, and its loader was implemented by modifying the code of Valgrind~\cite{nethercote2007valgrind}.
In addition, TaintCheck~\cite{newsome2005dynamic} performs instrumentation using Valgrind and
LIFT~\cite{qin2006lift} is based on the StarDBT dynamic binary translator~\cite{wang2007stardbt} while 
DFSan~\cite{dfsan} employs the LLVM framework~\cite{lattner2004llvm}.

Like other analyzers~\cite{newsome2005dynamic,ming2015taintpipe,she2020neutaint,ashouri2019hybrid,banerjee2019iodine}, these approaches
do not work properly with distributed software as they 
only track information flow in single threads/processes.

\subsection{Cross-process information flow analysis (IFA) \& taint analysis (TA)}
Among a number of prior techniques for tracking/checking dynamic information flow, only a few addressed the flow across processes.
One of those is Kakute~\cite{jiang2017kakute}, which 
tracks field-level data flow with unified APIs for reference propagation and tag sharing.
Since it is based on Phosphor~\cite{bell2015dynamic}, Kakute also needs to customize (instrument) its runtime platform (i.e., JVM).
Also, it targets big-data applications running on Spark~\cite{zaharia2012resilient}, not working with common distributed software.
This is similar to Pileus~\cite{sun2016pileus} targeting applications on a special cloud platform OpenStack~\cite{sefraoui2012openstack}.

Taint-Exchange~\cite{zavou2011taint} is a generic cross-process/host framework 
for taint tracking.
It builds on libdft~\cite{kemerlis2012libdft} to transfer taint information exchanged between hosts/processes through sockets and pipes.
Like Cloudfence~\cite{pappas2013cloudfence} and Cloudopsy~\cite{zavou2015information},
it relies on a customized runtime (the Pin dynamic instrumentation framework) and targets C/C++ software.

%% file: chapters/chapter5.tex
\chapter{{\seads}: Scalable and Cost-Effective Dynamic Dependence Analysis via Reinforcement Learning} \label{ch:seads}
A fundamental strategy for understanding and validating software behaviors is to model run-time interactions among program entities as
{\em dependencies} and then reason about program behaviors based on the dependence model~\cite{korel1988dynamic,jackson2000software,binkley06aug}.
Historically, dynamic dependence modeling and analysis~\cite{podgurski1990formal,horwitz1992use} supported software quality assurance,
ranging from fault diagnosis~\cite{faultlocsurvey2016} to security defense~\cite{attariyan2010automating,kemerlis2012libdft}.
Dynamic dependence analysis is important because many application techniques in software quality assurance rely on dynamic dependencies, such as performance monitoring, program optimization, security defense, software testing, vulnerability detection, and so on~\cite{kreindl2019towards}.
For instance, software testing is also crucial for software quality assurance, for which dynamic dependencies can be utilized to detect defects in the software by searching among the dependencies of the program entities where faulty outputs are observed.
Similarly, dependencies can be used to detect run-time sensitive data leaks to sinks from sources via the chains of the dependencies.
Compared with static approaches, dynamic dependence analysis has greater precision as it focuses on specific, concrete executions.
Thus, I developed a cost-effective dynamic dependence analysis framework and implement a dynamic slicer with respect to user budget constraints.

\section{Motivation}

In general, a fundamental challenge to dynamic analysis for distributed software is how to achieve an optimal balance between analysis overheads and the effectiveness of the analysis algorithm (e.g., maximal analytical accuracy with minimal cost).
Early dependence modeling and analysis for parallel and distributed software have been attempted~\cite{cheng1997dependence,korel1992dynamic,duesterwald1993distributed,kamkar1995dynamic,mohapatra2006distributed}
with unbalanced costs and the lack of scalability/effectiveness.

For example, I attempted to apply an existing state-of-the-art dependency analysis for distributed programs~\cite{cai2021d2abs} to a real-world system~\cite{voldemort}.
The analysis, running with an 8-processor 2.7 GHZ CPU and 512 GB DRAM, did not finish even in 12 hours.
And I did not find any existing dependence analysis approach which is able to provide a scalable solution for industry-scale distributed systems in the real world.

However, more and more industry-scale software systems are becoming distributed systems in nature.
Thus, there is an urgent requirement for tool support for distributed systems, and for which dynamic dependence analysis is a fundamental, enabling approach.
I believe that online analysis is a better option than offline analysis for distributed systems.
The online solution analyzes the system during the run,
while offline approaches compute dependencies after program executions.
Thus, online analysis would be much faster, albeit with slight runtime overhead.
In the continuous infinite execution, execution traces are unnecessary for an online approach.

By contrast, offline techniques require traces, whose storage costs may be expensive.
If the execution is interminable, the requirement for trace storage space would also be unlimited.
However, this is impractical, and hence offline analyses are not proper for distributed systems.

\section{Approach}
I developed {\seads} (Short for \underline{S}calable and cost-\underline{E}ffective dynamic dependence \underline{A}nalysis of \underline{D}istributed \underline{S}ystems)~\cite{fu2020seads}, a {\em cost-effective}
dynamic dependence analysis framework that can scale the analysis to real-world distributed systems.
The analysis itself is {\em distributed} and {\em online} to overcome the problem with unbounded execution traces 
while analyzing continuously running systems.
Moreover, given a user-specified time budget,
the analysis automatically adjusts itself by varying configurations for better cost-effectiveness tradeoffs (than otherwise), according to the analysis time cost(s).
The core idea is using a reinforcement learning method to decide which configurations to adjust to according to the current configuration and corresponding analysis cost with respect to a given user budget.
\subsection{Overview}
\begin{figure*}[tp]
  \centering
  \caption{An overview of the {\seads} architecture and workflow, including its input, output, and key modules}
  \includegraphics[width=0.9\textwidth]{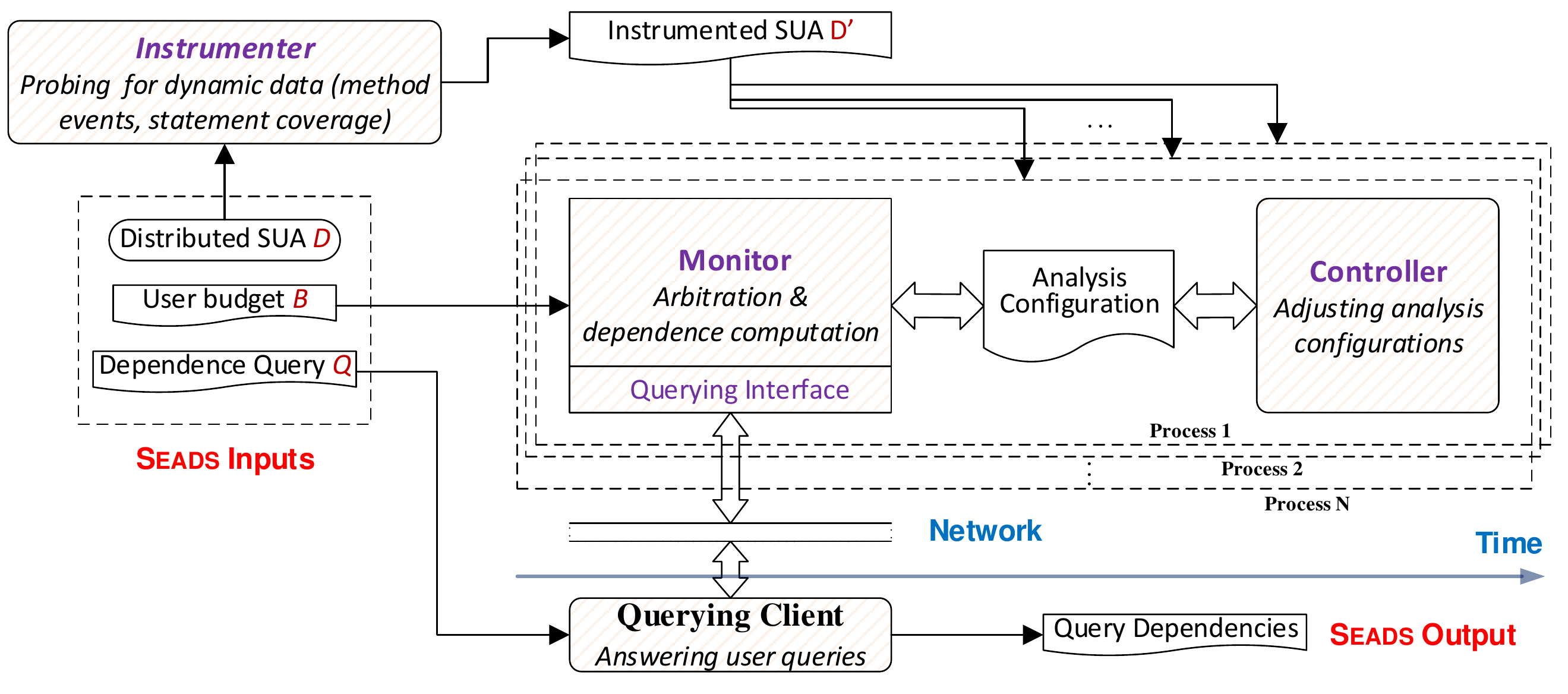}
  \label{fig:seadsoverview}
\end{figure*}
The overall architecture of {\seads} is shown in Figure~\ref{fig:seadsoverview}, consisting of an {\em instrumenter}, {\em monitors} each for a process of the system under analysis (SUA), {\em controllers} each for a process of the SUA, and a {\em querying\_client}.
{\seads} takes three user {\bf inputs}: the distributed SUA $D$, a user budget $B$, and a dependence query $Q$.
In particular, $B$ is a response time constraint for the dependence analysis, while $Q$ is a method name imputed by the user for s dependencies of the method, such as {\em org.apache.zookeeper.server.ZooKeeperServerMain: void main(java.lang.String[])}.

The {\em instrumenter} first inserts probes, which will monitor {\em entry} and {\em returned-into} method execution events and/or monitor the executed statement coverage,
into $D$ to produce the instrumented system $D'$.
Then, as {\em time} goes on, $D'$ continuously runs and {\seads} continually adjusts itself in its analysis configurations through per-process monitors and controllers.
In particular, during the execution of $D'$, I suppose that there are $N$ processes (e.g., a server and $N-1$ clients).
In each of its $N$ processes,
a {\em monitor} performs arbitration (deciding the adjustment) and dependence computation,
and a 
{\em controller} adjusts analysis configurations.
In short, the overall dynamic dependence analysis is performed in a distributed manner with per-process analysis configurations being adjusted independently of those in other processes.

The {\em querying\_client} receives query $Q$ from the user and then sends it to the {\em querying\_interface} that directly communicates with the 
{\em monitor} in each process, through the {\em network} facility.
After the dependence computation in a process has finished, resulting dependencies for $Q$ are delivered to the {\em querying\_interface} attached to the monitor in that process,
from which the {\em querying\_client} receives the queried dependencies for the process.
After the distributed dependence analyses in all processes are finished,
the {\em querying\_client} receives/merges all per-process dependence sets for $Q$
and then performs an interprocess analysis to compute the overall query dependencies
as the final {\bf output} of {\seads}.

\subsection{Configuration}
The key idea for achieving scalability and cost-effectiveness is to continually adjust analysis configurations according to (1) the user budget,
(2) the current and previous configurations, and (3) time costs of dependence computations.
In {\seads}, analysis configurations play a key role, each of whose parameters uniquely contributes to cost-effectiveness.
As a hybrid approach, {\seads} leverages various combinations of static and dynamic analysis techniques along with varied static/dynamic data
(e.g., the static dependence graph, method events, and the statement coverage) while using different static and dynamic configuration parameters.
I first present the parameters (i.e., configuration items) considered in the static and dynamic analysis separately (referred to as {\em static configurations} and {\em dynamic configurations}, respectively), and then describe the holistic (hybrid) configuration encodings.

\vspace{2pt}
\noindent
\textbf{Static configuration.}
There are two static configuration parameter dimensions: {\em data\_selection} and {\em sensitivity}.
In the {\em data\_selection} dimension, one parameter {\em staticGraph} concerns whether static data is used and determines whether {\seads} uses static dependencies to compute the dynamic dependencies of the given query, via traversing the per-component
static dependence graphs to infer precise dependencies with a high time cost.

The {\em sensitivity} dimension includes two parameters({\em context sensitivity} and {\em flow sensitivity}), expected to achieve a relatively high level of precision, as explained below.
\begin{enumerate}
\item {\em Context-sensitivity} concerns the effects of different calling contexts of the same variable/object (e.g., method).
A context-insensitive analysis uses a single abstraction for a variable/object for all calling contexts.
On the contrary, a context-sensitive analysis distinguishes different calling contexts of a variable/object and computes separate information for these calling contexts \cite{spath2019context,hammer2009flow}.
For instance, if a method is called multiple times at different callsites, a context-sensitive analysis would distinguish these callsites and computes separate analysis facts (e.g., dependencies) for them.

There are two (the functional and call-string) approaches implementing context-sensitive analyses, varying in the context abstraction and the analysis algorithm.
In the functional approach using methods' arguments, there is a summary function mapping each context to the corresponding method effect on the analysis facts.
In the call-string approach using the strings of call sites, the analysis facts are tagged with context, and the analysis propagates the tagged facts along the flow graph of the system under analysis \cite{sharir1978two}.
Context-sensitive analyses using call site strings as context are often called {\em k}-CFA, where CFA is short for control flow analysis and {\em k} is an integer limit on the (calling) lengths of the strings, referring to a hierarchy for call sites \cite{shivers1988control,might2010resolving,lhotak2008evaluating}.
In particular, 0-CFA, where the 0 indicates context insensitivity, works quickly.
However, its analysis results may be imprecise \cite{lncfa}.
\item
 {\em Flow-sensitivity} considers control flow reachability.
A flow-insensitive analysis does not consider the execution order (i.e., the control flow) of entities (e.g., statements),
while a flow-sensitive analysis approach takes into account the order, when computing the information (e.g., dependencies among entities).
For example, if a variable was defined twice and then used once in a program,
a flow-insensitive analysis only considers the second definition and the use.
However, a flow-sensitive analysis differentiates the order of these two definitions and the use during the execution(s).
Therefore, the relevant information (e.g., a dependence between the first definition and the use) is only reported by the flow-sensitive analysis \cite{ryder2003dimensions}.
A flow-sensitive analysis needs additional (time) costs to analyze the entity execution order and thus is more expensive but much more precise than a flow-insensitive analysis.
\end{enumerate}
\vspace{2pt}
\noindent
\textbf{Dynamic configuration.}
{\seads} considered two dimensions of dynamic configuration parameters: {\em data\_selection} and {\em data\_granularity}.
The {\em data\_selection} dimension concerns which types of data used, including {\em methodEvent} and {\em statementCoverage} parameters.
\begin{enumerate}
\item The parameter {\em methodEvent} decides whether {\seads} uses method ({\em entry} and {\em returned-into}) events to compute dependencies.
If the parameter is enabled, {\seads} infers more precise dependencies from dynamic data (e.g., method events) with additional costs.
Otherwise, with the disabled parameter {\em methodEvent}, {\seads} coarsely but quickly computes dependence sets without method events.
\item The parameter {\em statementCoverage} determines if {\seads} prunes the static dependence graphs using statement coverage information.
This means that with enabled {\em statementCoverage}, {\seads} only considers statements covered in the execution and dismissed other statements.
And thus, with other statements dismissed while referring to the static dependencies, {\seads} obtains more precise results than otherwise.
\end{enumerate}

Only one parameter {\em MethodInstanceLevel} is in the {\em data\_granularity} dimension, concerning the granularity of the dynamic data (i.e., method events) used in the dependence computation. 
It is about whether {\seads} uses all method event instances to compute dependencies.
If the parameter is enabled, {\seads} utilizes all instances of ({\em entry} and {\em returned-into}) events to compute dependencies more precisely with a larger level of costs for monitoring and utilizing a greater amount of dynamic data.
Otherwise, only the first {\em entry} and last {\em returned-into} method events gathered/used, the computation is faster but rougher (i.e., more imprecise).

\vspace{2pt}
\noindent
\textbf{Configuration encoding.}
The holistic configuration of {\seads} consists of both the static and dynamic configurations described above.
Three bits encode the three parameters in the static configuration, and other bits are for dynamic configuration parameters.
The first to third bits are encoded as the static configuration parameters,
while the fourth to sixth bits are used as the dynamic configuration parameters.
The binary number 1 or 0 means that corresponding parameter is enabled or disabled, respectively.
Therefore, 
the holistic (i.e., hybrid) configuration, including static and dynamic configuration parameters, is encoded as a 6-bit binary number ranging from 000000 through 111111.
As shown in Table~\ref{tab:seadstcn}, in the configuration encoding, the hybrid configuration parameters' ordering is:
{\em staticGraph}, {\em context-sensitivity}, {\em flow-sensitivity}, {\em methodEvent}, {\em statementCoverage}, and {\em methodInstanceLevel}.

\setlength{\tabcolsep}{0.5pt}
\begin{table}[tp]
  \centering
  \caption{Hybrid (Static/Dynamic) Configuration Encoding}
\resizebox{1\columnwidth}{!}{%
    \begin{tabular}{|c|c|c|c||c|c|c|}
    \multicolumn{1}{r}{} & \multicolumn{1}{r}{} & \multicolumn{1}{r}{} & \multicolumn{1}{r}{} & \multicolumn{1}{r}{} & \multicolumn{1}{r}{} & \multicolumn{1}{r}{} \\
    \hline
    \multirow{4}[6]{*}{Encoding} & \multicolumn{3}{c||}{Static Configuration} & \multicolumn{3}{c|}{Dynamic Configuration} \\
\cline{2-7}          & Data Selection & \multicolumn{2}{c||}{Sensitivity} & \multicolumn{2}{c|}{Data Selection} & Data Granularity \\
\cline{2-7}          & \multirow{2}[2]{*}{StaticGraph} & Context & Flow  & Method & Statement & Method \\
          &       & Sensitivity & Sensitivity & Event & Coverage & InstanceLevel \\
    \hline
    000000 & Disabled (0)  & Disabled (0)  & Disabled (0)  & Disabled (0)  & Disabled (0)  & Disabled (0)  \\
    \hline
    000001 & Disabled (0)  & Disabled (0)  & Disabled (0)  & Disabled (0)  & Disabled (0)  & Enabled (1) \\
    \hline
    \multicolumn{7}{|c|}{......} \\
    \hline
    111110 & Enabled (1) & Enabled (1) & Enabled (1) & Enabled (1) & Enabled (1) & Disabled (0)  \\
    \hline
    111111 & Enabled (1) & Enabled (1) & Enabled (1) & Enabled (1) & Enabled (1) & Enabled (1) \\
    \hline
    \end{tabular}%
}
  \label{tab:seadstcn}%
\end{table}%
Some of all possible 64 ($2^6$) hybrid configurations are invalid and cannot be used in {\seads}.
The reason is that certain configuration parameters are dependent on others, and they are meaningful only with other parameters enabled.
For example, three parameters ({\em context-sensitivity}, {\em flow-sensitivity}, {\em statementCoverage}) depend on the parameter {\em staticGraph}.
If the parameter {\em staticGraph} is disabled, meaning that {\seads} does not use the static data (i.e., the static dependence graph),
then the three relevant parameters ({\em context-sensitivity}, {\em flow-sensitivity}, and {\em statementCoverage}) are meaningless---statement coverage data is only used for pruning the static dependence graph in {\seads}.
Thus, configurations 001xxx, 010xxx, 011xxx, and 0xxx1x are invalid, where 'x' is a bit which can be 0 or 1.
Another parameter {\em methodInstanceLevel} depends on the parameter {\em methodEvent} and thus configurations xxx0x1 are invalid. 
Configuration 000000 is also invalid, meaning no data is used in the analysis.
In sum,  there are 38 invalid configurations and 26 valid configurations in {\seads}.

\subsection{Instrumenter}
The {\em instrumenter} inserts probes into the system $D$ to generate the instrumented version $D'$ that will continuously run.
During the execution, these probes monitor and record the {\em entry} and {\em returned-into} events of all executed method to infer the method happens-before relations and further approximate dynamic dependencies among the methods within and across processes~\cite{cai14diver}.

The instrumenter also probes statement branches for efficiently inferring statement coverage~\cite{cai2016diapro}, another kind of dynamic data considered.
In particular, besides the branches related to explicit predicates, the method entries are also treated as special branches (i.e., {\em entry branches}), whose {\tt true} edges lead to the corresponding method (execution) entries.

\subsection{Monitor}
\begin{figure}[tp]
  \centering
  \caption{The monitor and controller run along with the instrumented subject}
  \includegraphics[width=0.8\textwidth]{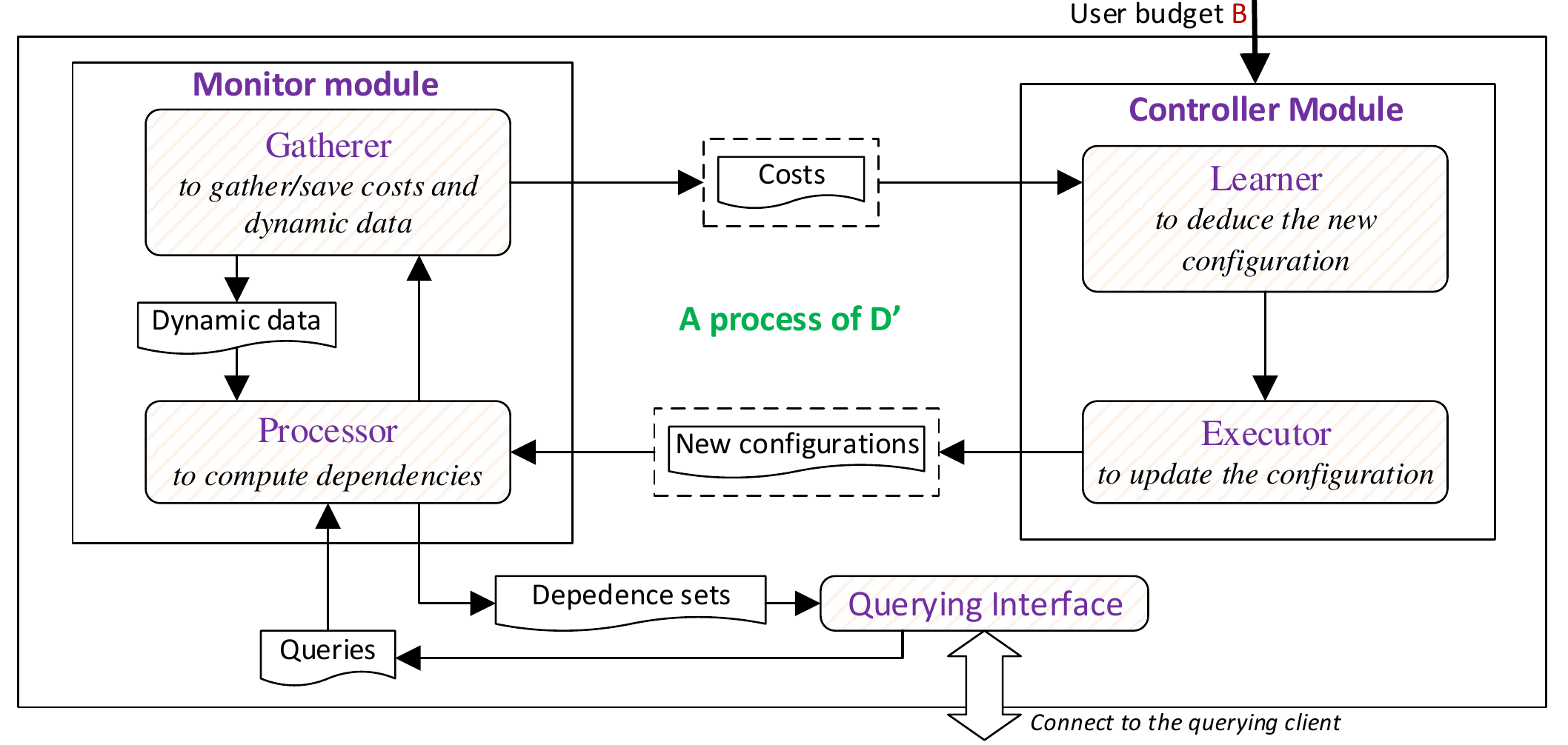}
  \label{fig:seadsinsystem}
\end{figure}

After the instrumentation, with instrumented system $D'$ launched and continuously running in its $N$ distributed processes, the monitor and controller in each process of $D'$ also start and continuously run along with the process, as shown in Figure~\ref{fig:seadsinsystem}.
During the execution of the process, as the core component of {\seads},
the monitor decides {\em when} the analysis configuration needs to be adjusted (i.e., {\em arbitration}) and computes dynamic dependencies with a current configuration (i.e., {\em dependence computation}).
The module consists of two sub-modules: a gatherer collecting dynamic data (method execution events and/or statement coverage) and a processor receiving the data for the dependence computation.
More specifically, the processor computes/updates the dynamic dependencies for all possible queries (i.e., methods exercised) when
(i) the time gap since the previous dependence computation exceeds a threshold (e.g., 15 minutes) and
(ii) the number of method-execution events accumulated since the previous dependence computation exceeds another threshold (e.g., 1000).
As part of {\seads} settings, both thresholds are customized by users.

\vspace{2pt}
\noindent
\textbf{Arbitration.}
\begin{algorithm}[htbp]
\begin{flushleft}
\small{
    \caption{Arbitration}
    \label{algo:tc}
    \begin{algorithmic}[1]
        \State {Set \emph{$gCounter$} = 0, {$LastT$} = 0, {$QU$} = $\emptyset$, $TCN$ =111111, $oldTCN$ =None}
        \State {Assign $sgc\_T$, $sgl\_T$, and $d\_T$ from $B$}
        \While {true}
            \If {event($method$)==entry}
                \State $gCounter$++
                \State Add (-$method$) to $QU$
            \EndIf
            \If {event($method$)==returnInto}
                \State $gCounter$++
                \State Add $method$ to $QU$
                \If {$gCounter$ $>$ $TC$ $and$ ($CurrentTime$ - $LastT$) $>$ $TT$} 
                    \State Read current configuration parameters
                    \State { $isTimeOut$ = false }
                    \If {$Configuration\_staticGraph$} 
                        \If {$TCN$/$oldTCN$'s static configuration parameters are different}
                            \State Construct a new static (dependence) graph
                            \If { (Not $isTimeOut$) $and$ ($constructTime$ $>$ $sgc\_T$) }
                                \State Cancel the static graph construction, and set $isTimeOut$ = true
                            \EndIf
                        \EndIf
                        \If {the static graph $exists$ }
                            \State Load the static graph
                            \If { (not $isTimeOut$) $and$ ($loadTime$ $>$ $sgl\_T$) }
                                \State Cancel the static graph loading, and set $isTimeOut$ = true
                            \EndIf
                        \EndIf
                    \EndIf
                    \If{not $isTimeOut$}
                        \State Call the processor to compute dependencies with $TCN$
    					\If { (not $isTimeOut$) $and$ ($ComputeTime$ $>$ $d\_T$)}
    							\State Cancel the dependence computation, and set $isTimeOut$ = true
    					\EndIf
                    \EndIf
                    \State \emph{$gCounter$} = 0, $LastT$ = $CurrentTime$
                    \State Call the gatherer to record the time cost of the analysis
					\State Call the controller to compute new configuration $newTCN$
                    \State $oldTCN$ = $TCN$
                    \State $TCN$ = $newTCN$
                \EndIf
            \EndIf
        \EndWhile
    \end{algorithmic}
}
\end{flushleft}
\end{algorithm}
Algorithm~\ref{algo:tc} shows the arbitration pseudo-code which decides when and how to trigger dependence computations and configuration adjustments.
In this algorithm, I set some variables as inputs, letting $method$ be the executed method, $gCounter$ be the number counter of method events, $LastT$ be the time of the last computation,
$B$ be the user budget, $QU$ be the method event queue, $isTimeOut$ be the boolean value to record timeouts, $TC$/$TT$ be the thresholds of event number and analysis time interval, $sgc\_T$/$sgl\_T$/$d\_T$ be timeouts of constructing/loading the static graph and computing dependencies, $TCN$/$oldTCN$ be current and immediately previous configurations, respectively.

{\seads} first initiates $gCounter$, $QU$, and $LastT$, and sets the current configuration $TCN$ as 111111 for the most precise but possibly the slowest analysis to start with (line 1).
The variable $oldTCN$ is initialized as {\tt None} and will be used to keep the previous configuration.
The variables $sgc\_T$, $sgl\_T$, and $d\_T$ are timeouts for constructing/loading the static dependence graph, and computing dependencies, respectively, 
and their values are empirically allocated from the total given user budget $B$ (line 2).
For example, I can set and allocate the user budget as 30 seconds $sgc\_T$, $sgl\_T$, and $d\_T$ as 21s, 6s, and 3s, respectively.

Then, there is an infinite loop, arbitrating dependence computations and configuration adjustments (lines 3--40),
via invoking (collaborating with) the controller module for the same process as is this monitor.
During the execution, in each $method$ {\em entry} event,
{\seads} increments $gCounter$ by one and adds minus $method$ (id) to $QU$ (lines 4--6).
For example, $gCounter$ is 1, and minus $method$ (id) (e.g., -571) is added to $QU$.
In each {\em returned-into} event,
{\seads} also increments $gCounter$ by one but adds $method$ (id) to $QU$ (lines 8--10).
For example, $gCounter$ is 2, and $method$ (id) (e.g., -571) is added to $QU$.

If $gCounter$ is greater than $TC$ and the time span (between the current time and the last computation time $LastT$) is greater than $TT$ (i.e., both conditions (i) and (ii) are satisfied), {\seads} will start a new round of analysis (e.g., updating dynamic dependencies for all possible queries) as detailed below.
To start with, {\seads} reads the current configuration and set $isTimeOut$ as false (lines 11--13).
For example, after 1000 events occurred and 3 minutes passed, $gCounter$ $>$ $TC$ (e.g., 1000) and (the current time - $LastT$) $>$ $TT$ (e.g., 3 minutes).

Then, {\seads} reads the current configuration TCN = 111111, and $isTimeOut$ = false.
If the {\em staticGraph} parameter is enabled and at least one of the static analysis parameters (i.e., the first to third bits of the configuration) varies between the current and immediately previous configurations,
{\seads} constructs a new static dependence graph (lines 14--16) using the static configuration.
For example, with TCN = 111111 now, the first to third bits (three static parameters) are all 1 (enabled).
Thus, {\seads} constructs the new static dependence graph with both context sensitivity and flow sensitivity applied.
When the static dependence graph is ready, {\seads} loads it (lines 21--22).
The static analysis reuses the relevant algorithms of {\diver}~\cite{cai14diver} and {\diveronline}~\cite{cai2017hybrid}. 

Moreover, the processor is invoked to compute dependencies with the current configuration $TCN$ (line 29), as detailed in Algorithm~\ref{algo:cd}.
When $isTimeOut$ is false, if any part of the static or dynamic analysis---(1) constructing and (2) loading the static dependence graph and (3) computing dynamic dependencies (costing {\em constructTime}, {\em loadTime}, and {\em computeTime} respectively), runs timeout, 
{\seads} would cancel the respective part of the analysis and set $isTimeOut$ as true (lines 17--18, 23--24, 28--31).
For example, with the current configuration TCN = 111111, {\seads} has not finished constructing the static graph in $sgc\_T$ time (e.g., 21 seconds) and thus cancel it
the construction of the static graph.

Then, $isTimeOut$ is true 
and hence {\seads} skips the static graph loading and dependence computation.
After resetting $gCounter$ and $LastT$,
{\seads} calls the gatherer to collect the time costs of above analyses
(i.e., constructing/loading the static graph and computing dependencies) under the current configuration and then calls the controller 
to compute the next configuration $newTCN$ (lines 34--36),
as detailed later in Algorithm~\ref{algo:ca}. 
For example, now $gCounter$ = 0, the time cost = 22 ($>$ 21) seconds, and $newTCN$ = 000101.

In the last place, the algorithm updates the current ($TCN$) and previous configuration ($oldTCN$) accordingly for the next arbitration iteration (lines 37--38).
For example, I obtained $oldTCN$ = 111111, and $TCN$ = 000101 now.

\vspace{2pt}
\noindent
\textbf{Dependence computation.}
When {\seads} calls the processor to compute dependencies,
the online analysis based on {\diveronline}~\cite{cai2017hybrid} is adopted, avoiding execution tracing to economize analysis costs,
such as storage and I/O costs.
Algorithm~\ref{algo:cd} gives the pseudo-code of the online algorithm to compute dependencies.
In Algorithm~\ref{algo:cd}, $QU$ is the same method event queue as in Algorithm~\ref{algo:tc},
and $DS(m)$ is the dependence set for the method $m$.
\begin{algorithm}[tp]
\begin{flushleft}
\small{
    \caption{Computing dependencies}
    {Let $Configuration\_staticGraph$ be $staticGraph$ parameter of current configuration }\\
    {Let $Configuration\_methodEvent$ be $methodEvent$ parameter of current configuration }\\
    {Let $Configuration\_statementCoverage$ be $statementCoverage$ parameter of current configuration }\\
    {Let $Configuration\_methodInstanceLevel$ be $methodInstanceLevel$ parameter of current configuration }\\
    {Let $DS(m)$ be dependence set for method $m$}\\
     \label{algo:cd}
    \begin{algorithmic}[1]
        \State Read current configuration parameter settings
        \If {$Not$ $Configuration\_methodInstanceLevel$}
            \State $QU$=getFirstLastInstances($QU$)
        \EndIf
        \If {$Configuration\_staticGraph$ and $Configuration\_statementCoverage$}
            \State Prune the static graph  with the statement coverage
        \EndIf
        \For{{\bf each} method event $e\in QU$}
            \State $m$=abs($e$), {{$DS(m)$} = $\emptyset$}
            \If {$Configuration\_methodEvent$}
                \If {$e$ $<$ 0}
                    \State $DS(m)$ $\cup=$ $\{m\}$
                \EndIf
                \If {$Configuration\_staticGraph$}
                    \If {$e$ $<$ 0}
                        \State AddDSEntry($m$)
                    \Else
                        \State AddDSReturnInto($m$)
                    \EndIf
                \Else
                    \If {$e$ $<$ 0}
                        \For{{\bf each} last returned-into event $e'$ that happens after $e\in Q$}
                            \State $m'$=abs($e'$)
                            \State $DS(m)$ $\cup=$ $\{m'\}$
                        \EndFor
                    \EndIf
                \EndIf
            \Else
                \If {$Configuration\_staticGraph$}
                    \State AddDSEntry($m$)
                    \State AddDSReturnInto($m$)
                \EndIf
            \EndIf
        \EndFor
    \end{algorithmic}
}
\end{flushleft}
\end{algorithm}

First, four configuration parameter variables (i.e., $Configuration\_staticGraph$, 

\noindent
$\_methodEvent$, $\_statementCoverage$, and $\_methodInstanceLevel$) are read from the current configuration (line 1).
For example, from the current configuration 000101, I have these variables assigned 0, 1, 0, 1, respectively.
If the parameter {\em methodInstanceLevel} is disabled, {\seads} filters the first {\em entry} and the last {\em return-into} events from the event sequence in $QU$ (lines 2--3).
For example, since $methodInstanceLevel$ is enabled, {\seads} skips the filtering.
If $staticGraph$ and $statementCoverage$ are both enabled,
the static dependence graph is pruned according to the statement coverage (lines 5--6).
For example, since both $staticGraph$ and $statementCoverage$ are disabled, {\seads} skips the pruning.
Then, {\seads} traverses $QU$ to compute dynamic dependencies corresponding to the method events in $QU$ (lines 8--34).

For each event $e$ in $QU$, I let $m$ be the corresponding method of $e$ and $DS(m)$ be the dependence set of $m$ which is empty initially (line 9).
For example, for $e$ = -571, $DS(e)$ is empty now, where 571 is the id of method {\em org.apache.zookeeper.server.ZooKeeperServerMain:} {\em void main(java.lang.String[])}.
If the parameter {\em methodEvent} is enabled and the value of $e$ is negative (i.e., $e$ is an {\em entry} event) and $m$ is executed,
{\seads} adds $m$ itself into the dependence set $DS(m)$ (lines 8--10).
For example, since the parameter {\em methodEvent} is enabled and $e$ = -571 ($<$ 0), 
the method (of id = 571) is added to $DS(m)$.

If both parameters {\em methodEvent} and {\em staticGraph} are enabled,
{\seads} adds dependencies via calling a function {\tt AddDSEntry} for the negative $e$ value ({\em entry} event)
or calling another function {\tt AddDSReturnInfo} for the positive $e$ value ({\em returned-into} event) (lines 14--18).
If the parameter {\em staticGraph} is not enabled, {\seads} skips both subroutines.
I leveraged {\diveronline}~\cite{cai2017hybrid} to develop these two subroutines, in which {\seads} traverses the static dependence graph to add dependencies into $DS(m)$, using different dependence propagation rules for the {\em entry} and {\em returned-into} event of the method ($m$), respectively.
If the parameter {\em methodEvent} is enabled yet the parameter {\em staticGraph} is disabled, upon each negative $e$ (i.e., an {\em entry} event),
{\seads} adds all methods whose last (returned-into) event in $QU$ happened after $e$, into the dependence set $DS(m)$ (lines 20--24).
For example, with enabled {\em methodEvent} and disabled {\em staticGraph}, {\seads} adds all methods whose last (returned-into) event in $QU$ happened after $e$ (-571) into the dependence set $DS(m)$.

If {\em methodEvent} is disabled yet {\em staticGraph} is enabled,
{\seads} simply calls these two functions {\tt AddDSEntry} \& {\tt AddDSReturnInfo} (lines 28--31) to add dependencies to $DS(m)$
by traversing the static dependence graph without utilizing any dynamic data.
If {\em methodEvent} is enabled and {\em staticGraph} is disabled,
{\seads} skips both functions here.

It is worth noting that for each process, the monitor only computes the run-time dependencies {\em within} the process (referred to as {\em intraprocess dependencies}).
Dependencies across processes (i.e., {\em interprocess dependencies}) will be computed for a given query $Q$ by the {\em querying\_client} after receiving the intraprocess dependencies of $Q$ from each process.

\vspace{2pt}
\noindent
\textbf{Querying interface.}
For interacting with the user, the monitor module for each process includes a {\em querying\_interface} that
receives the dependence query $Q$ from and sends corresponding dependence sets back to the {\em querying\_client} module, both
through the {\em network} facility (see Figure~\ref{fig:seadsoverview}).
When a query is received at (the {\em querying\_interface} of) the {\em monitor} module, there are two situations that should be handled differently:
(1) If the {\em monitor} is in the middle of computing/updating the dependence sets for all possible queries. In this situation, the {\em querying\_interface} will wait until the dependence computation/updating is completed to run $Q$'s dependence set.
And (2) if the {\em monitor} is performing arbitration functionalities, but not computing dependencies now after the previous round of dependence computation, and waiting for the next round. In this situation, the {\em querying\_interface} will immediately return the most recently computed dependence set of the query $Q$.

While the {\em querying\_interface}  (attached to the monitor) for each process computes intraprocess run-time dependencies,
the {\em querying\_client} module derives interprocess dependencies hence produce the final dependence set, while merging all the per-process intraprocess dependence sets, for the user-supplied query $Q$.
Once it has received $Q$, the {\em querying\_client} sends it to the {\em querying\_interface} for each process, and then
waits for all the per-process interfaces to return their respective intraprocess dependence sets for
all possible queries.
The reason is that all these dependence sets are needed for computing interprocess dependencies for the query $Q$.

More specifically, the {\em querying\_client} will identify the process $P_i$ where the query $Q$ was executed first (i.e., where the earliest first entry event of $Q$ occurred). 
\begin{enumerate}
\item 
If no process exercised $Q$, an empty dependence set would be returned.
\item Otherwise, the final dependence set of $Q$, noted as $fDS(Q)$, is initialized as the intraprocess dependence set of $Q$ returned from the {\em querying\_interface} for $P_i$ (noted as $intraDS(Q,i)$).
\end{enumerate} 

Then, for each other process $P_j$,if $P_j$ also exercised $Q$, $intraDS(Q,j)$ is straightforwardly merged into $fDS(Q)$;
otherwise,for each method $m$ exercised in $P_j$, $intraDS(m,j)$ is merged into $fDS(Q)$ if the last returned-into event happens after the first entry event of $Q$.
This merging process implicitly derives and adds to $fDS(Q)$ the interprocess dependencies for $Q$ according to the happens-before relationships
among method execution events across all processes of the system.

\subsection{Controller}
For each process of the system, {\seads} utilizes a controller to adjust its analysis configurations, as shown in Figure~\ref{fig:seadsinsystem}.
The controller takes the costs of the current configuration and user budget $B$ as inputs to determine which next configuration the analysis should use in order to achieve a better cost-effectiveness 
(than with the current configuration)
with respect to the user budget (i.e., the total analysis time constraint under the budget).

Each controller module consists of two sub-modules: a learner and an executor.
The learner uses the data from the gatherer (i.e., the analysis costs under the current configuration) while referring to the user budget, to adjust the configuration.
The resulting configuration may be the same as or different from the current configuration.
Then, the executor updates {\seads} to the new configuration from the learner's output and simply transfer the new configuration to the collaborating monitor.
Next, I elaborate the learner's inner workings for configuration adjustment.

{\seads} decides new analysis configurations using a reinforcement learning methodology, in particular the Q-learning method.
Since supervised learning needs a large training set, it is not suitable for the configuration adjustment in {\seads}.
There are not enough data for training when {\seads} starts with a particular system.
Meanwhile, since the dynamics of execution may vary widely across different systems, learning from other systems beforehand may not be effective either.
Thus, given the unpredictably changing {\em environment} during the execution of a system, reinforcement learning, which is not subject to those constraints, 
is more suitable.
Moreover, as a special type of reinforcement learning, Q-learning is particularly appropriate for configuration adjustments in {\seads},
because dependence computation time costs constantly vary during the execution without an existing policy or model of the adjustments~\cite{wawrzynski2004model}.
Therefore, as an off-policy and model-free learning strategy, Q-learning was employed.

\begin{figure*}[tp]
  \centering
  \caption{The interactions between the agent and environment of Q-learning}
  \includegraphics[width=0.5\textwidth]{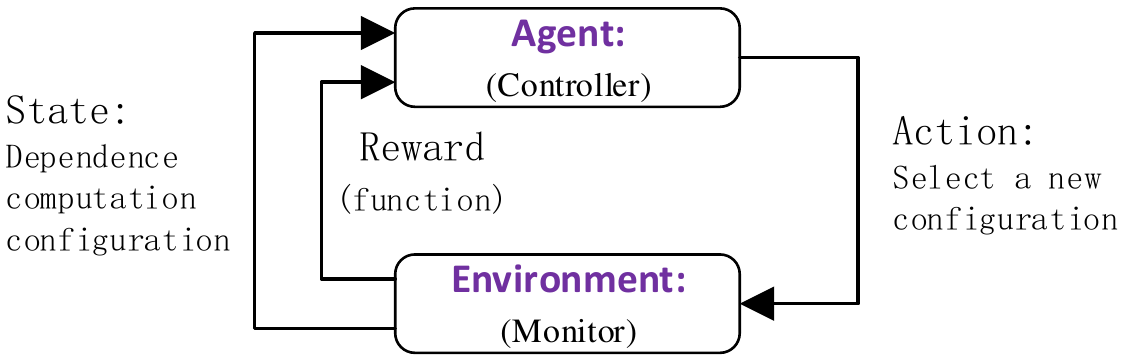}
  \label{fig:ql2}
\end{figure*}

In Q-learning as applied in {\seads}, an agent receives a state (the current analysis configuration) from the environment and takes an action (i.e., choosing a new configuration), either from a Q-table or by a random exploration of possible actions.
Subsequently, the agent receives feedback in terms of a reward computed according to the action's performance.
As shown in Figure~\ref{fig:ql2},
a state represents the current dependence computation configuration, 
and the {\em monitor} is the environment while the {\em controller} is the agent.
With the user budget and time cost of the dependence computation, 
a reward is computed and sent back to the agent as feedback.
For a positive reward, the corresponding action is encouraged/reinforced.
Otherwise, the action is discouraged.
Q-learning uses the reward to update the Q-table whose largest value will be presumably chosen as the future action~\cite{fuchida2010study}.
In other words, the larger the reward is, the more likely the corresponding configuration will be chosen.

Q-learning updates the Q-table according to the Bellman equation to find optimal policies~\cite{huang1992optimal}.
With this equation, 
the value in the Q-table (i.e., $Vq$) for the next state is calculated as follows with two parameters ($\gamma$ and $\alpha$):
\vspace{-5pt}
\begin{equation}
\begin{tabular}{ll}
$Vq = Vq + \alpha * \left[reward + \gamma * \max \{q|q\in Q\mhyphen table\} - Vq\right]$
 \end{tabular}
\end{equation}
\vspace{-5pt}
Here $\gamma$ is a discount factor between 0 to 1 related to the importance of future rewards.
If $\gamma$ is 0, the Q-learning agent only considers current rewards; if $\gamma$ is 1, the agent strives for a long-term high reward.
And $\alpha$ is the learning rate between 0 to 1 to control how much the difference between the previous and new $Vq$ values considered.
If $\alpha$ is 0, the agent only utilizes prior knowledge; if $\alpha$ is 1, the agent ignores prior knowledge and only considers the most recent information to explore possibilities.


Algorithm~\ref{algo:ca} shows the learning pseudo-code for configuration adjustment.
First, {\seads} initiates Q-learning components/variables (e.g., $learner$, $Agent$, $Qtable$, $actions$, $rewards$, $\epsilon$, and $states$) and Bellman equation's parameters: $\gamma$ and $\alpha$ (lines 1--2).
For example, I set the values in the Q-table as all zeros and the parameters $\gamma$, $\alpha$, and $\epsilon$ as 0.9, 0.9, and 0.2, respectively.
Because $\gamma$=0.9 is slightly lower than 1, the Q-learning agent prefers for a long-term high reward rather than the current reward.
Due to $\alpha$ (0.9), the agent prefers for the most recent data (e.g., reward, values in the Q-table).

$\epsilon$ is used to control the agent taking an action (i.e., selecting a new configuration) from the Q-table or by a random exploration of possible actions. If a randomly calculated variable value $<=$ (1 - $\epsilon$), the agent uses the epsilon greedy strategy~\cite{wunder2010classes} to take the best action according to the largest $Vq$ in the Q-table. 
Otherwise, the agent randomly selects an action~\cite{even2003learning}.  (lines 8--9).
Now $\epsilon$ = 0.2 and (1 - $\epsilon$) =  0.8 = 80\%.
Thus, the possibility that the agent takes the best action according to the largest value in the Q-table is 80\%, 
while the possibility of a random selection is 20\%.

\begin{algorithm}[t]
\begin{flushleft}
    \caption{Configuration Adjustment using Q-learning}
\scriptsize{
Let $TCN$ and $newTCN$ be current and new configurations \\
Let $Qtable$, $actions$, $rewards$, $states$, $\gamma$, $\alpha$, and $\epsilon$ be components and parameters used by Q-learning \\
Let $T$ be overall dynamic dependence analysis time cost with the current configuration \\
Let $B$ be the user budget \\
Let $probability$ be the possibility to select the action according to the largest value in $Qtable$ \\
}
    \label{algo:ca}
    \begin{algorithmic}[1]
        \State Initiate Q-learning components: $learner$, $Agent$, $Qtable$, $actions$, $rewards$, and $states$
        \State Set Bellman equation parameters $\gamma$, $\alpha$, and $\epsilon$ between 0 and 1
        \State Update $reward$ = 1/($B$ - $T$) * 1000.
        \State Update $Qtable$ using the Bellman equation
    	\State $probability$=random(0,1)
    	\If {$probability$ $<$ = (1 - $\epsilon$)}
    			\State Take the best action according to the largest value in $Qtable$
    	\Else
    			\State Randomly take an action
    	\EndIf
    	\State Return a new configuration $newTCN$ 
        \end{algorithmic}
\end{flushleft}
\end{algorithm}

Next, the $reward$ and $Qtable$ are updated (lines 3--4).
The $reward$ is defined as 1/(the user budget $B$ - the current dependence analysis time cost $T$) * 1000---the rationale is that the closer the analysis cost is to the user budget (the learning goal here), the higher the reward.
For example, I suppose the user budget $B$ is 60000 (ms) and the current analysis time cost $T$ is 40000 ms.
Then the $reward$ is 1/(60000 - 40000) * 1000 = 0.05 that is used to update the Q-table.
In the algorithm, the action means the transfer from the current state (configuration) to the next state (configuration).
If the randomly calculated $probability$ value equals or is less than (1 - $\epsilon$),
the Q-learning $Agent$ uses the epsilon greedy strategy~\cite{auslander2008recognizing} to take the best action according to the largest value in the $Qtable$ (lines 5--7).
For example, suppose
{\seads} randomly calculated the $probability$ value as 0.803 ($>$ 0.8).

Then, {\seads} skips taking the best action according to the largest value in $Qtable$.
Otherwise, the $Agent$ randomly selects an action (lines 8--9).
For example,
The $Agent$ may randomly select an action (selecting the next configuration (e.g., 000101).
Eventually, a new configuration $newTCN$ is returned (line 11).

\subsection{Querying client}\label{sec:ui}
{\seads} uses {\em querying\_client} to interact with the user(s).
Through this module, 
the user sends a dependency query $Q$ to the querying interface within each process of the instrumented system $D'$
and waits for their responses.
When the intraprocess dependence computation in a process has finished, resulting dependencies
are delivered back to the querying interface in the process,
from which the {\em querying\_client} receives these intraprocess dependencies computed.

Once the intraprocess analyses in all processes are completed, the {\em querying\_client} performs an interprocess analysis,
which splices all the intraprocess dependence sets received according to the timestamps associated with the partially ordered method events while leveraging
basic message-passing semantics to reduce false positives~\cite{fu2020seads}.

\subsection{Limitations}
During the instrumentation, {\seads} needs to insert probes into the bytecode of the system under analysis (SUA) to monitor method events.
If the administrator does not allow to modify the SUA, {\seads} cannot work.


{\seads} attempts to provide the most cost-effective result (i.e., dependence set) achieved within a response time constraint (i.e., the user budget).
However, the current controller (Q-learning algorithm) might not be optimal.
The configuration with which the dynamic dependence computation does not necessarily have the optimal cost-effectiveness tradeoff possible with respect to {\seads}).
For example, the Q-learning algorithm might take a wrong action at certain steps (especially when the selection is random, as shown in Algorithm~\ref{algo:ca}).
As a result, the dynamic dependence analysis may not always be the most cost-effective.

In addition,
if the user sets an improper budget (e.g., one that is far off the typical response time for a particular system), the analysis configuration adjustment by the controller can be even less effective (i.e., leading the analysis in {\seads} to be further away from optimal cost-effectiveness balances).
On the other hand, when the user does not specify a budget,
{\seads} would have to use a default budget, which may not be desirable to the user or not suitable for the given system.

\begin{figure*}[htbp]
  \centering
  \caption{An overview of {\dads} architecture}
  \includegraphics[width=1\textwidth]{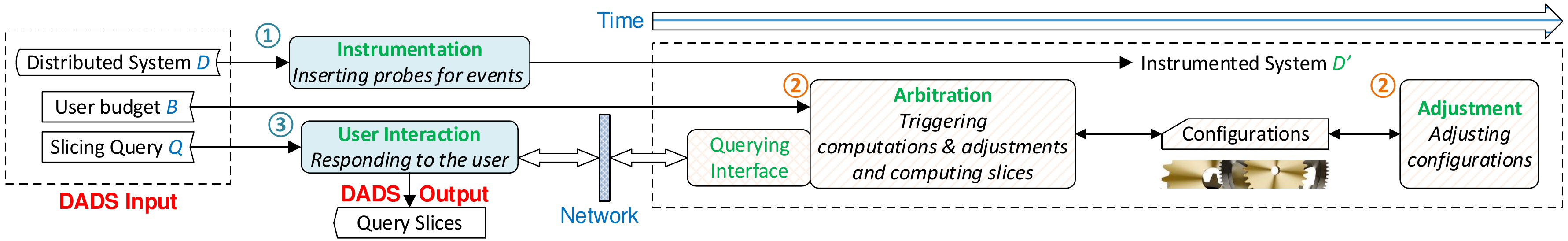}
  \label{fig:dadsoverview}
\end{figure*}
\section{Tool Implementation}

To avoid tracing and hence relevant time/space costs when analyzing continuously-running distributed software, I developed {\dads}~\cite{fu2020dads,fu2019towards}, an {\em online}, {\em scalable}, and {\em cost-effective} dynamic slicer with respect to user-specified budget constraints, as shown in Figure~\ref{fig:dadsoverview}.
{\dads} is distributed by design to utilize parallel and distributed computing/storage resources.
Furthermore, {\dads} continually and automatically adjusts its analysis configurations on the fly via reinforcement learning,
to achieve and maintain practical scalability and cost-effectiveness tradeoffs according to given budgets.
Adapted against eight Java distributed systems in the real world, 
{\dads} demonstrated its scalability and cost-effectiveness advantages over a similar slicer without the capabilities of configuration learning and adjustment. Through the dynamic slices within the given budget, {\dads} can benefit software maintenance, testing, and security tasks with respect to time budget constraints.

\section{Evaluation}         \label{sec:seadseval}

\subsection{Experiment setup} \label{ss:seadses}
I applied {\seads} against eight Java distributed systems that typically run continuously, as shown in Table~\ref{tab:seadssubjects}.
The sizes of these subjects are measured as the number of methods defined in the subject source code (\emph{\#Method}) and the number of Java source code lines excluding blank lines and code comments (\emph{\#SLOC}).
The last column (\emph{Test Type}) shows the test types, including integration, load, and system tests.
I selected these subjects covering different architectures, domains, and scopes.

\setlength{\tabcolsep}{0.5pt}
\begin{table}[tp]
  \centering
  \caption{Experimental subjects}
    \begin{tabular}{|l|r|r|c|}
    \hline
    \multicolumn{1}{|c|}{\textbf{Subject (Version)}} & \multicolumn{1}{c|}{\textbf{\#Method}} & \multicolumn{1}{c|}{\textbf{\#SLOC}} & \multicolumn{1}{c|}{\textbf{Test Type}} \\
    \hline
    NioEcho (r69) & 27    & 412   & Integration \\
    \hline
     MultiChat (r5) & 37    & 470   & Integration \\
    \hline
    OpenChord (v1.0.5) & 736   & 9,244 & Integration \\
    \hline
     Thrift (v0.11.0) & 1,941 & 14,510 & Integration \\
    \hline
    xSocket (v2.8.15) & 2,209 & 15,760 & Integration \\
    \hline
     ZooKeeper (v3.4.11) & 5,383 & 62,194 & Integration, Load, System \\
    \hline
    Netty (v4.1.19) & 12,389 & 167,961 & Integration \\
    \hline
     Voldemort (v1.9.6) & 20,406 & 115,310 & Integration, Load, System \\
    \hline

    \end{tabular}%
  \label{tab:seadssubjects}%
\end{table}%

In order to show the impact and merits of this innovation, due to the absence of a directly comparable peer technique,
I created and used an online version (referred to as {\doda}) of {\textsc{D$^2$Abs}}~\cite{cai2021d2abs}, the state-of-the-art dynamic dependence analysis for distributed programs to the best of the knowledge, as the baseline in my evaluation.
In terms of implementation, {\doda} is essentially a version of {\seads} that does not change its analysis configurations on the fly but constantly uses a fixated configuration (111111, for the possibly highest precision but expensive analysis).

\subsection{Results and analysis}
In this section, I report and discuss the empirical results.
Table~\ref{tab:seadscosts} shows the major results,
including the cost and precision measures of {\seads} versus the baseline {\doda} against the 12 subject executions (in the first column).
The third to fifth and sixth to eighth columns list the total run time ({\bf Run Time}),
run-time slowdowns ({\bf SlowDown}), and average query response time ({\bf Response Time}), for each instrumented subject execution with {\doda} and {\seads}, respectively.
The normal run time (the second column) for each subject execution was the total execution time of the subject against the same sequence of run-time inputs driving the corresponding instrumented subject execution.
For each subject execution, the run-time inputs to the subject were exactly the same during the analysis by both techniques, in addition to feeding them with the same sequence of queries.


%
%
%

\setlength{\tabcolsep}{1pt}
\renewcommand{\arraystretch}{1}
\begin{table*}[tp]
  \centering
  \caption{Time (in seconds) \& storage costs (in MB) \& precision (ratios) of {\seads} versus {\doda} as the baseline}
\scriptsize
    \begin{tabular}{|l|r||r|r|r|r|r|r||r|r|}
    \hline
    \multicolumn{1}{|c|}{\multirow{3}[4]{*}{Execution}} & \multicolumn{1}{c|}{Normal} & \multicolumn{3}{c|}{\doda} & \multicolumn{3}{c|}{\seads} & \multicolumn{1}{c|}{\multirow{3}[4]{*}{Precision}} & \multicolumn{1}{c|}{\multirow{3}[4]{*}{Storage}} \\
\cline{3-8}      & \multicolumn{1}{c|}{Run} & \multicolumn{1}{c|}{Run} & \multicolumn{1}{c|}{Slow} & \multicolumn{1}{c|}{Response} & \multicolumn{1}{c|}{Run} & \multicolumn{1}{c|}{Slow} & \multicolumn{1}{c|}{Response} &       &  \\
          & \multicolumn{1}{c|}{Time} & \multicolumn{1}{c|}{Time} & \multicolumn{1}{c|}{Down} & \multicolumn{1}{c|}{Time} & \multicolumn{1}{c|}{Time} & \multicolumn{1}{c|}{Down} & \multicolumn{1}{c|}{Time} &       &  \\
    \hline
    NioEcho & 158.37 & 228.17 & \textcolor[rgb]{ .267,  .447,  .769}{44.07\%} & \textcolor[rgb]{ 1,  0,  0}{14.39} & 214.15 & \textcolor[rgb]{ .267,  .447,  .769}{35.22\%} & \textcolor[rgb]{ 1,  0,  0}{13.71} & 100.00\% & 2.00 \\
    \hline
    MultiChat & 148.96 & 241.89 & \textcolor[rgb]{ .267,  .447,  .769}{62.39\%} & \textcolor[rgb]{ 1,  0,  0}{15.12} & 223.67 & \textcolor[rgb]{ .267,  .447,  .769}{50.15\%} & \textcolor[rgb]{ 1,  0,  0}{14.32} & 100.00\% & 2.00 \\
    \hline
    Openchord & 233.78 & 606.87 & \textcolor[rgb]{ .267,  .447,  .769}{159.59\%} & \textcolor[rgb]{ 1,  0,  0}{51.36} & 359.37 & \textcolor[rgb]{ .267,  .447,  .769}{53.72\%} & \textcolor[rgb]{ 1,  0,  0}{25.33} & 77.44\% & 14.00 \\
    \hline
    Thrift & 199.87 & 573.49 & \textcolor[rgb]{ .267,  .447,  .769}{186.93\%} & \textcolor[rgb]{ 1,  0,  0}{45.23} & 345.19 & \textcolor[rgb]{ .267,  .447,  .769}{72.71\%} & \textcolor[rgb]{ 1,  0,  0}{23.82} & 90.68\% & 25.00 \\
    \hline
    xSocket & 380.59 & 1,817.61 & \textcolor[rgb]{ .267,  .447,  .769}{377.58\%} & \textcolor[rgb]{ 1,  0,  0}{170.82} & 772.38 & \textcolor[rgb]{ .267,  .447,  .769}{102.94\%} & \textcolor[rgb]{ 1,  0,  0}{65.37} & 83.25\% & 21.00 \\
    \hline
    Netty & 589.39 & 4,218.85 & \textcolor[rgb]{ .267,  .447,  .769}{615.80\%} & \textcolor[rgb]{ 1,  0,  0}{409.56} & 1,226.16 & \textcolor[rgb]{ .267,  .447,  .769}{108.04\%} & \textcolor[rgb]{ 1,  0,  0}{115.16} & 87.12\% & 105.00 \\
    \hline
    Zookeeper &       &       & \textcolor[rgb]{ .267,  .447,  .769}{} & \textcolor[rgb]{ 1,  0,  0}{} &       & \textcolor[rgb]{ .267,  .447,  .769}{} & \textcolor[rgb]{ 1,  0,  0}{} &       &  \\
    Integration & 598.34 & 3,543.47 & \textcolor[rgb]{ .267,  .447,  .769}{492.22\%} & \textcolor[rgb]{ 1,  0,  0}{343.19} & 1,139.25 & \textcolor[rgb]{ .267,  .447,  .769}{90.40\%} & \textcolor[rgb]{ 1,  0,  0}{103.21} & 66.29\% & 96.00 \\
    \hline
    Zookeeper &       &       & \textcolor[rgb]{ .267,  .447,  .769}{} & \textcolor[rgb]{ 1,  0,  0}{} &       & \textcolor[rgb]{ .267,  .447,  .769}{} & \textcolor[rgb]{ 1,  0,  0}{} &       &  \\
    Load  & 616.16 & 3,804.37 & \textcolor[rgb]{ .267,  .447,  .769}{517.43\%} & \textcolor[rgb]{ 1,  0,  0}{368.43} & 1,209.77 & \textcolor[rgb]{ .267,  .447,  .769}{96.34\%} & \textcolor[rgb]{ 1,  0,  0}{111.97} & 65.55\% & 96.00 \\
    \hline
    Zookeeper &       &       & \textcolor[rgb]{ .267,  .447,  .769}{} & \textcolor[rgb]{ 1,  0,  0}{} &       & \textcolor[rgb]{ .267,  .447,  .769}{} & \textcolor[rgb]{ 1,  0,  0}{} &       &  \\
    System & 598.17 & 3,676.91 & \textcolor[rgb]{ .267,  .447,  .769}{514.69\%} & \textcolor[rgb]{ 1,  0,  0}{355.56} & 1,183.97 & \textcolor[rgb]{ .267,  .447,  .769}{97.93\%} & \textcolor[rgb]{ 1,  0,  0}{107.94} & 63.28\% & 96.00 \\
    \hline
    Voldemort &       &  & \textcolor[rgb]{ .267,  .447,  .769}{} & \textcolor[rgb]{ 1,  0,  0}{} &       & \textcolor[rgb]{ .267,  .447,  .769}{} & \textcolor[rgb]{ 1,  0,  0}{} &       &  \\
    Integration & 398.06 & \multicolumn{1}{c|}{-} & \multicolumn{1}{c|}{\textcolor[rgb]{ .267,  .447,  .769}{-}} & \multicolumn{1}{c|}{\textcolor[rgb]{ 1,  0,  0}{-}} & 791.37 & \textcolor[rgb]{ .267,  .447,  .769}{98.81\%} & \textcolor[rgb]{ 1,  0,  0}{69.62} & \multicolumn{1}{c|}{-} & 200.00 \\
    \hline
    Voldemort &       &  & \textcolor[rgb]{ .267,  .447,  .769}{} & \textcolor[rgb]{ 1,  0,  0}{} &       & \textcolor[rgb]{ .267,  .447,  .769}{} & \textcolor[rgb]{ 1,  0,  0}{} &       &  \\
    Load  & 194.44 & \multicolumn{1}{c|}{-} & \multicolumn{1}{c|}{\textcolor[rgb]{ .267,  .447,  .769}{-}} & \multicolumn{1}{c|}{\textcolor[rgb]{ 1,  0,  0}{-}} & 731.71 & \textcolor[rgb]{ .267,  .447,  .769}{276.32\%} & \textcolor[rgb]{ 1,  0,  0}{67.58} & \multicolumn{1}{c|}{-} & 200.00 \\
    \hline
    Voldemort &       &  & \textcolor[rgb]{ .267,  .447,  .769}{} & \textcolor[rgb]{ 1,  0,  0}{} &       & \textcolor[rgb]{ .267,  .447,  .769}{} & \textcolor[rgb]{ 1,  0,  0}{} &       &  \\
    System & 355.78 & \multicolumn{1}{c|}{-} & \multicolumn{1}{c|}{\textcolor[rgb]{ .267,  .447,  .769}{-}} & \multicolumn{1}{c|}{\textcolor[rgb]{ 1,  0,  0}{-}} & 719.38 & \textcolor[rgb]{ .267,  .447,  .769}{102.20\%} & \textcolor[rgb]{ 1,  0,  0}{66.96} & \multicolumn{1}{c|}{-} & 200.00 \\
    \hline
    Average: & 372.66 & 2,079.07 & \textcolor[rgb]{ .267,  .447,  .769}{330\%} & \textcolor[rgb]{ 1,  0,  0}{197.07} & 743.03 & \textcolor[rgb]{ .267,  .447,  .769}{99\%} & \textcolor[rgb]{ 1,  0,  0}{65.41} & 81.5\% & 88.08 \\
    \hline
    \end{tabular}%
   \label{tab:seadscosts}%
\end{table*}%

In the ninth column ({\em Precision}), I report the average relative precision of {\seads} versus the baseline: for each query, the measure was computed as the ratio of the size of query dependence set computed by {\doda} to the size of the query dependence set computed by {\seads}.
I computed and reported the relative measure for two reasons:
(1) There are no ground-truth dependence sets available for the sample queries, and no existing tools which scale to and work with the subject systems to compute such ground truths;
and (2) the evaluation goal is mainly to validate {\seads}' scalability and cost-effectiveness merits over conventional analysis which has a configuration.

The last column ({\em Storage}) lists the total storage costs (disk space taken) of {\seads}, ranging from 2MB for the two smallest subjects (NioEcho and MultiChat) to 200MB on the largest system (Voldemort), and including an average cost of 88MB across all the 12 executions.
These spaces were mainly used to store the static dependence graph for each subject component and the instrumented subject versions.
%
Baseline storage costs were almost the same as {\seads} ones and hence were omitted.

Results for {\doda} against the Voldemort executions are unavailable (hence missing from the table) because the baseline did not scale to the system: I killed the analysis after running it for 12 hours. For this reason, the relative precision of {\seads} for these executions is missed also.
Next, I discuss major findings and observations from these results, so as to answer the research questions.

\vspace{2pt}
\noindent
\textbf{Efficiency: response time.}
The response time, as shown in the fifth and eighth columns,
is the user's waiting time since a dependence query is sent to each dependence analysis technique until the user receives the dependence set in return.
{\seads} took 65 seconds on average to respond to random user queries with random intervals and user-specified budgets, over all subject executions.
For individual executions, {\seads} took the shortest response time (14 seconds) on average against the NioEcho execution, due to its smallest size.
On the other hand, {\seads} took the longest average response time (115 seconds) on Netty, the largest system among the subjects.
However, looking at this efficiency measure across all the subject executions reveals no consistent correlation between subject sizes and the average response time.
One reason is that the source size of a subject is not the only factor that affects this efficiency measure.
For example, the complexity of the execution analyzed is another major factor here.
Other factors (e.g., the time cost of network communication) could also affect the response time experienced by the user.
In fact, there is noticeably different average response time with {\seads} in the three subject executions for the same subject, Voldemort.

{\doda} took 197 seconds on average over the 12 subject executions for the same query requests as sent to {\seads}, with the average response time for individual subject executions ranging from 14 seconds on NioEcho to 410 seconds on Netty.
In particular, for Voldemort, {\doda} could not answer any query within 12 hours.
In practice, most users (either a human or an application using the dynamic dependence results) do not want to wait even longer than 12 hours for querying a dependence set.
In other words, {\doda} might suffer a serious scalability issue hampering its practical adoption to large distributed systems in the real world.
Conversely, for the three executions of Voldemort considered, {\seads} took short time to respond, demonstrating
its scalability and efficiency advantages over the conventional dependence analysis.
%

The efficiency advantage in terms of mean response time of {\seads} over {\doda} was generally more significant with larger-scale subject executions. As shown,
the gap in this efficiency measure between the two techniques tended to increase when the subject grows in source size and execution complexity.
In particular, for the two smallest subjects, the average response time of {\doda} was very close to that of {\seads} (the difference was less than one second on average for each query); for a medium-scale subject such as Thrift, {\seads} was about 2x faster; and for the largest subjects, {\seads} was over 3x faster.
This further means that the scalability and efficiency advantages of {\seads} over {\doda} are especially significant for large distributed systems.

\vspace{2pt}
\noindent
\textbf{Efficiency: analysis overheads.}
Table~\ref{tab:seadscosts} shows that the run-time slowdowns of {\seads} ranged from 35\% (NioEcho) to 276\% (Voldemort load test), with 99\% on average over the 12 subject executions.
Meanwhile, the run-time overheads of {\doda} ranged from 44\% (NioEcho) to 616\% (Netty), with 330\% on average.
Generally, the slowdown of either technique was greater against larger subjects with more complex executions, as expected.

Since {\doda} did not answer any dependence query for each of three Voldemort executions within 12 hours, the corresponding slowdown measures were not meaningful and hence omitted ($>$ (12 * 3600 - 398.06 / 398.06) = 4,319,900\%)
For the other 9 subject executions, {\seads} was consistently more efficient than the baseline.

Similar to their contrasts in average response time, the advantage of {\seads} over {\doda} was increasingly significant for subjects growing size and execution complexity.
For instance, for the two smallest and simplest subjects, the slowdowns of both techniques were close; for a medium-scale subject xSocket, {\doda} incurred over 2x greater slowdown than {\seads}; for two large-scale subjects (Netty and ZooKeeper), {\doda}'s slowdown was 4x to 5x larger.
Overall, {\seads} was more than 3x as efficient as the conventional approach to dynamic dependence analysis in terms of the slowdown measure, further describing the advantages of the on-the-fly analysis configuration adjustments through reinforcement learning in {\seads}.

\vspace{2pt}
\noindent
\textbf{Cost-effectiveness: precision-cost ratios.}
To evaluate the cost-effectiveness of {\seads}, 
I need to first compute the precision by comparing the average sizes of query dependence sets computed by {\doda}
and {\seads}.
The precision for each subject execution, as shown in the ninth column ({\em Precision}) of Table~\ref{tab:seadscosts}, is the average ratio of the size of the dependence set for each query computed by {\doda} to the size of that computed by {\seads} for the same query.
For each dependence query, I also compared the content of both dependence sets, and found that {\em {\seads}'s dependence set always subsumed the dependence set given by {\doda}} for any of the queries involved in our evaluation.
This confirms that although sacrificing precision by adjusting analysis configuration to achieve higher scalability and efficiency, {\seads} had no loss in recall.
These relative effectiveness measures essentially treated the baseline results as ground truths.
Thus, given the equally 100\% recall of both techniques, I only considered the relative precision of {\seads} (with {\doda} precision as constantly 100\%) when computing the cost-effectiveness of both techniques as the ratio of cost-effectiveness.

For the 9 subject executions for which the baseline dependence sets were available to enable the relative precision measurement for {\seads}, the precision achieved by {\seads} ranged from 63\% to 100\%, for an overall average of 82\%.
In the best cases, for the two smallest subject and simplest subject executions (i.e., NioEcho and MultiChat), {\seads} did not lose any precision relative to the baseline. The reason was mainly because the online analysis by {\seads} constantly incurred time costs lower than the user budgets even with the most precise analysis configuration for these subjects. 
Thus, {\seads} did not need to leave the highest-precision configuration.

Likewise, {\seads} had the lowest precision of 63.28\% for Zookeeper system test, most plausibly because {\seads} experienced the most aggressive and frequent adjustments of its analysis configurations in order to maintain scalability and efficiency with respect to given user budgets.
The average relative precision (over the ten queries) achieved by {\seads} for a subject execution had to do with the size and complexity of the subject execution was reflected in part in the two efficiency metrics: response time and run-time slowdown.

\begin{figure*} [tp]
  \centering
  \caption{Comparisons ($y$ axis) of the cost-effectiveness expressed as the ratios of the precision to the response time of {\doda} and {\seads} per execution ($x$ axis). The higher the ratio, the more cost-effective}
    \includegraphics[width=1\textwidth]{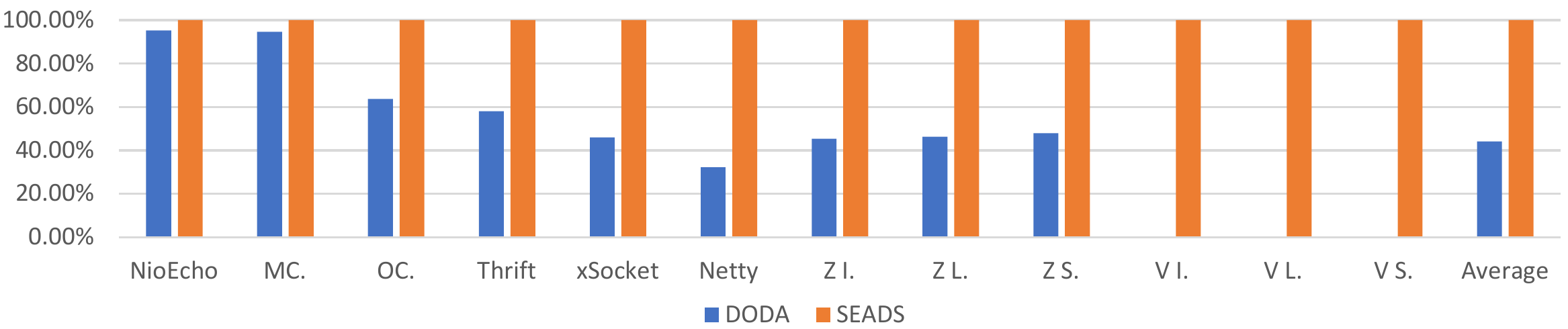}
  \label{fig:ce1}
\end{figure*}

\begin{figure*} [tp]
  \centering
 \caption{Comparisons ($y$ axis) of the cost-effectiveness expressed as the ratios of the precision to the run-time slowdown of {\doda} and {\seads} per execution ($x$ axis). The higher the ratio, the more cost-effective}	
    \includegraphics[width=1\textwidth]{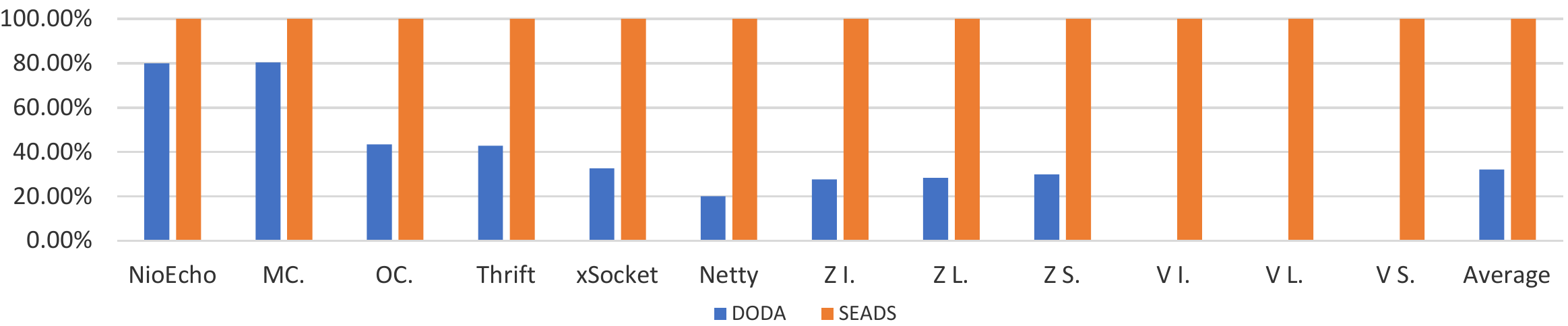}
   
  \label{fig:ce2}
\end{figure*}

To compute the cost-effectiveness of {\seads} and {\doda}, for each subject execution, I calculated the ratio of the average precision (over the ten queries) to one of the two cost measures I considered: average response time (over the ten queries), and run-time slowdown (overall during the entire execution across the ten queries).
Accordingly, I had two measures, each with respect to one of the two cost measures, in our cost-effectiveness assessment and comparison between the two techniques, shown in Figure~\ref{fig:ce1} (with average response time as the cost factor) and Figure~\ref{fig:ce2} (with the run-time slowdown as the cost factor), respectively.

For ease of presentation with respect to space constraints, I used abbreviations in both figures as follows (on the $x$ axis): MC. for MultiChat, OC. for OpenChord, V for Voldemort, Z for ZooKeeper, I. for integration test, L. for load test, and S. for system test.
In particular, to highlight the advantages of {\seads} over {\doda}, either figure only shows comparisons in terms of the cost-effectiveness per subject execution, as the percentage of {\doda}'s cost-effectiveness over the cost-effectiveness of {\seads}.
Thus, both figures show the bars for {\seads} constantly corresponding to 100\%, whereas the cost-effectiveness of {\doda} is shown as the fractions of that of {\seads}.
Since {\doda} could not be applied to Voldemort, the corresponding cost-effectiveness measures of {\doda} for the three Voldemort executions were zero.
Consider storage costs were not considered in computing the cost-effectiveness because they were almost negligible (only 88MB on average and 200MB at most).

In Figure~\ref{fig:ce1},
the cost-effectiveness (with respect to response time) of {\doda} was about 44\% of that of {\seads} on average.
For individual subject executions, {\doda} and {\seads} had very close cost-effectiveness against the two smallest and simplest subjects, NioEcho and MultiChat.
In previous sections, I observed that the efficiency and scalability advantages of {\seads} over the baseline were more prominent when applied to larger and more complex systems.
This comparative trend applies here also: the cost-effectiveness merits of {\seads} were more substantial with subjects of large scale and more complex executions, compared to {\doda}.
For example, {\doda} cost-effectiveness was about 60\% of that of {\seads} for medium-scale subjects like Open chord and Thrift, while the ratio went down to less than 45\% for larger systems like ZooKeeper and further down to 30\% for even the larger subject Netty.
As a trend extreme, {\doda}'s cost-effectiveness was zero for Voldemort, the most challenging subject in the project.

Figure~\ref{fig:ce2} shows the cost-effectiveness in the alternative measure, concerning the run-time slowdown as the cost factor.
Overall, the cost-effectiveness with respect to this cost measure of {\doda} was only about 32\% of that of {\seads} on average, substantially lower than the cost-effectiveness with respect to response time as shown in Figure~\ref{fig:ce1}.
It was obvious that the efficiency and scalability advantages of {\seads} over {\doda} in terms of run-time slowdown were greater than those in terms of average response time.
On the other hand, compared with the cost-effectiveness variations (with respect to response time)
shown in
Figure~\ref{fig:ce1}, the variations in the cost-effectiveness with respect to run-time slowdown were similarly associated with the scale of the subject executions:
The cost-effectiveness advantages of {\seads} were greater against larger and more complex subjects.

In sum, {\seads} was substantially more cost-effective than {\doda}, regardless of the cost factor concerned with.
This suggests that the precision sacrifices of {\seads} were well paid off by the gains in efficiency and scalability, resulting in the ultimate advantages in cost-effectiveness overall.
Thus, the methodology of adjusting analysis configurations on the fly appeared to be a scalable and cost-effective solution to the dynamic dependence analysis of large-scale distributed systems in the real-world.

\vspace{22pt}
\noindent
\textbf{Configuration learning speed.}
{\seads} uses Q-learning to learn analysis configurations for obtaining and maintaining scalability and high cost-effectiveness of dynamic dependence analysis. 
However, its controller module takes time to learn and hence start producing reasonably good decisions (i.e., the next configuration to switch to).
In fact, as an iterative learning method widely used in approximate dynamic programming for Markov decision processes (MDPs), Q-Learning computes an optimal MDP policy through multiple iterations such that the averaged dynamics can be desired with convergence properties~\cite{peng1994incremental}.
To demonstrate how fast {\seads} can learn cost-effective configurations, I have collected the numbers of Q-learning iterations and learning time before {\seads} started computing cost-effective results (i.e., when the controller started stably choosing the next configuration that led to cost-effective dynamic dependence computation) in my empirical evaluation. 

\setlength{\tabcolsep}{1 pt}
\begin{table}[tp]
  \centering
  \caption{The number of iterations and learning time  (in seconds) of Q-learning}
   \begin{tabular}{l||r|r}
    \hline
    \multicolumn{1}{c||}{\textbf{Execution}} & \multicolumn{1}{c|}{\textbf{\#Iteration}} & \multicolumn{1}{c}{\textbf{Time}} \\
    \hline
    NIOEcho & 1     & 36.73 \\
    \hline
    MultiChat & 1     & 41.38 \\
    \hline
    OpenChord  & 2     & 327.19 \\
    \hline
    Thrift  & 2     & 316.70 \\
    \hline
    xSocket & 3     & 639.37 \\
    \hline
     ZooKeeper\_integration & 4     & 1013.73 \\
    \hline
    ZooKeeper\_load & 4     & 1097.26 \\
    \hline
         ZooKeeper\_system & 4     & 1053.82 \\
    \hline
    Netty & 4     & 1132.98 \\
    \hline
    Voldemort\_integration  & 3     & 697.35 \\
    \hline
         Voldemort\_load  & 3     & 632.42 \\
    \hline
    Voldemort\_system  & 3     & 621.94 \\
    \hline
    \textbf{Overall Average:} & \textbf{3} & \textbf{634.24} \\
   \hline
    \end{tabular}%
  \label{tab:seadsitime}%
\end{table}%

Table~\ref{tab:seadsitime} shows the number of learning iterations ({\em \#Iteration}) and learning time ({\em Time}) in seconds
for each of our 12 subject executions ({\em Execution}).
Generally, {\seads} tended to take more iterations to learn better configurations for more complex system executions, intuitively due to the greater variations in the dynamics of these executions.
Because the learning time included the time cost of dependence computations, 
as shown in Table~\ref{tab:seadscosts},
{\seads} took longer time to compute dependencies and hence responded more slowly to
dependence queries for more complex system executions.
This explains the observation here that the learning time was generally longer as well for those system executions.
On overall average,
{\seads} needed 3 rounds of learning and 634 seconds before it started to achieve cost-effective results.

\subsection{Threats to validity} \label{ss:seadsthreats}

\vspace{2pt}
\noindent
\textbf{Internal validity.}
The major threat to internal validity concerns potential mistakes in the implementation of {\seads}, {\doda}, and our experimental procedure.
Errors in any of these implementations would compromise the validity of our empirical results and our conclusions drawn based on the results.
However, {\seads} is based on Soot~\cite{lam11oct}, a framework that has matured over a decade. Many of the key components of {\seads} and {\doda},
including the code for static instrumentation, static dependence analysis, run-time monitoring/profiling, and hybrid dependence computation, were drawn from several tools (e.g., {\diver}~\cite{cai14diver}, {\diapro}~\cite{cai2017hybrid}, {\distea}~\cite{cai2016distea}) that have been debugged and tuned for years.

To minimize the threats concerning the implementation scripts and newly developed components of {\seads} (e.g., the controller), 
I carefully reviewed code and manually inspected against simple samples (e.g., the two smallest subjects) and cases (e.g., queries with relatively small dependence sets) to ensure the functional correctness of {\seads}.

\vspace{2pt}
\noindent
\textbf{External validity.}
One threat to the external validity lies in the representativeness of the subject subjects and their executions used in the evaluation study.
The subjects selected may not well represent all real-world distributed systems that {\seads} could apply to,
and the executions considered for each chosen subject might not have exercised all typical subject behaviors.
If the differences between the sample subject executions and other distributed system executions are significant,
{\seads} users may experience different performance and merits from what I reported.

I have attempted to reduce this threat by considering subject subjects covering various size, architecture, and application domains, as well as varied execution scenarios.
Nevertheless, to minimize such threats, I would need to use real operations scenarios of real-world distributed systems in their actual deployment settings.

\vspace{2pt}
\noindent
\textbf{Construct validity.}
The major threat to construct validity is that the evaluation
baseline selection may be not ideal.
To avoid potential biases, I used a state-of-the-art online dynamic dependence analysis tool {\doda} (developed by my advisor and me) that at least works with some (if not able to scale to all) real-world distributed systems.
Moreover, to reduce this threat, I ensured that both {\doda} and {\seads} share underlying analysis infrastructure and some utilities.

\vspace{2pt}
\noindent
\textbf{Conclusion validity.}
%
%
The main conclusion validity threat is the generalizability of the evaluation results and conclusions.
Meanwhile, the method-level results of {\seads} versus the baseline for the dynamic dependence analysis may not generalize to finer-grained levels (e.g., statement level).
The reason is that statement-level dynamic analysis would face much great efficiency and scalability challenges.
Due to the limited number and diversity of subject subjects and subject executions considered,
I cannot claim that the results generalize to all real-world distributed systems in all scenarios.

\section{Related Work}
In this section, I discuss prior works closely related to {\em Dependence analysis for distributed software} and  {\em analysis with variable cost-effectiveness}.

\subsection{Dependence analysis for distributed software}
Earlier approaches attempted to extend traditional dependence analysis algorithms to concurrent
programs~\cite{xu2005brief,xiao2005improved,Nanda2006ISM,giffhorn2009precise}.
Most of these approaches are limited to analyzing dependencies within single processes. 
In~\cite{krinke2003context}, a slicing algorithm was proposed which incorporates dependencies due to socket-based message passing.
Since it is purely static, those dependencies are approximated in an overly-conservative manner.

Several dynamic slicing algorithms have been developed too, first for procedural programs only~\cite{korel1992dynamic,cheng1997dependence,goswami2000dynamic,duesterwald1993distributed,kamkar1995dynamic} and later for object-oriented systems as well~\cite{mohapatra2006distributed,barpanda2011dynamic}.
In particular, the approach in~\cite{barpanda2011dynamic} defines varied kinds of dependencies induced by interprocess
communication. However, the approach was not implemented to work on real-world distributed software, and its algorithmic nature
implies scalability barriers.

Reasoning about happens-before relations by addressing global timing via partial ordering based on logic clocks
is a standard technique
in concurrent program analysis.
This technique has been used in testing concurrent programs and 
distributed systems.
For example, DCatch detects concurrency bugs by checking a distributed execution against
a set of happens-before relation rules~\cite{liu2017dcatch}.

\subsection{Analysis with variable cost-effectiveness}
Balancing cost-effectiveness has been a long-standing common challenge to program analysis techniques in general. 
To tackle this challenge,
{\diapro}~\cite{cai2016diapro} offers variable cost-effectiveness tradeoffs for dynamic dependence analysis to satisfy diverse user requirements.
It does so by unifying several previously proposed dependence analysis techniques (e.g., {\pieas}~\cite{apiwattanapong2005efficient}, {\diver}~\cite{cai14diver}) each providing a single, unique cost-effectiveness tradeoff.

In addition, {\textsc{D$^2$Abs}}~\cite{cai2021d2abs} aims at practical scalability, offering various levels of cost-effectiveness tradeoffs in the dynamic dependence analysis for distributed programs.
To achieve several cost-effectiveness tradeoffs,
{\textsc{D$^2$Abs}} provides four versions (e.g., basic version, msg+ version, csd+ version, and scov+ version) to users for
enabling and disabling different analysis steps.


%% file: chapters/chapter6.tex
\chapter{Conclusion and Future Work}  \label{ch:fwk}
\section{Summary}
In my dissertation, I have presented three associated projects ({\distmeasure}, {\flowdist}, and {\seads}) for my overall doctoral research goal:
offering scalable and cost-effective data flow analysis to support quality assurance for common distributed software.
To achieve this goal, I have solved the scalability, cost-effectiveness, applicability, and portability challenges first for a particular security application (i.e, information flow analysis, which is essentially a targeted data flow analysis), and then for more general problems (e.g., self-adaptation, dependence analysis) to support wider applications, such as software maintenance, quality assurance, and so on.

First, in project \textbf{{\distmeasure}} (the expansion of \cite{fu2019measuring}), I defined interprocess communications (IPC) metrics to measure the coupling and cohesion of distributed systems and then to predict and understand their quality.

Second, after understanding distributed systems through IPC metrics, I developed and evaluated a refinement-based dynamic information flow analysis framework, \textbf{{\flowdist}}~\cite{fu2021flowdist} (with two alternative designs), which could overcome applicability, portability, and scalability challenges through a multi-phase analysis strategy.

Lastly, I developed a distributed, online, continuous, and cost-effective dynamic dependence analysis framework \textbf{{\seads}}~\cite{fu2020seads} for continuously running distributed systems to achieve practical scalability and cost-effectiveness through automatically adjusting analysis configurations.
It offers self-tuning cost-effectiveness tradeoffs via a reinforcement learning method.

\section{Future Work}
There are many directions in which I can extend my research work. 
I list three major directions ones below, including {\em fuzzing techniques}, {\em analysis for cloud systems}, and {\em analysis for internet of things (IoT) systems}.   
\subsection{Fuzzing techniques}
Software vulnerabilities have caused serious security problems and significant losses to many industrial areas,
such as aviation, energy, finance, etc~\cite{li2018fuzzing}.
However, my approach {\flowdist} still has a problem that it depends on test inputs to discover vulnerabilities.
If some vulnerabilities are not covered during the execution with all inputs, these vulnerabilities would not be found.
Thus, I will develop techniques to generate more test inputs of {\flowdist} (and other approaches).

\begin{figure*}[htbp]
	\centering
	\caption{An overview of the fuzzing testing workflow}
	\includegraphics[width=1\textwidth]{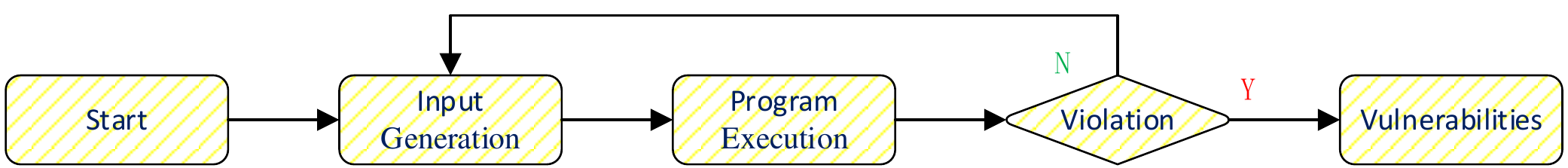}
    \vspace{-15pt}
	\label{fig:fuzzoverview}
\end{figure*}
To discover and then fix vulnerabilities, researchers have presented several techniques, among which fuzzing has been greatly evolved and is becoming the most popular one.
Fuzzing testings first generate a large number of abnormal and normal inputs (i.e., test cases) to target programs, and attempt to find and monitor exceptions (e.g., code assertion failures, crashes, data leaks) by feeding the inputs to the programs~\cite{li2018fuzzing}, as shown in Figure~\ref{fig:fuzzoverview}.
Thus, I will apply fuzzing techniques (e.g., AFL (short for American Fuzzy Lop)~\cite{gutmann2016fuzzing}) to information flow analysis for detecting more vulnerabilities in common distributed software.

\subsection{Analysis for cloud systems}
Cloud system are the applications of distributed systems~\cite{wang2010distributed}.
They are rapidly increasing for offering the platforms and software as services~\cite{nassif2021machine} provided by Amazon Web Service (AWS), Google Cloud Platform, IBM Cloud Services, Microsoft Azure, Oracle Cloud, Salesforce, SAP, etc.
Various services and applications have been deployed in these cloud systems, such as e-commerce, storage, high-performance computing, and so on~\cite{wang2010distributed}.
Users only need to send the requests to and then wait for the results to be returned from cloud providers, without knowing how cloud services/platforms run~\cite{li2012neural}.

However, there are some challenges preventing cloud systems and applications, such as security, performance, availability, and so on~\cite{dillon2010cloud}.
Analysis techniques/tools will focus on dealing with these challenges.
For example, various ways will be used to detect/prevent security gaps/attacks in cloud~\cite{li2012neural}.
I will explore techniques that analyze and/or can be deployed on cloud systems, addressing these challenges.

\subsection{Analysis for Internet of Things (IoT) systems}
Using embedded technologies, Internet of Things (IoT) defines an environment where various physical objects can interact and cooperate with each other~\cite{dorodchi2016trust}. 
IoT systems unify real-world objects to help people perceive/control of relevant things.
 
There are lot of the development and applications of IoT over the past years, such as Connected Cars, Smart Cities, Smart Grids, Smart Factories, etc.
And many IoT systems/platforms have been developed and deployed, such as AWS IoT, IBM Watson IoT Platform, Microsoft Azure IoT Hub, OpenMTC, SiteWhere, and so on~\cite{madakam2015internet}.

In recent years, academic research focuses on address the security challenges (e.g., confidentiality, integrity, key management, policy enforcement, privacy) for IoT systems.
Traditional technologies (e.g., cryptography) and new methods (e.g., Blockchain, Software Defined Network (SDN)) are implemented to solve current IoT security challenges~\cite{hassan2019current}.
Besides IoT security, there are some future research directions for IoT, such as data analysis, performance modelling, and monitoring.
The IoT can transform, aggregate, store, and analyze its data in the cloud that provides several benefits such as advanced security, low-cost resources, measurement service, resource consolidation, self-service access, and so on~\cite{el2019performance}.
Therefore, I plan to develop analysis approaches to IoT systems.